\documentclass[12pt,pdftex]{article}
\usepackage{graphicx,psfrag,epsfig}
\usepackage{subfigure}
\usepackage{colortbl}
\graphicspath{{figures/} }
\usepackage{nccmath}
\usepackage{float}
\usepackage{natbib}
\usepackage{comment}
\usepackage{multirow}
\usepackage{tikz}
\usepackage{epsfig}
\usepackage{graphicx}
\usepackage{setspace}
\usepackage{amsfonts}
\usepackage{hyperref}
\usepackage{latexsym}
\usepackage{amsmath,amssymb,amsthm}
\usepackage{bm}
\usepackage{bbm}
\usepackage{booktabs}
\usepackage{parskip}
\usepackage{mathtools}
\usepackage{cleveref}
\usepackage{rotating}
\newcommand\reallywidehat[1]{\arraycolsep=0pt\relax%
\begin{array}{c}
\stretchto{
  \scaleto{
    \scalerel*[\widthof{\ensuremath{#1}}]{\kern-.5pt\bigwedge\kern-.5pt}
    {\rule[-\textheight/2]{1ex}{\textheight}} 
  }{\textheight} %
}{0.5ex}\\           
#1\\                 
\rule{-1ex}{0ex}
\end{array}
}
\newcommand{\reals}{\mathbb{R}}
\newcommand{\E}{\mathrm{E}}
\newcommand{\Var}{\mathrm{Var}}

\newcommand{\bsomega}{\boldsymbol{\omega}}

\newcommand{\bsTheta}{\boldsymbol{\Theta}}

\newcommand{\bsbeta}{\boldsymbol{\beta}}
\newcommand{\bseta}{\boldsymbol{\eta}}
\newcommand{\bsalpha}{\boldsymbol{\alpha}}
\newcommand{\bspsi}{\boldsymbol{\psi}}
\newcommand{\bsphi}{\boldsymbol{\phi}}

\newcommand{\bszeta}{\boldsymbol{\zeta}}
\newcommand{\bsgamma}{\boldsymbol{\gamma}}

\newcommand{\bfJ}{\mbox{\boldmath $J$}}
\newcommand{\bfz}{\mbox{\boldmath $z$}}
\newcommand{\bfx}{\mbox{\boldmath $x$}}

\newcommand{\bfy}{\mbox{\boldmath $y$}}

\newcommand{\bfY}{\mbox{\boldmath $Y$}}

\newcommand{\bfI}{\mbox{\boldmath $I$}}

\newcommand{\bfzeta}{\mbox{\boldmath $\zeta$}}

\newcommand{\bfphi}{\mbox{\boldmath $\phi$}}

\newcommand{\bfomega}{\mbox{\boldmath $\omega$}}

\newcommand{\bfpsi}{\mbox{\boldmath $\psi$}}

\newcommand{\norm}[1]{\left\lVert#1\right\rVert}

\theoremstyle{plain} 
\theoremstyle{plain} 
\theoremstyle{definition} 
\theoremstyle{plain} 
\theoremstyle{definition} 
\mathchardef\mhyphen="2D

\title{A Bayesian Approach to Spherical Factor Analysis for Binary Data}
\author{Xingchen Yu, Abel Rodriguez\\
Department of Statistics\\
University of California, Santa Cruz\\
}
\begin{document}

\maketitle


\abstract

Factor models are widely used across diverse areas of application for purposes that include dimensionality reduction, covariance estimation, and feature engineering. Traditional factor models can be seen as an instance of linear embedding methods that project multivariate observations onto a lower dimensional Euclidean latent space.  This paper discusses a new class of geometric embedding models for multivariate binary data in which the embedding space correspond to a spherical manifold, with potentially unknown dimension.  The resulting models include traditional factor models as a special case, but provide additional flexibility.  Furthermore, unlike other techniques for geometric embedding, the models are easy to interpret, and the uncertainty associated with the latent features can be properly quantified.  These advantages are illustrated using both simulation studies and real data on voting records from the U.S.\ Senate.

\newpage
\section{Introduction}

Factor analysis \citep{child2006essentials,mulaik2009foundations} has a long history that dates back at least to the pioneering work of Charles Spearman on intelligence during the first decade of the 20\textsuperscript{th} century.  Factor analysis is widely used across all sorts of disciplines within the social and natural sciences to account for and explain the correlations observed across multivariate responses.  In traditional factor analysis, which in the case of normally distributed data includes principal component analysis as a special case, (the mean of) each response variable is represented as a linear combination of a common set of subject-specific \textit{factors}.  These factors can be interpreted as providing a linear embedding of the original high-dimensional data into a low-dimensional Euclidean space.  These embeddings can be useful for various purposes, including data description, sparse estimation of covariance matrices, and/or as inputs to predictive models.

Linear embeddings on Euclidean spaces are relatively easy to compute and interpret.  However, they might not be appropriate in all circumstances, particularly when the underlying  data-generation mechanism involves highly non-linear phenomena.  Over the last 30 years, a rich literature has developed covering the use of nonlinear embeddings and data-driven, non-Euclidean embedding manifolds.  Auto-encoders (e.g., see \citealp{kramer1991nonlinear}, \citealp{xu2018spherical}, \citealp{davidson2018hyperspherical} and \citealp{kingma2019introduction}) are a natural generalization of principal component analysis.  They explain the structure in the data through the composition of two non-linear functions (often represented as neural networks). The first function (usually called the \textit{encoder}) embeds the input data into the low-dimensional latent variables, while the second function (called the \textit{decoder}) maps those latent variables into the original space of the data in a way that minimizes the reconstruction error.   Alternatively, locally-linear embeddings \citep{roweis2000nonlinear} reconstruct each input as a weighted average of its neighbours, and subsequently map the data into a low-dimensional embedding using those same weights. Isomap \citep{tenenbaum2000global} extends the classical multidimensional scaling method into a general, data-driven manifold using geodesic distances measured on a neighbourhood graph. The goal is to preserve the intrinsic geometry of the manifold in which the data lies. Laplacian eigenmaps \citep{belkin2002laplacian} also seek to preserve the intrinsic geometry of local neighborhoods.  The  embedding eigenvectors are obtained as the solution of a generalized eigenvalue problem based on the Laplacian matrix of the neighbourhood graph.  Local tangent space alignment \citep{Zhang04principalmanifolds} finds the local geometry through the tangent space in the neighbourhood of inputs from which a global coordinate system for a general manifold is constructed. The inputs are embedded into a low-dimensional space through such global coordinate system. Finally, Gaussian process latent variable models \citep{lawrence2005probabilistic} use Gaussian process priors to model the embedding function. These various approaches are very flexible in capturing the shape of the relationship between latent variables and outcomes, as well as the geometry of the underlying manifold on which the data lives.  As a consequence, they can produce very compact representations of high-dimensional data using a very small number of latent dimensions. However, interpreting and quantifying the uncertainty associated with these embeddings can be quite difficult because of identifiability issues.  Indeed, when performing non-linear embeddings, invariance to affine transformations is not enough to ensure identifiability of the latent features.  When the goal is prediction or sparse estimation of covariance matrices, this does not matter as these quantities are (usually) identifiable functions of the unidentified latent factors.  However, when interest lies in the embedding coordinates themselves (as is the case in many social sciences applications), the lack of identifiability means that we are in the presence of irregular problems in which standard techniques for generating confidence/credible intervals are not applicable.  Similarly, interpretation of the latent positions is dependent on the exact geometry of the embedding space, which is in turn only partially specified unless the lack of identifiability is addressed.

An alternative to the nonparametric embedding methods discussed above is to consider more flexible (but fixed) embedding spaces for which identifiability constraints can be easily derived.  Such approach trades off some of the flexibility of the nonparametric methods for interpretability and the ability to properly quantify the uncertainty associated with the embedding.  Along these lines, this manuscript proposes a general framework for embedding multivariate binary data into a general Riemannian manifold by exploiting the random utility formulation underlying binary regression models \citep{marschak1960binary,mcfadden1973conditional}.  To focus ideas, and because of their practical appeal, we place a special emphasis on spherical latent spaces.
Most of our attention in this paper is focused on two key challenges.  
The first one is eliciting prior distribution for model parameters that do not degenerate as the dimension of the embedding space grows.  Well calibrated priors, in the previous sense, are key to enable dimensionality selection.  The second challenge is designing efficient computational algorithms for a model in which some of the parameters live on a Riemannian manifold.

It is worthwhile noting that the literature has considered generalizations of data-reduction techniques such as principal components and factor analysis to situations in which the observations live on a manifold.  Examples include principal Geodesic Analysis \citep{fletcher2004principal,sommer2010differential,NIPS2013_5133} and principal nested spheres \citep{jung2012analysis}.  This literature, however, is only marginally relevant to us since it is the parameters of our model, and not the data itself, that are assumed to live on a spherical manifold.

The remainder of the paper is organized as follows:  Section \ref{se:euclideanmodel} reviews traditional factor analysis models for binary data, including their random utility formulation. Section \ref{se:spherical} introduces our spherical factor models and discusses the issues associated with prior elicitation.  Section \ref{se:comput} discusses our computational approach to estimating model parameters, which is based on Geodesic Hamiltonian Monte Carlo algorithms.  Section \ref{se:illustration} illustrates the behavior of our model on both simulated and real datasets.  Finally, Section \ref{se:conclusion} concludes the paper with a discussion of future directions for our work.

\section{Bayesian factor analysis models for binary data:  A brief overview}\label{se:euclideanmodel}

Consider data consisting of independent multivariate binary observations $\bfy_1, \ldots, \bfy_I$ associated with $i$ \textit{subjects}, where $\bfy_i = (y_{i,1}, \ldots, y_{i,J})^T$ is a vector in which each entry is associated with a different \textit{item}.  In the political science application we discuss in Section \ref{se:illustration}, $y_{i,j}$ represents the vote of legislator (subject) $i$ in question (item) $j$ (with $y_{i,j}=1$ corresponding to an affirmative vote, and $y_{i,j}=0$ corresponding to a negative one).  On the other hand, in marketing applications, $y_{i,j}$ might represent whether consumer (subject) $i$ bought product (item) $j$ (if $y_{i,j}=1$) or not (if $y_{i,j}=0$).

A common approach to modeling this type of multivariate data relies on a generalized bilinear model of the form
\begin{align}\label{eq:IDEALglm}
\Pr \left( y_{i,j} = 1 \mid \mu_j,\bsalpha_j, \bsbeta_i \right) = G \left(\mu_j+\bsalpha_j^T\bsbeta_i \right),
\end{align}
where $G$ is the (known) link function, and the intercepts $\mu_1, \ldots, \mu_J$ as well as the bilinear terms $\bsalpha_1, \ldots, \bsalpha_J \in \reals^K$ and $\bsbeta_1, \ldots \bsbeta_I \in \reals^K$ are all unknown and need to be estimated from the data (e.g., see \citealp{bartholomew1984scaling}, \citealp{legler1997latent}, \citealp{ansari2000bayesian} and \citealp{jackman2001multidimensional}).   Different choices for the link function $G$ lead to well-known classes models, such as logit and probit models.  Furthermore, given a distribution over the latent factors $\bsbeta_1, \ldots, \bsbeta_I$, integrating over them (which, in the case of binary data, can be done in closed form only in very special cases) yields a wide class of correlation structures across items.

The class of factor analysis models for binary data in \eqref{eq:IDEALglm} can be derived through the use of random utility functions \citep{marschak1960binary, mcfadden1973conditional}.  To develop such derivation, we interpret the latent trait $\bsbeta_i \in \reals^K$ as  representing the preferences of subject $i$ over a set of unobserved item characteristics, and associate with each item two positions, $\bspsi_j\in \reals^K$ (corresponding to a positive responsive, i.e., $y_{i,j}=1$) and $\bszeta_j\in \reals^K$ (corresponding to a negative one, i.e., $y_{i,j}=0$).  Assuming that individuals make their choice for each item based on the relative value of two random quadratic utilities,  
\begin{align}\label{eq:utlitiesEuclidean}
U_{+}(\bspsi_j, \bsbeta_i)& = -\norm{\bspsi_j-\bsbeta_i}^2 + \epsilon_{i,j}, &  
U_{-}(\bszeta_j, \bsbeta_i) &= -\norm{\bszeta_j-\bsbeta_i}^2 + \nu_{i,j}, 
\end{align}
where $\epsilon_{i,j}$ and $\nu_{i,j}$ represent random shocks to the utilities, and $\upsilon_{i,j} = \nu_{i,j} - \epsilon_{i,j}$ are independently distributed for all $i$ and $j$ and have cumulative distribution function $G_j(x)=G(x/\sigma_j)$, 
it is easy to see that 
$$
\Pr\left( y_{i,j} = 1 \mid \bspsi_j, \bszeta_j, \bsbeta_i, \sigma_j \right) = \Pr\left( U_{+}\left(\bspsi_j, \bsbeta_i\right) > U_{-}\left(\bszeta_j, \bsbeta_i\right) \right) = G\left(\mu_j+\bsalpha_j^T\bsbeta_i\right) ,
$$
where $\bsalpha_j=2(\bspsi_j-\bszeta_j)/\sigma_j$ and $\mu_j=(\bszeta_j^T\bszeta_j-\bspsi_j^T\bspsi_j)/\sigma_j$.  This formulation, which is tightly linked to latent variable representations used to fit categorical models (e.g., see \citealp{albert1993bayesian}), makes it clear how the factor model embeds the binary observation into a Euclidean latent space.  Furthermore, this construction also highlights a number of identifiability issues associated with the model.  In particular, note that the utility functions in \eqref{eq:utlitiesEuclidean} are invariant to affine transformations.  Therefore, the positions $\bspsi_j$, $\bszeta_j$ and $\bsbeta_i$ are only identifiable up to translations, rotations and rescalings, and the scaling $\sigma_j$ cannot be identified separately from the scale of the latent space.  These identifiability issues are usually dealt with by fixing $\sigma_j=1$ for all $j=1,\ldots,J$, and by
either fixing the position of $K+1$ legislators (e.g., see \citealp{clinton2004statistical}), or by fixing the location and scale of the ideal points, along with constraints on the matrix of discrimination parameters (e.g., see \citealp{geweke1981maximum}).  In either case, identifiability constraints can be enforced either a priori, or a posteriori using a parameter expansion approach (e.g., see \citealp{liu1999parameter}, \citealp{rubin2001using} and \citealp{hoff-2002}).

Bayesian inference for this class of models requires the specification of prior distributions for the various unknown parameters.  For computational simplicity, it is common to use (hierarchical) priors for the  $\mu_j$s, $\bsalpha_j$s and $\bsbeta_i$s, so that computation can proceed using well-established Markov chain Monte Carlo algorithms (e.g., see \citealp{albert1993bayesian} and \citealp{bafumi13andrew}).  However, the hyperparameters of these priors need to be selected carefully.  For example, when the dimension $K$ of the embedding space is unknown and needs to be estimated from the data, it is important to ensure that the prior variance on $\theta_{i,j} = G(\mu_j+\bsalpha_j^T\bsbeta_i)$ induced by these priors remains bounded as $K \to \infty$ if one is to avoid Bartlett's paradox \citep{bartlett1957comment}.  One approach to accomplish this goal is to select the prior covariance matrix for the $\bsbeta_i$s to be diagonal, and to have the marginal variance of its components decrease fast enough as more of them are added.\footnote{As an example, consider a probit model (i.e., $G(x) = \Phi(x)$) where, a priori, $\mu_j \sim N(0,1/2)$, $\alpha_{j,k} \sim N(0,1/2)$, and $\beta_{j,k} \sim N(0,6/[\pi
k]^2)$. In this case, $\theta_{i,j} = \Phi\left(\mu_j + \sum_{k=1}^K \alpha_{j,k}  \beta_{i,k} \right)$ converges in distribution to a uniform distribution on $[0,1]$ as $K\to \infty$ (see the online supplementary materials.}  This type of construction, which is related but distinct from the one implied by the Indian Buffet process \citep{griffiths2011indian}, provides a prior that is consistent as the number of components $K$ grows, and which can be interpreted as a truncation of an infinite dimensional model.  Assigning prior distributions to the $\bspsi_j$s and $\bszeta_j$s (which, in turn, imply priors on the $\bsalpha_j$s and $\mu_j$s) is also a possibility, but it is rarely done in practice. In that setting, ensuring a satisfactory behavior as the number of dimensions grows typically requires that the variances of all three latent positions decrease as the number of dimensions of the latent space increases.  This observation will be important to some of the developments in Section \ref{se:sphericalprior}.
\section{Bayesian factor models for binary data on spherical spaces} \label{se:spherical}

\subsection{Likelihood formulation}\label{se:likelihood}

The formulation of classical factor models for binary data in terms of random utility functions described in Section \ref{se:euclideanmodel} lends itself naturally to extensions to more general embedding spaces.  In particular, we can generalize the model by embedding the positions $\bsbeta_i$, $\bspsi_j$ and $\bszeta_j$ into a connected Riemannian manifold $\mathcal{D}$ equipped with a metric $\rho : \mathcal{D} \times \mathcal{D} \to M \subseteq \reals^+$, and then rewriting the utility functions in \eqref{eq:utlitiesEuclidean} as
\begin{align}\label{eq:utlities}
U_{+}(\bspsi_j, \bsbeta_i)& = -\left\{ \rho\left( \bspsi_j, \bsbeta_i \right) \right\}^2 + \epsilon_{i,j}, &  
U_{-}(\bszeta_j, \bsbeta_i) &= -\left\{ \rho\left( \bszeta_j, \bsbeta_i \right) \right\}^2 + \nu_{i,j} .
\end{align}
Assuming that the errors $\epsilon_{i,j}$ and $\nu_{i,j}$ are such that their differences $\upsilon_{i,j} = \nu_{i,j} - \epsilon_{i,j} $ are independent with cumulative distribution function $G_j$, this formulation leads to a likelihood function of the form
$$
p\left(\bfY \mid \{ \bspsi_j \}_{j=1}^{J}, \{ \bszeta_j \}_{j=1}^{J}, \{ \bsbeta_i \}_{i=1}^{I} \right)= \prod_{i=1}^{I}\prod_{j=1}^{J} \theta_{i,j}^{y_{i,j}} \left(1 - \theta_{i,j} \right)^{1-y_{i,j}},
$$
where $\bfY$ is the $I \times J$ matrix of observations whose rows correspond to $\bfy_{i}^T$, and 
\begin{align}\label{eq:difdist}
\theta_{i,j} &= G_{j}\left( e\left(\bspsi_j, \bszeta_j, \bsbeta_i\right) \right), & e\left(\bspsi_j, \bszeta_j, \bsbeta_i\right) &= \left\{ \rho\left( \bszeta_j, \bsbeta_i \right) \right\}^2-\left\{ \rho\left( \bspsi_j, \bsbeta_i \right) \right\}^2 .
\end{align}

In this paper, we focus on the case in which $\mathcal{D}$ corresponds to the unit hypersphere in $K+1$ dimensions, $\mathcal{S}^{K}$, which is equipped with its geodesic distance $\rho_K$, the shortest angle separating two points measured over the great circle connecting them.  There are two alternative ways to describe such a space.  The first one uses a hyperspherical coordinate system and relies on a vector of $K$ angles, $\bfphi = (\phi_1, \ldots, \phi_{K}) \in [-\pi, \pi] \times [-\pi/2, \pi/2]^{K-1}$.  The second one uses the fact that $\mathcal{S}^{K}$ is a submanifold of $\reals^{K+1}$, and relies on a vector of $K+1$ Cartesian coordinates $\bfx = (x_{1}, \ldots, x_{K+1})$ subject to the constraint $\left\|\bfx\right\| = 1$.  The two representations are connected through the transformation
\begin{equation}
\begin{aligned}\label{eq:sph_coord}
x_1=&\cos \phi_{1} \cos\phi_{2} \cos\phi_{3} \cdots  \cos\phi_{K-1},\\
x_2=&\sin\phi_{1} \cos\phi_{2} \cos\phi_{3} \cdots \cos\phi_{K-1},\\
x_3=& \sin\phi_{2} \cos \phi_{3} \dots \cos\phi_{K-1}\\
\vdots\\
x_{K}=&\sin \phi_{K-1} \cos \phi_{K},\\
x_{K+1}=& \sin \phi_{K}.
\end{aligned}
\end{equation}

While the hyperspherical coordinate representation tends to be slightly more interpretable and directly reflects the true dimensionality of the embedding space, Cartesian coordinates often result in more compact expressions. For example the distance metric $\rho_K$ can be simply written as $\rho_K (\bfx, \bfz)=\arccos\left(\bfx^T \bfz\right)$.  Furthermore, the Cartesian coordinate representation will be key to the development of our computational approaches.  Notation-wise, in the sequel, we adopt the convention of using Greek letters when representing points in $\mathcal{S}^{K}$ through their angular representation, and using Roman letters when representing them through embedded Cartesian coordinates. 

Lastly, we discuss the selection of the link function. $G_j$ must account for the fact that, when $\mathcal{S}^K$ is used as embedding space and $\rho_K$ as the distance metric, the function $e\left(\bspsi_j, \bszeta_j, \bsbeta_i\right)$ introduced in \eqref{eq:difdist} has as its range the interval $[-\pi^2, \pi^2]$.  While various choices are possible, we opt to work with the cumulative distribution of a shifted and scaled beta distribution,
\begin{align}\label{eq:linkcircular}
G_{j}(z) = G_{\kappa_j}(z) & =\int_{-\pi^2}^{z}\frac{1}{2\pi^2} \frac{\Gamma(2\kappa_j)}{\Gamma(\kappa_j)\Gamma(\kappa_j)}\Bigg(\frac{\pi^2+z}{2\pi^2}\Bigg)^{\kappa_j-1}\Bigg(\frac{\pi^2-z}{2\pi^2}\Bigg)^{\kappa_j-1}, & z &\in [-\pi^2,\pi^2].
\end{align}
Note that $1/\sqrt{\kappa_j}$, which controls the dispersion of the density $g_j$ associated with $G_j$, plays an analogous role to the one that $\sigma_j$ played in Section \ref{se:euclideanmodel}.  In fact, note that
\begin{align}\label{eq:linklimit}
\lim_{\kappa_j \to \infty} \frac{g_{\kappa_j}(z)}{ 
 \sqrt{\frac{2\kappa_j +1}{2\pi^5}} \exp\left\{- \frac{2\kappa_j +1}{\pi^4} z^2 \right\} } = 1  .
\end{align}
This observation, together with the fact that $\rho_K (\bfx, \bfz)=\arccos\left(\bfx^T \bfz\right) \approx \left\| \bfx - \bfz \right\|$ for small values of $\left\| \bfx - \bfz \right\|$, make it clear that the projection of the likelihood of a spherical model in $\mathcal{S}^{K}$ on its tangent bundle is very close to a version of the Euclidean factor model in $\reals^K$ described in \eqref{eq:IDEALglm} in which the link function is the probit link.  We expand on this connection in Section \ref{se:coneceuclid}

Finally, a short note on the connection between the likelihood functions associated with two models defined in $\mathcal{S}^{K}$ and $\mathcal{S}^{K-1}$, respectively.  Let $\bsbeta, \bspsi \in \mathcal{S}^{K}$ and $\bsbeta^{*}, \bspsi^{*} \in \mathcal{S}^{K-1}$ be (the angular representations of) two pairs of points in spherical spaces of dimension $K$ and $K-1$, respectively.  It is easy to verify that, when $\beta_{K} = \psi_{K} = 0$ as well as $\beta_k = \beta_k^{*}$ and $\psi_k = \psi_k^{*}$ for all $k=1, \ldots, K-1$, then $\rho_{K}(\bsbeta, \bspsi) = \rho_{K-1}(\bsbeta^{*}, \bspsi^{*})$.  Hence, the likelihood function of a spherical factor model in which the latent positions are embedded in $\mathcal{S}^{K}$ and for which $\beta_{i,K} = \psi_{j,K} = \zeta_{j,K} = 0$ for all $i$ and $j$, is identical to the likelihood function that would be obtained by directly fitting a model in which the latent positions are embedded in $\mathcal{S}^{K-1}$.  This observation will play an important role in the next Section, which is focused on the defining prior distributions for the latent positions.

\subsection{Prior distributions}\label{se:sphericalprior}

As we discussed in the introduction, a key challenge involved in the implementation of the spherical factor models we just described is selecting priors for the latent positions $\bspsi_1, \ldots, \bspsi_J$ and $\bszeta_1, \ldots, \bszeta_J$ and $\bsbeta_1, \ldots, \bsbeta_J$.  This challenge is specially daunting in settings where estimating the dimension $K$ of the latent space is of interest.

To illustrate the issues involved, consider the use of the von Mises–Fisher family as priors for the latent positions. The Hausdorff density with respect to the uniform measure on the sphere for the von Mises–Fisher distribution on $\reals^d$ is given by
\begin{align*}
p_{\mathcal{H}}(\bfx \mid \bseta, \omega) &= \frac{\omega^{d/2-1}}{(2\pi)^{d/2} I_{d/2-1}(\omega)} \exp\left\{ \omega \bseta^T \bfx \right\}  ,  & \left\| \bfx \right\| &= 1 ,
\end{align*}
where $I_{\nu}(\cdot)$ is the modified Bessel function of order $\nu$, $\bseta$ satisfies $\left\| \bseta \right\|=1$ and represents the mean direction, and $\omega$ is a scalar precision parameter.  The von Mises–Fisher distribution is a natural prior in this setting because it is the spherical analogue to the multivariate Gaussian distribution with diagonal, homoscedastic covariance matrix that is sometimes used as a prior for the latent positions in traditional factor models.  Indeed, note that
$$
\lim_{\omega \to \infty} \frac{\frac{\omega^{d/2-1}}{(2\pi)^{d/2} I_{d/2-1}(\omega)} \exp\left\{ \omega \bseta^T \bfx \right\}}
{\left( \frac{\omega}{2 \pi}\right)^{d/2} \exp\left\{ - \frac{\omega}{2} (\bfx - \bseta)^T(\bfx - \bseta) \right\}} = 1 .
$$

\begin{figure}
\begin{tabular}{cccc}
    & $\omega = 0.5$ & $\omega = 2$ & $\omega = 10$ \\ 
    \begin{sideways} \;\;\;\;\;\;\;\;\;\;\;\;\;\;\;\;\;\;\; $\tau = 0.5$\end{sideways}&
    \includegraphics[width=.3\textwidth]{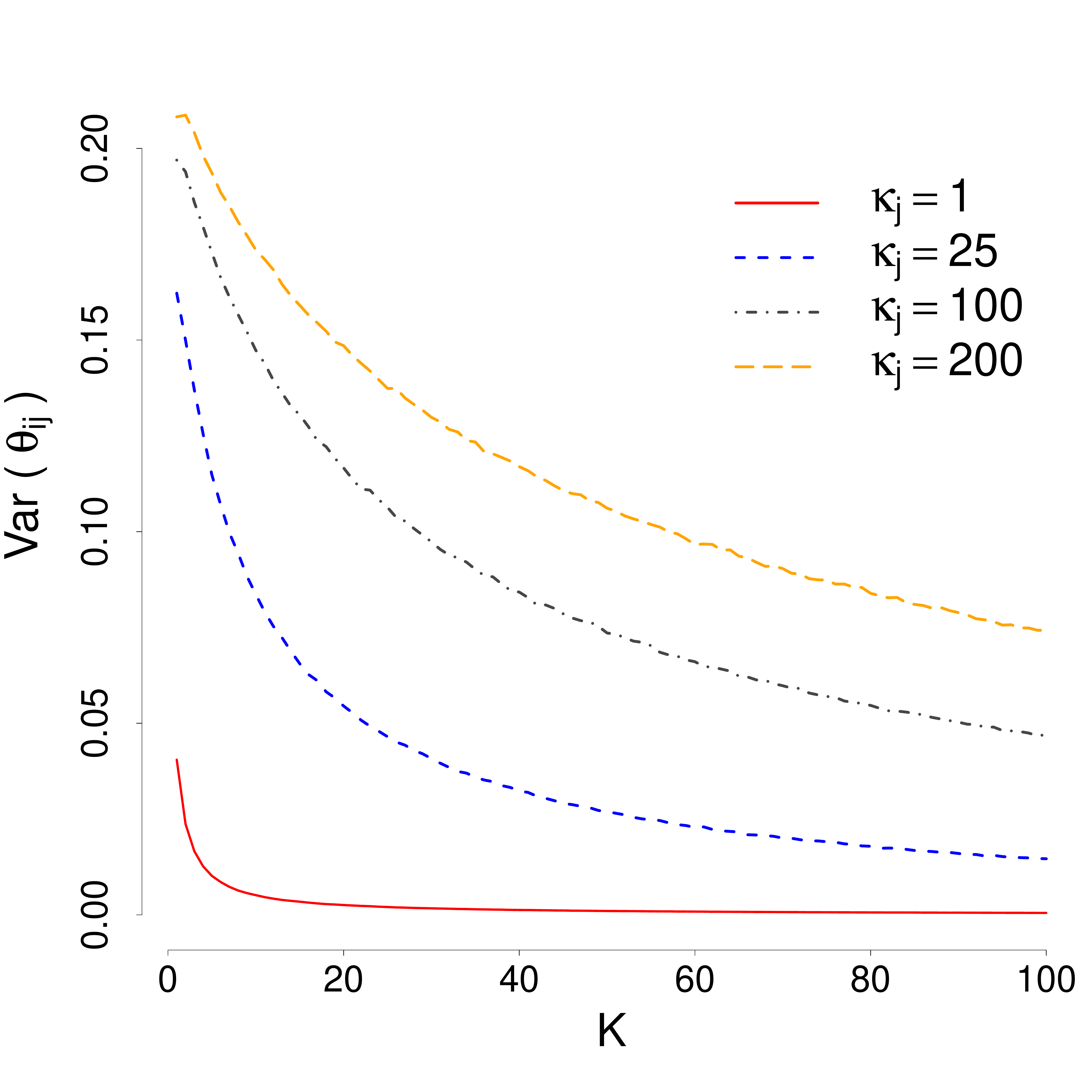} &
    \includegraphics[width=.3\textwidth]{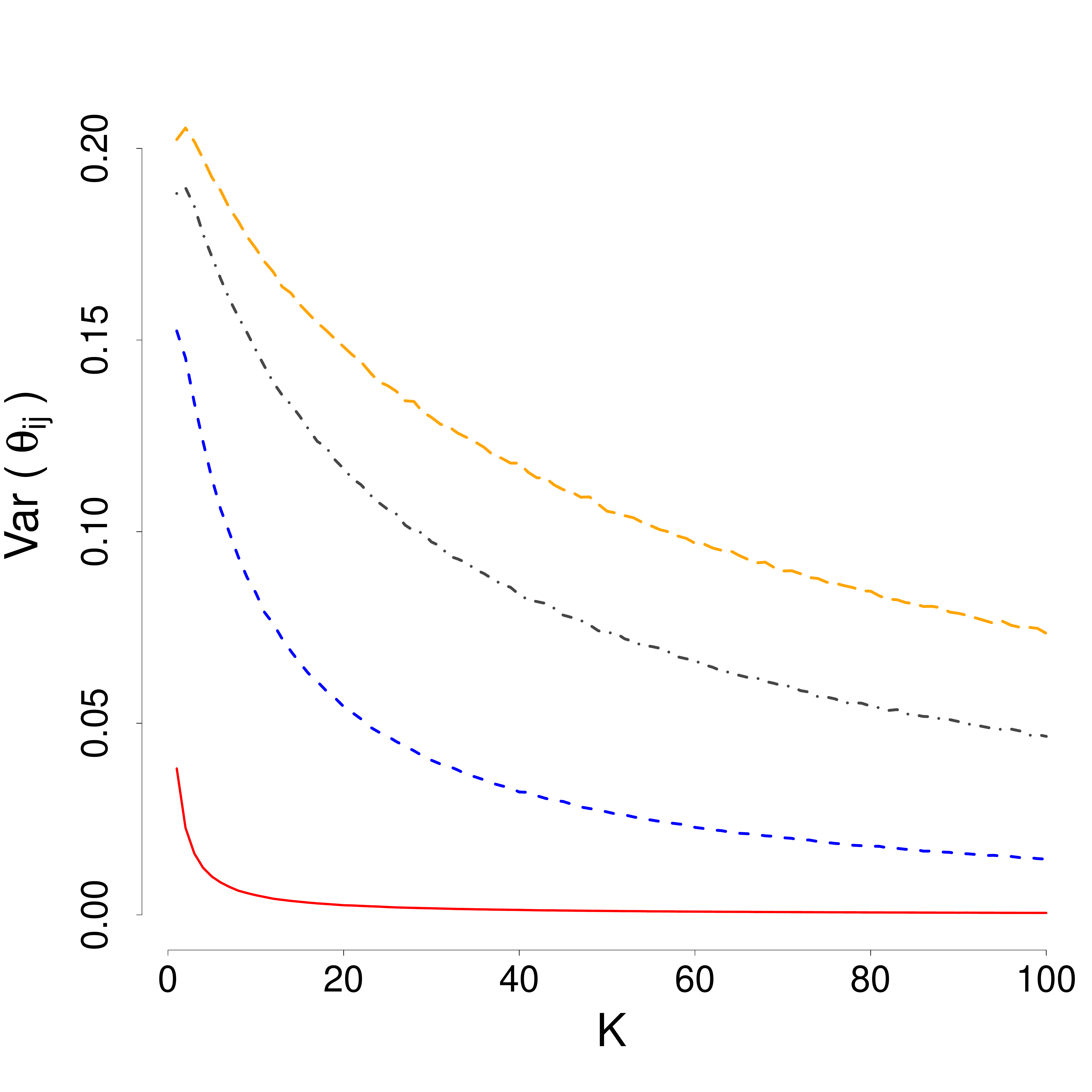} &
    \includegraphics[width=.3\textwidth]{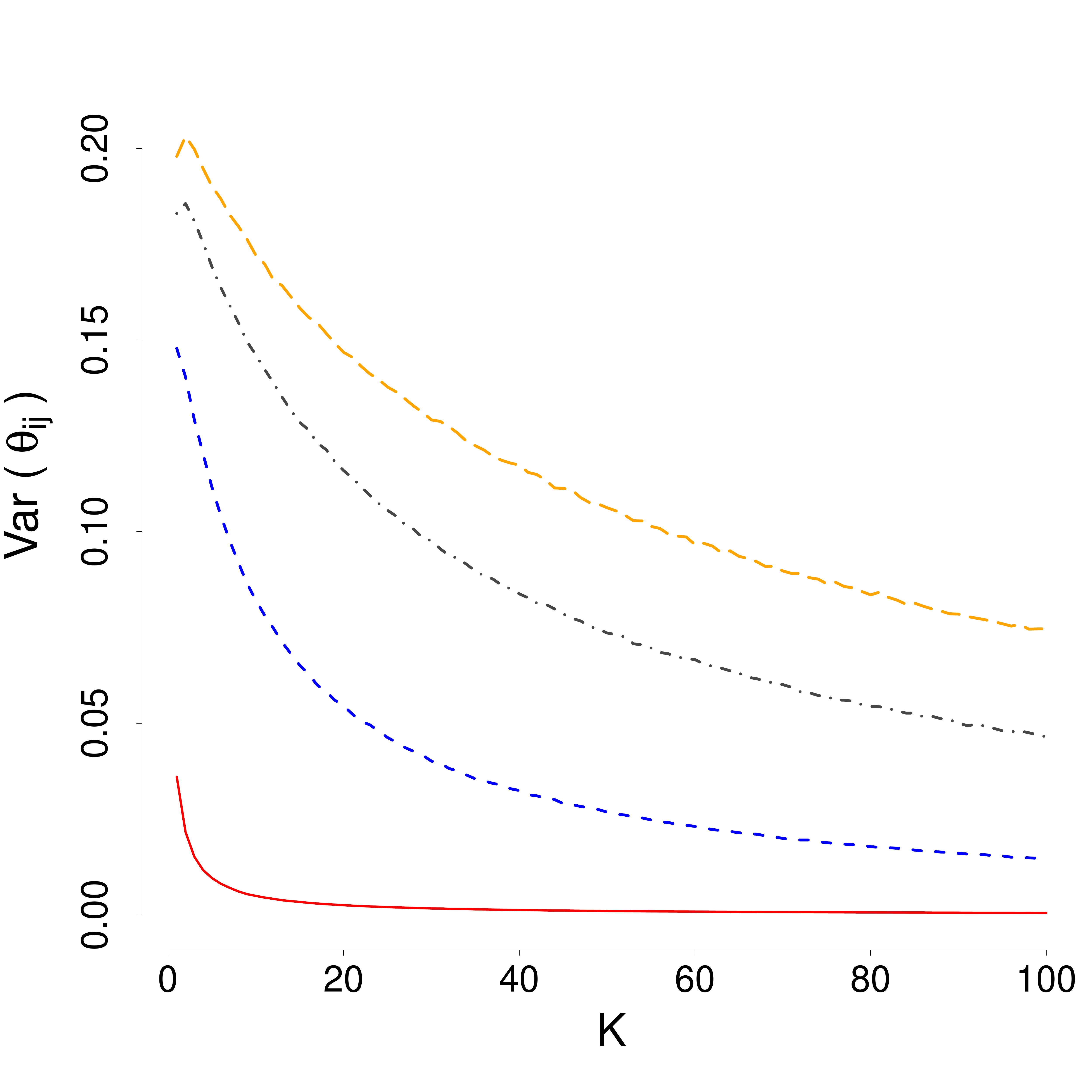} \\
    \begin{sideways} \;\;\;\;\;\;\;\;\;\;\;\;\;\;\;\;\;\;\;\;\, $\tau = 2$ \end{sideways} & 
    \includegraphics[width=.3\textwidth]{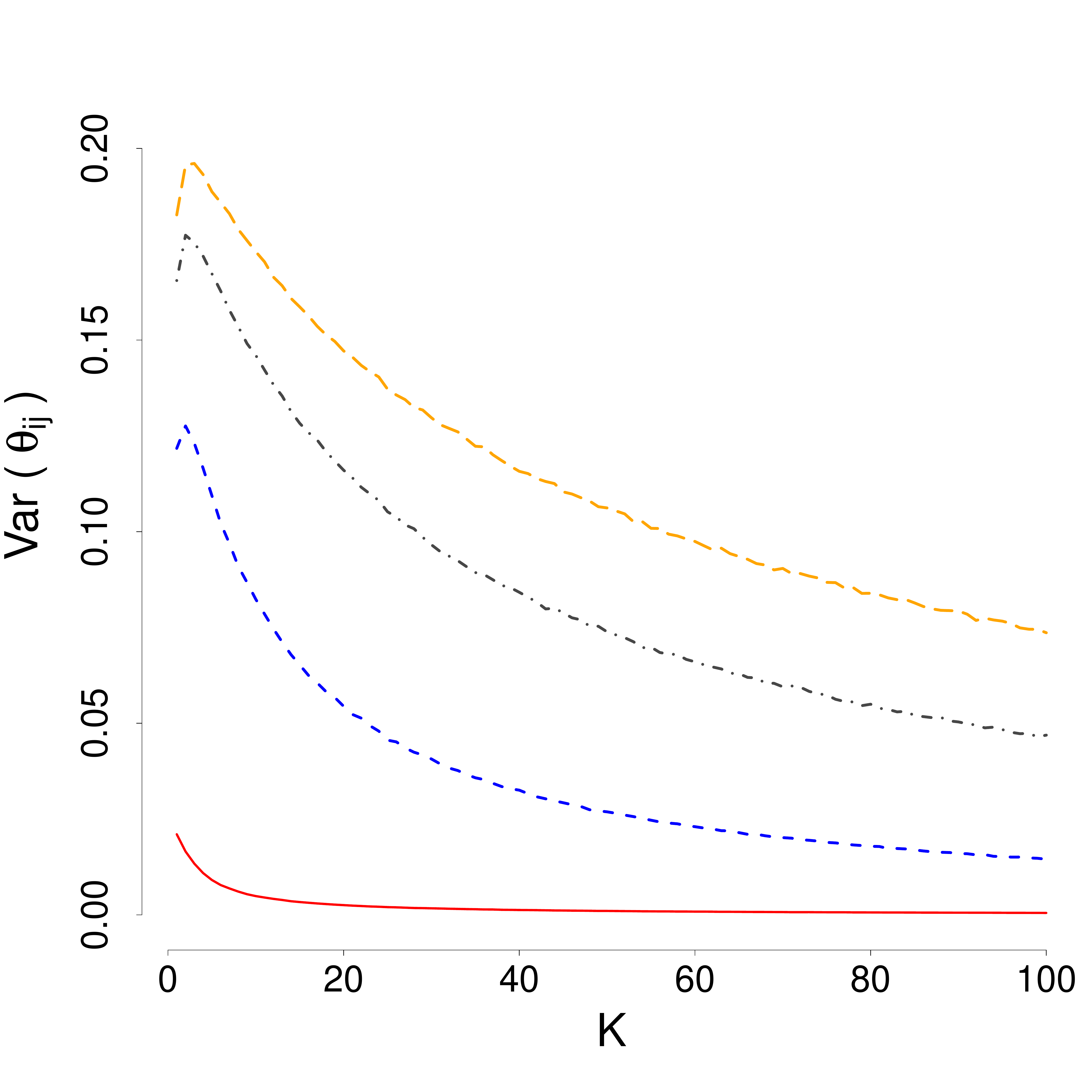} &
    \includegraphics[width=.3\textwidth]{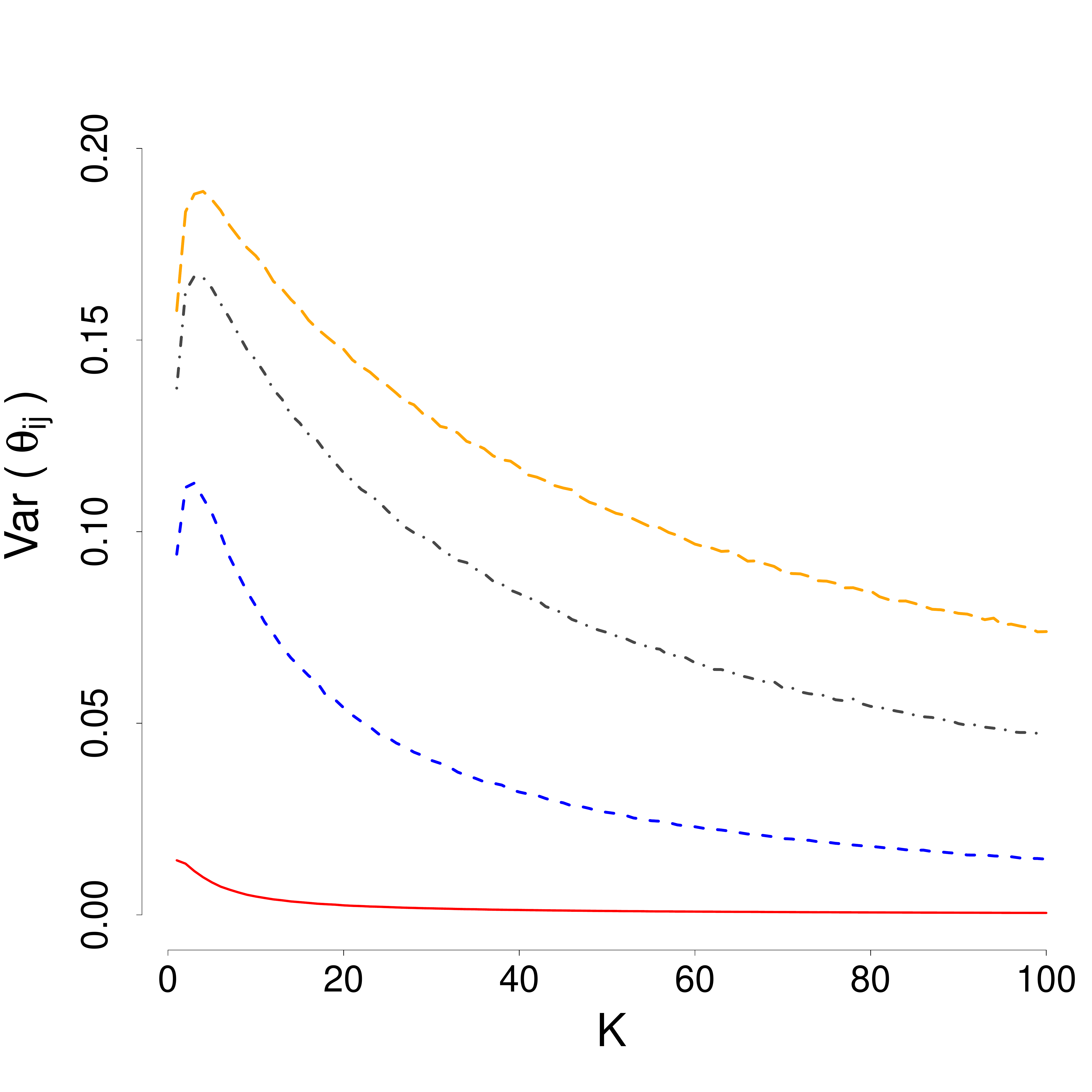} &
    \includegraphics[width=.3\textwidth]{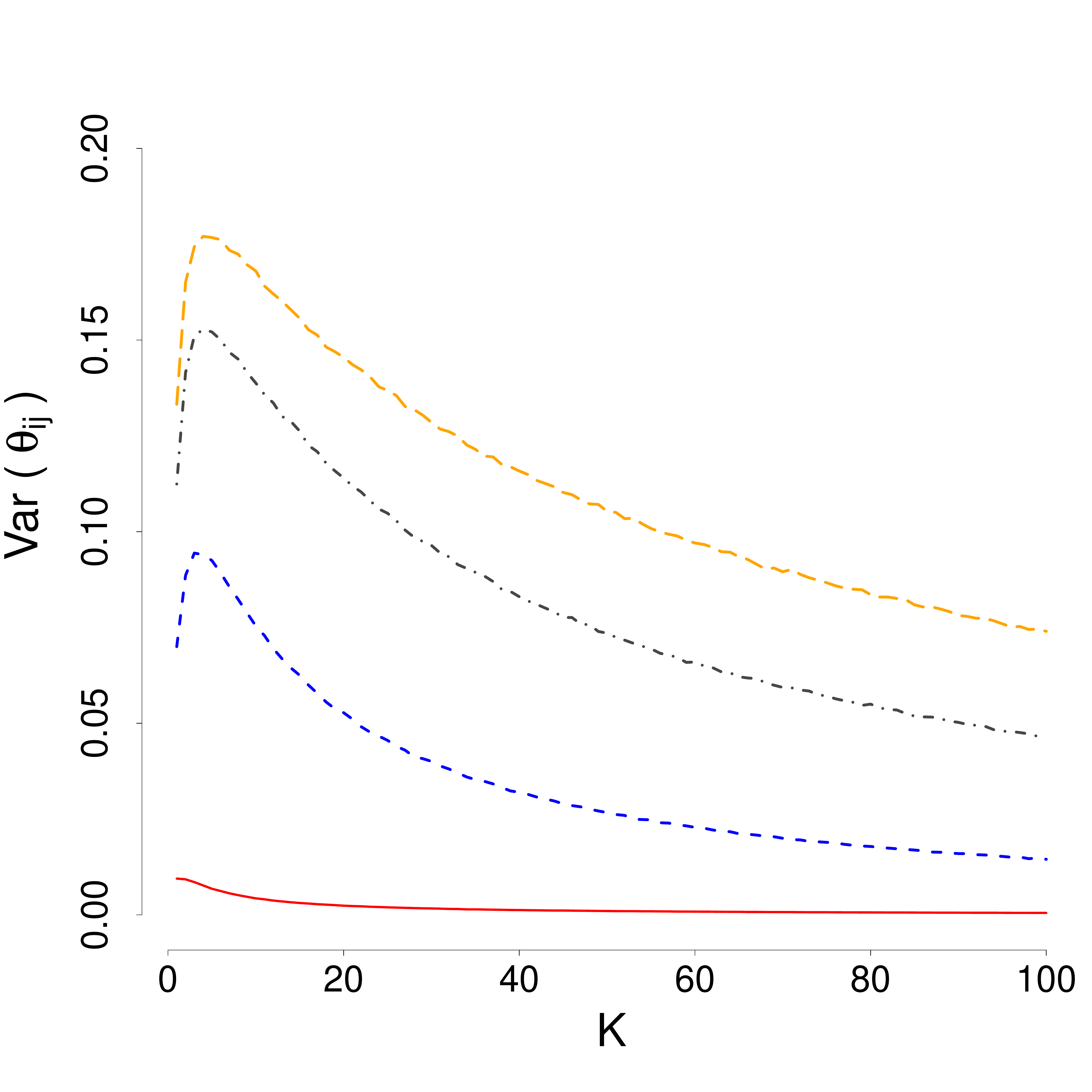} \\
    \begin{sideways} \;\;\;\;\;\;\;\;\;\;\;\;\;\;\;\;\;\;\; $\tau = 10$ \end{sideways} & 
    \includegraphics[width=.3\textwidth]{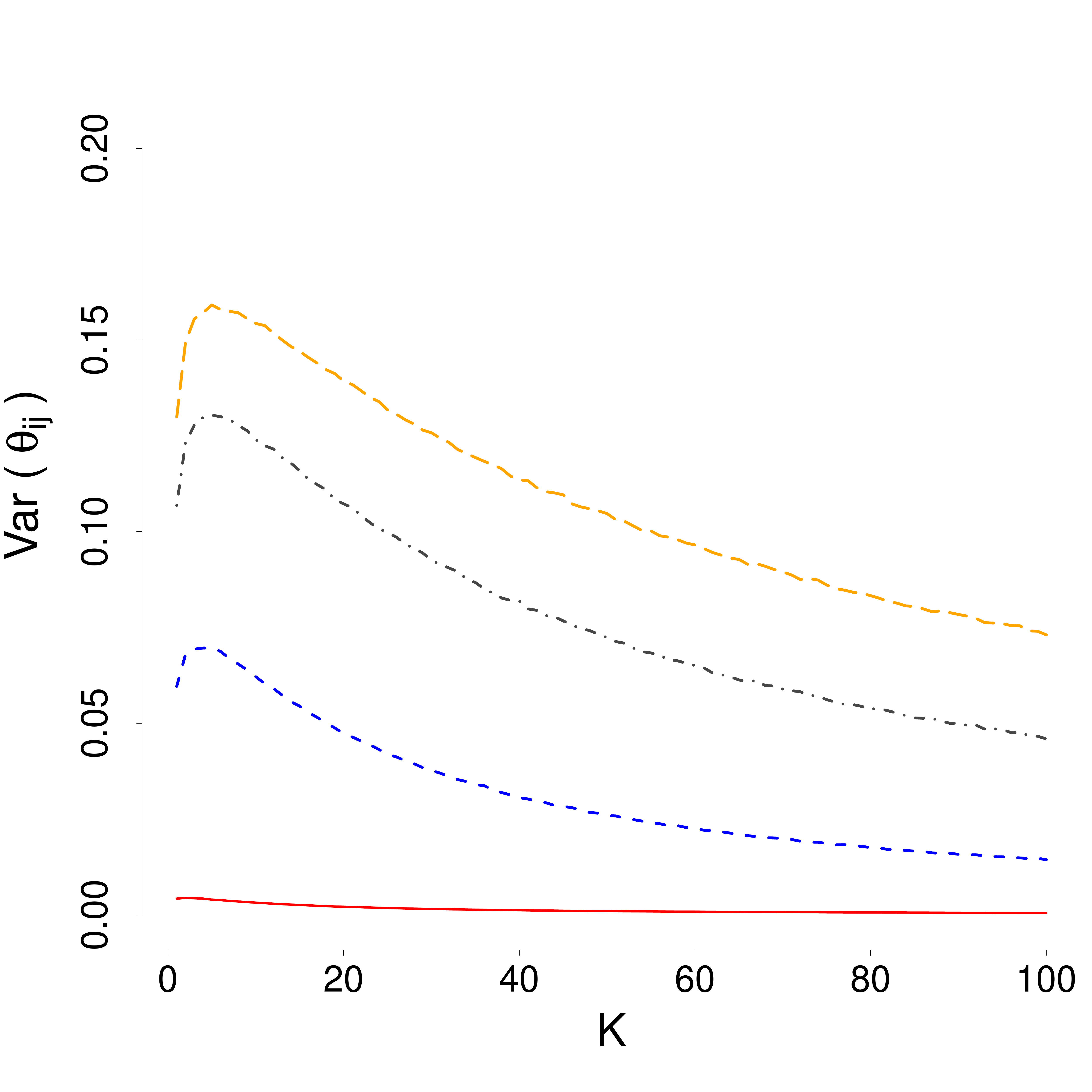} &
    \includegraphics[width=.3\textwidth]{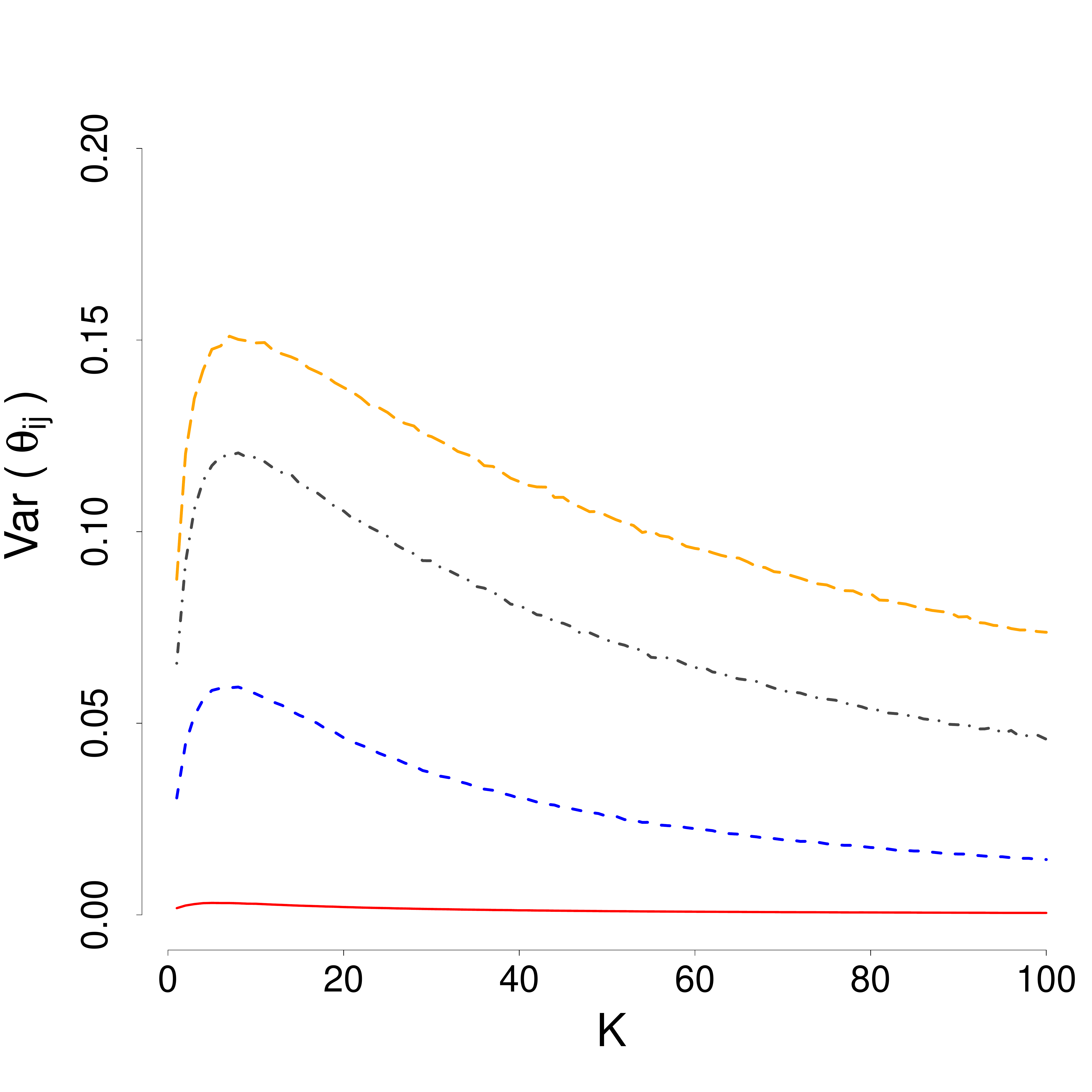} &
    \includegraphics[width=.3\textwidth]{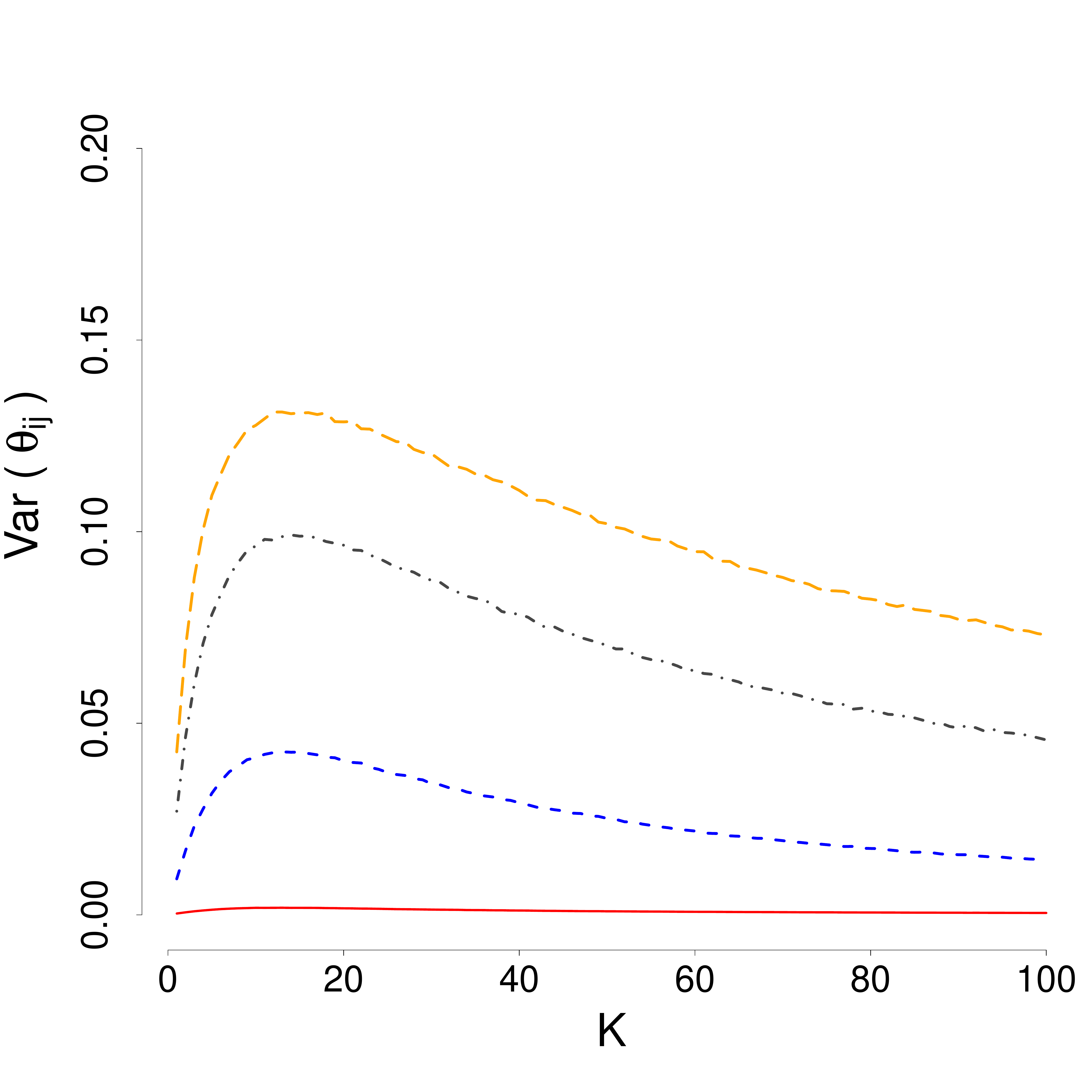} \\
\end{tabular}
  \caption{Prior variance of $\theta_{i,j}$ induced by von Mises–Fisher priors on the latent coordinates.}\label{fi:priorontheta}
\end{figure}

We are interested in the behavior of the prior on
$$
\theta_{i,j} = G_{\kappa_j}\left( \left\{ \rho_K\left( \bszeta_j, \bsbeta_i \right) \right\}^2-\left\{ \rho_K\left( \bspsi_j, \bsbeta_i \right) \right\}^2 \right)
$$
(recall Equation \eqref{eq:difdist}) induced by a von Mises–Fisher prior with mean direction $\mathbf{0}$ and precision $\omega$ for $\bsbeta_i$, and independent von Mises–Fisher priors with mean direction $\mathbf{0}$ and precision $\tau$ for both $\bspsi_j$ and $\bszeta_j$.   Because, $\bspsi_j$ and $\bszeta_j$ are assigned the same prior distributions, it is clear from a simple symmetry argument that $\E\left(\theta_{i,j}\right) = 1/2$ for all values of $\omega$, $\tau$ and $\kappa_j$.  On the other hand, Figure \ref{fi:priorontheta} shows the value $\Var \left( \theta_{i,j} \right)$ as a function of the embedding dimension $K$ for various combinations of $\omega$, $\tau$ and $\kappa_j$.  Note that, in every case, it is apparent that $\lim_{K \to \infty} \Var \left( \theta_{i,j} \right) \to 0$, which implies that $\theta_{i,j}$ converges in distribution to a point mass at $1/2$ as the number of latent dimensions grows.


This behavior, which is clearly unappealing, is driven by two features.  First, the von Mises–Fisher prior has a single scale parameter for all its dimensions.  Secondly, the surface area of the $K$ dimensional sphere, $\frac{2\pi^{K/2}}{\Gamma(K/2)}$, tends to zero as $K \to \infty$.  In fact, not only the von Mises-Fisher distribution, but also other widely used distributions on the sphere such as the Fisher-Bingham and the Watson distributions suffer from the same drawback. \cite{dryden2005} proposes to overcome this phenomenon by scaling the (common) concentration parameter of the von Mises-Fisher according to the number of dimensions.  However, this approach is unappealing in our setting for two reasons.  First, it does not appear to overcome the degeneracy issues just discussed.  Secondly, under this approach changes in $K$ would fundamentally change the nature and structure of the prior distribution. In the sequel, we discuss a novel class of flexible priors distributions on $\mathcal{S}^K$ that allow for different scales across various dimensions and can be used to construct priors for $\theta_{i,j}$ that do not concentrate on a point mass as the number of dimensions increase.

We build our new class of prior distributions in terms of the vector of angles $\bsphi = (\phi_1, \ldots, \phi_{K}) \in [-\pi,\pi]\times[-\pi/2,\pi/2]^{K-1}$ associated with the hyperspherical coordinate system.  In particular, we assign each component of $\bfphi$ a (scaled) von Mises distribution centered at 0, so that
\begin{align}\label{eq:newprior}
p(\bsphi \mid \bsomega) = \left(\frac{1}{2\pi}\right)^{K} 2^{K-1} \frac{1}{ I_0(\omega_1)} \exp\left\{ \omega_1 \cos \phi_1 \right\}
\prod_{k=2}^{K} \frac{1}{I_0(\omega_k)} \exp\left\{ \omega_k \cos 2\phi_k \right\},
\end{align}
where $\bsomega = (\omega_1, \ldots, \omega_{K})$ is a vector of dimension-specific precisions.  We call this prior the spherical von Mises distribution, and denote it by $\bfphi \sim \mbox{SvM}(\bfomega)$.

\begin{figure}
 \subfigure{\includegraphics[width=0.49\linewidth]{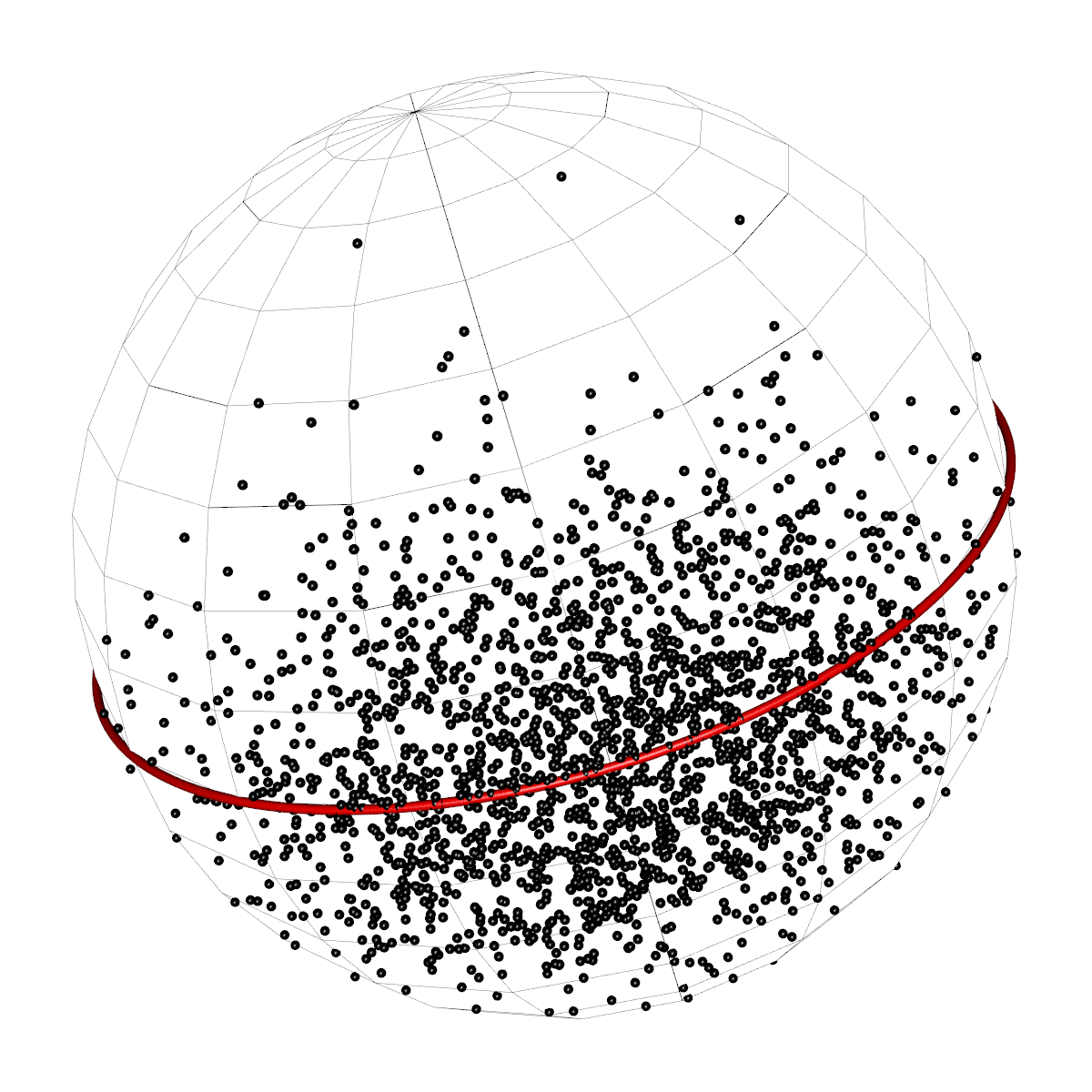}}
 \subfigure{\includegraphics[width=0.49\linewidth]{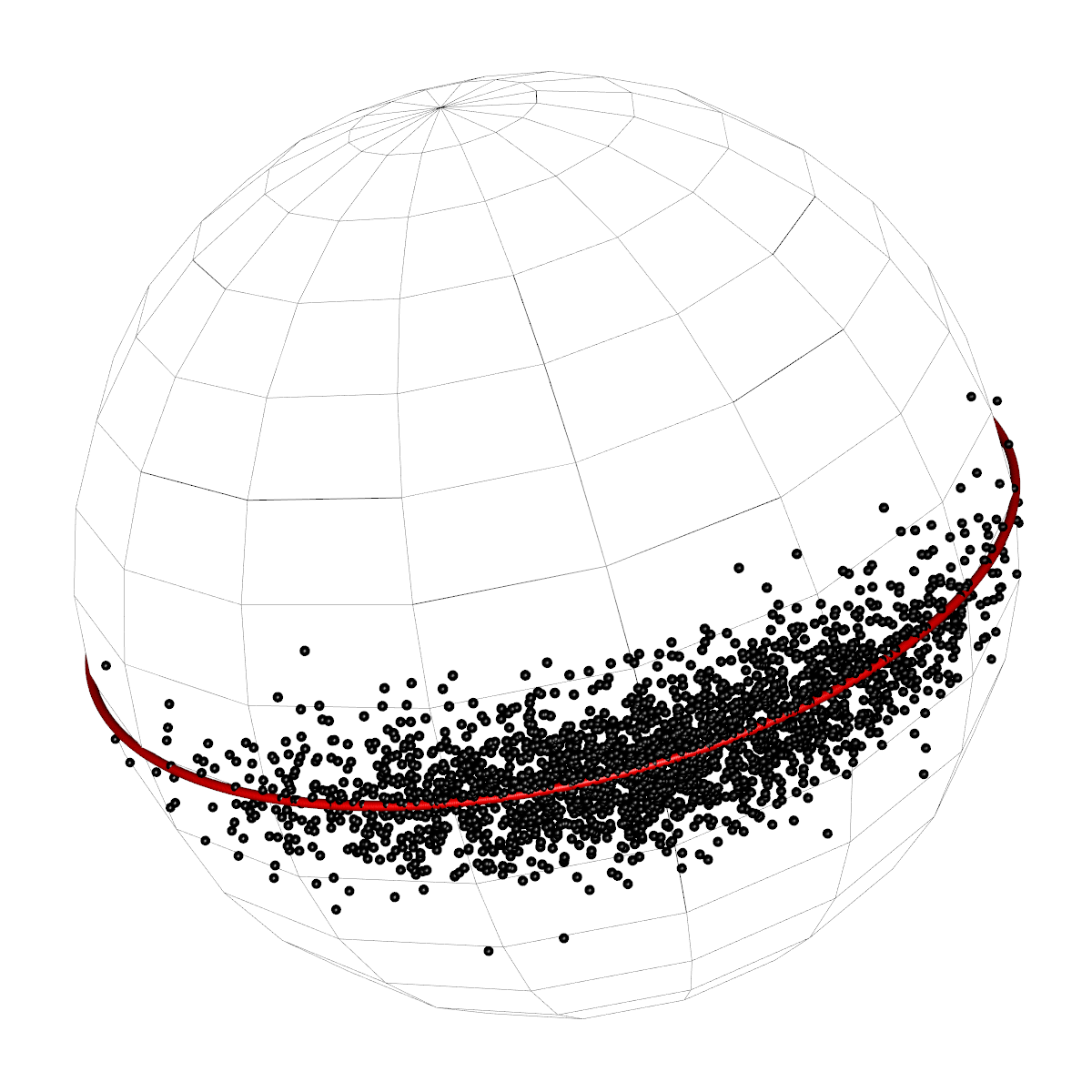}}
  \caption{Draws from two spherical von Mises distributions in $\mathcal{S}^3$.  The left panel corresponds to $\omega_1 = 7$ and $\omega_2 = 4$, while the right panel corresponds to $\omega_1 = 7$ and $\omega_2 = 30$.  Note that, as $\omega_2$ increases, the draws concentrate around a great circle.}\label{fi:drawsSvM}
\end{figure}

One key feature of this prior is that the parameters $\omega_1, \ldots, \omega_{K}$ control the concentration of each marginal distribution around their mean.  In particular, when $\omega_{K} \to \infty$, the marginal distribution of $\phi_{K}$ becomes a point mass at zero, and $p(\bsphi \mid \bsomega)$ concentrates all its mass in a greater nested hypersphere of dimension $K-1$ (see Figure \ref{fi:drawsSvM} for an illustration).  

A second key feature of this class of priors is that 
\begin{align}\label{eq:limitprior}
    \lim_{\bfomega \to \mathbf{\infty}} \frac
{\left(\frac{1}{2\pi}\right)^{K} 2^{K-1}\frac{1}{ I_0(\omega_1)} \exp\left\{ \omega_1 \cos \phi_1 \right\}
\prod_{k=2}^{K} \frac{1}{I_0(\omega_k)} \exp\left\{ \omega_k \cos 2\phi_k \right\}}
{\left(\frac{1}{2\pi}\right)^{K/2} 2^{K-1} \left\{ \prod_{k=1}^{K}\omega_k \right\}^{1/2} \exp\left\{ - \frac{1}{2} \left[ \omega_1\phi_1^2 + 4 \sum_{k=2}^{K} \omega_k \phi_k^2 \right] \right\}} = 1 .
\end{align}
Somewhat informally, this means that for large values of the precision parameters, the prior behaves like a zero-mean multivariate normal distribution with covariance matrix $\mbox{diag} \{ 1/\omega_1, 1/(4\omega_2), \ldots, 1/(4\omega_{K-1}) \}$.

Finally, we note that the density spherical von Mises distribution can be written in terms of the Cartesian coordinates $\bfx = (x_1, \ldots, x_{K+1})^T$ associated with $\bsphi$ (recall Equation \eqref{eq:sph_coord}).  More specifically, the density of the Hausdorff measure with respect to the uniform distribution on the sphere associated with \eqref{eq:newprior} is given by
\begin{multline}\label{eq:newpriorHausdorff}
p(\bfx \mid \bsomega) = \frac{1}{\prod_{k=1}^K\sqrt{\sum_{t=1}^{k+1} 
x_{t}^2}}   \frac{1}{2\pi  I_o(\omega_1)} \exp\left\{\omega_1 \frac{x_{1}}{\sqrt{x_{1}^2+x_{2}^2}}\right\} \\
\left\{\prod_{k=2}^K \frac{1}{\pi  I_o(\omega_k)}\right\}\exp\left\{ -\sum_{k=2}^K \omega_k \left(2 \frac{x_{k+1}^2}{ \sum_{t=1}^{k+1} x_{t}^2} -1 \right)\right\} 
,\quad \quad \bfx^T \bfx = 1.
\end{multline}
See the online supplementary materials for the derivation.

\begin{figure}
  \centering
\begin{tabular}{cccc}
    & $\omega = 0.5$ & $\omega = 2$ & $\omega = 10$ \\ 
    \begin{sideways} \;\;\;\;\;\;\;\;\;\;\;\;\;\;\;\;\;\;\; $\tau = 0.5$\end{sideways}&
    \includegraphics[width=.3\textwidth]{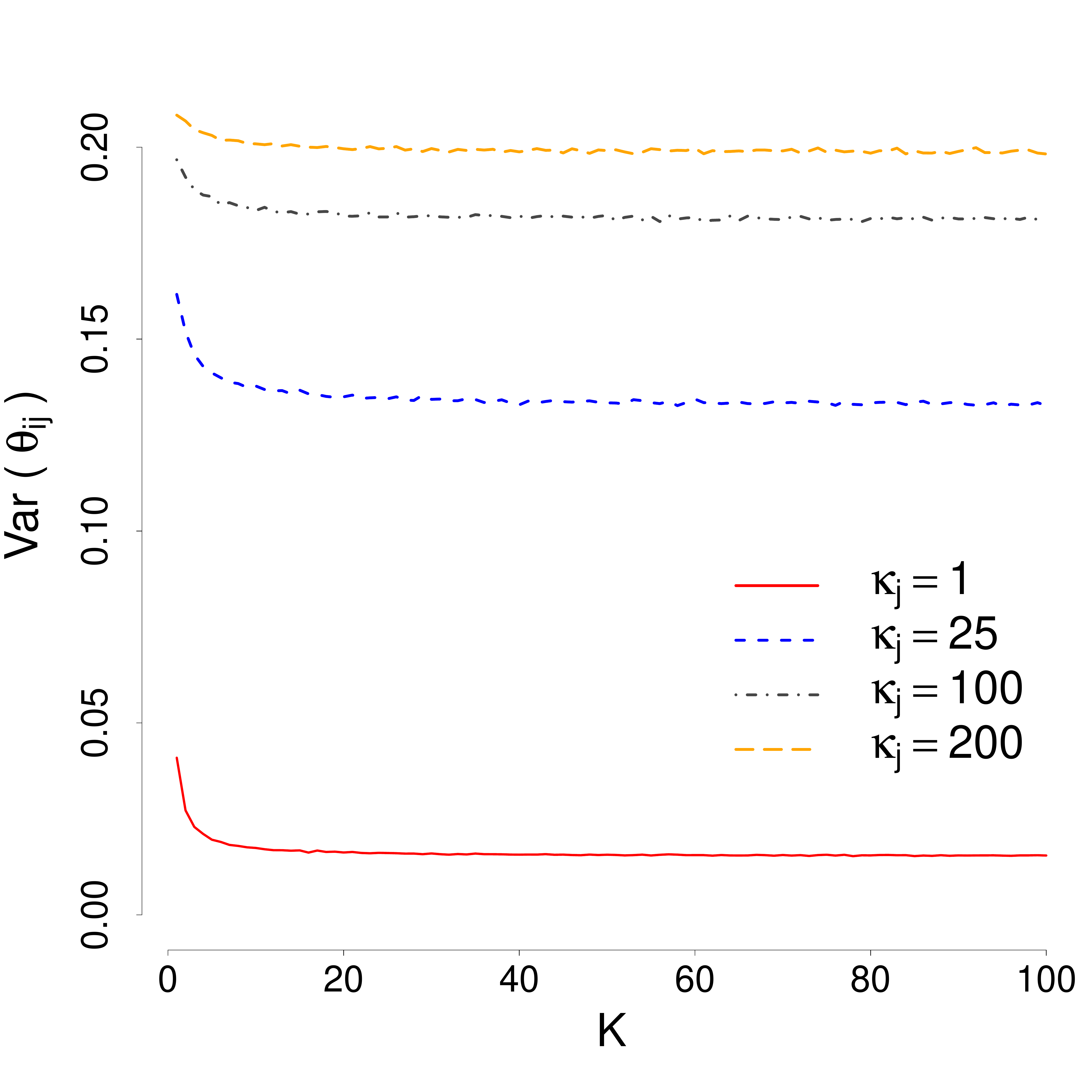} &
    \includegraphics[width=.3\textwidth]{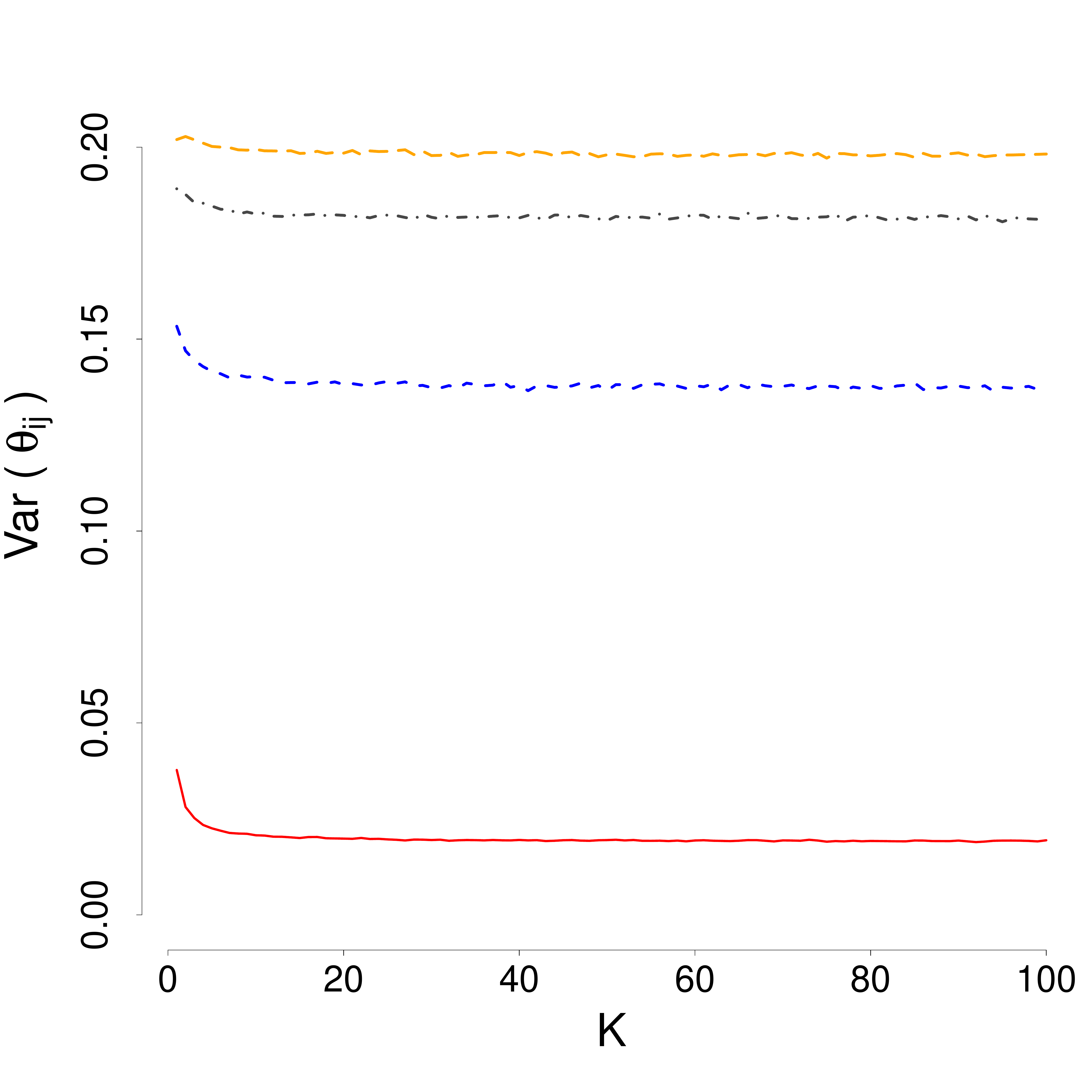} &
    \includegraphics[width=.3\textwidth]{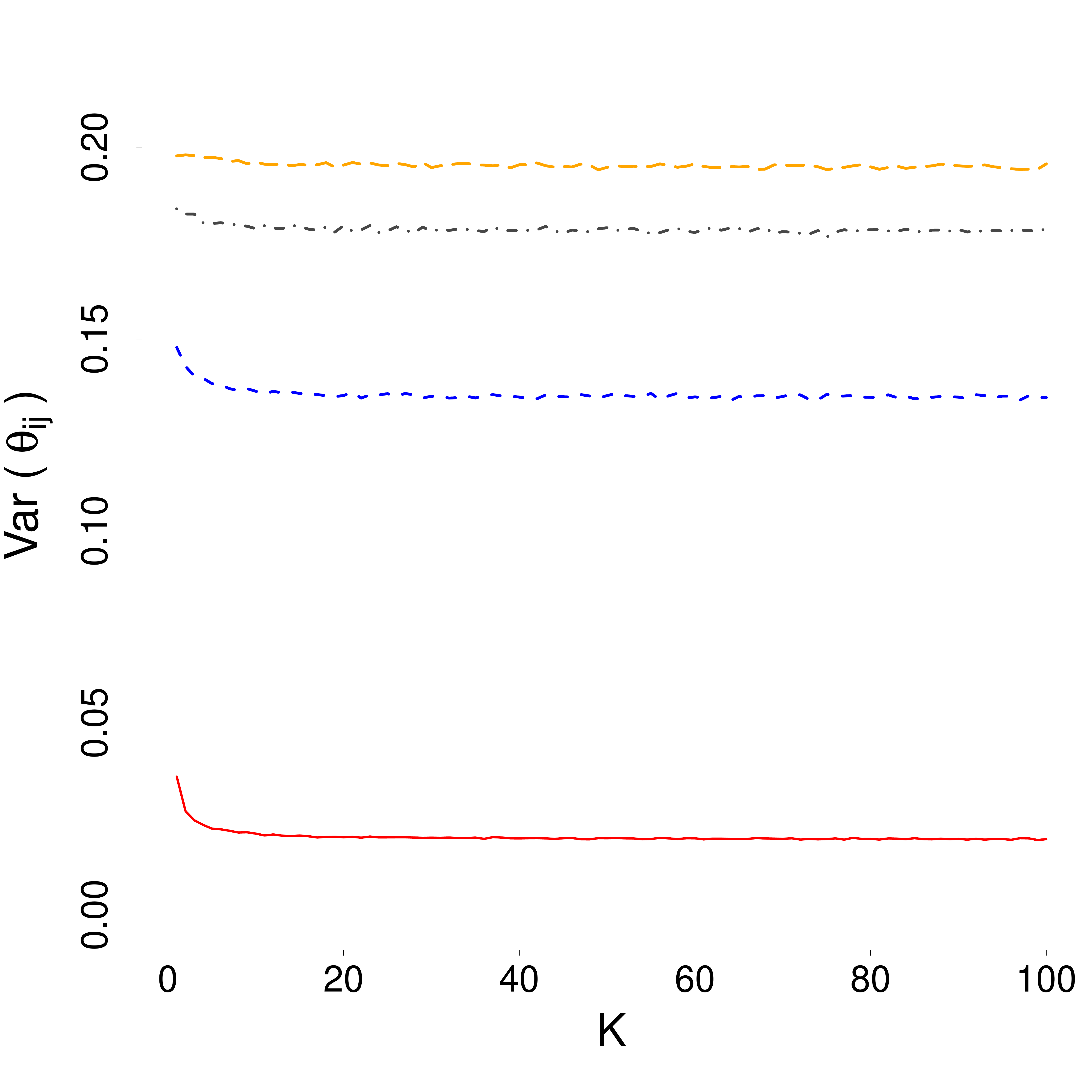} \\
    \begin{sideways} \;\;\;\;\;\;\;\;\;\;\;\;\;\;\;\;\;\;\;\;\, $\tau = 2$ \end{sideways} & 
    \includegraphics[width=.3\textwidth]{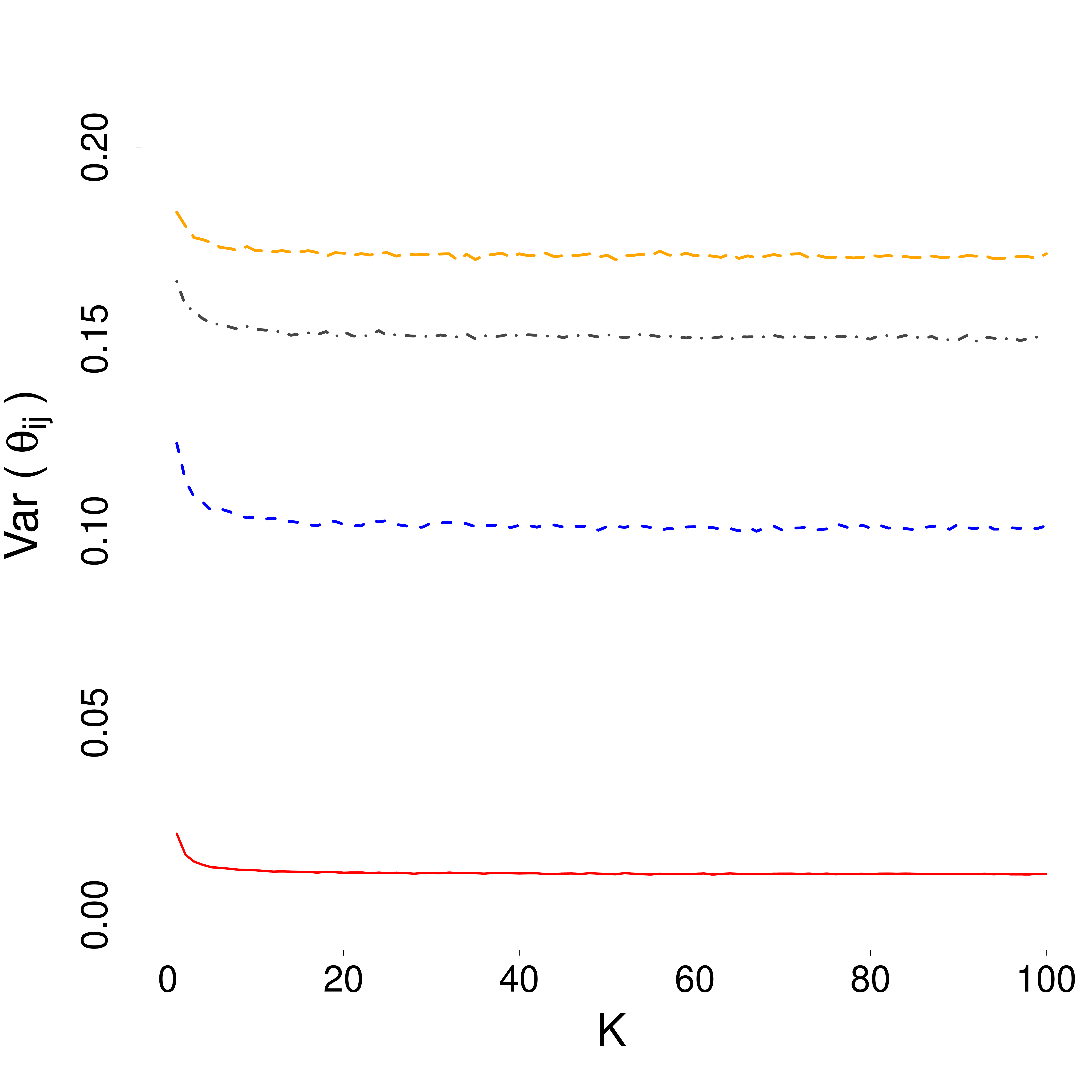} &
    \includegraphics[width=.3\textwidth]{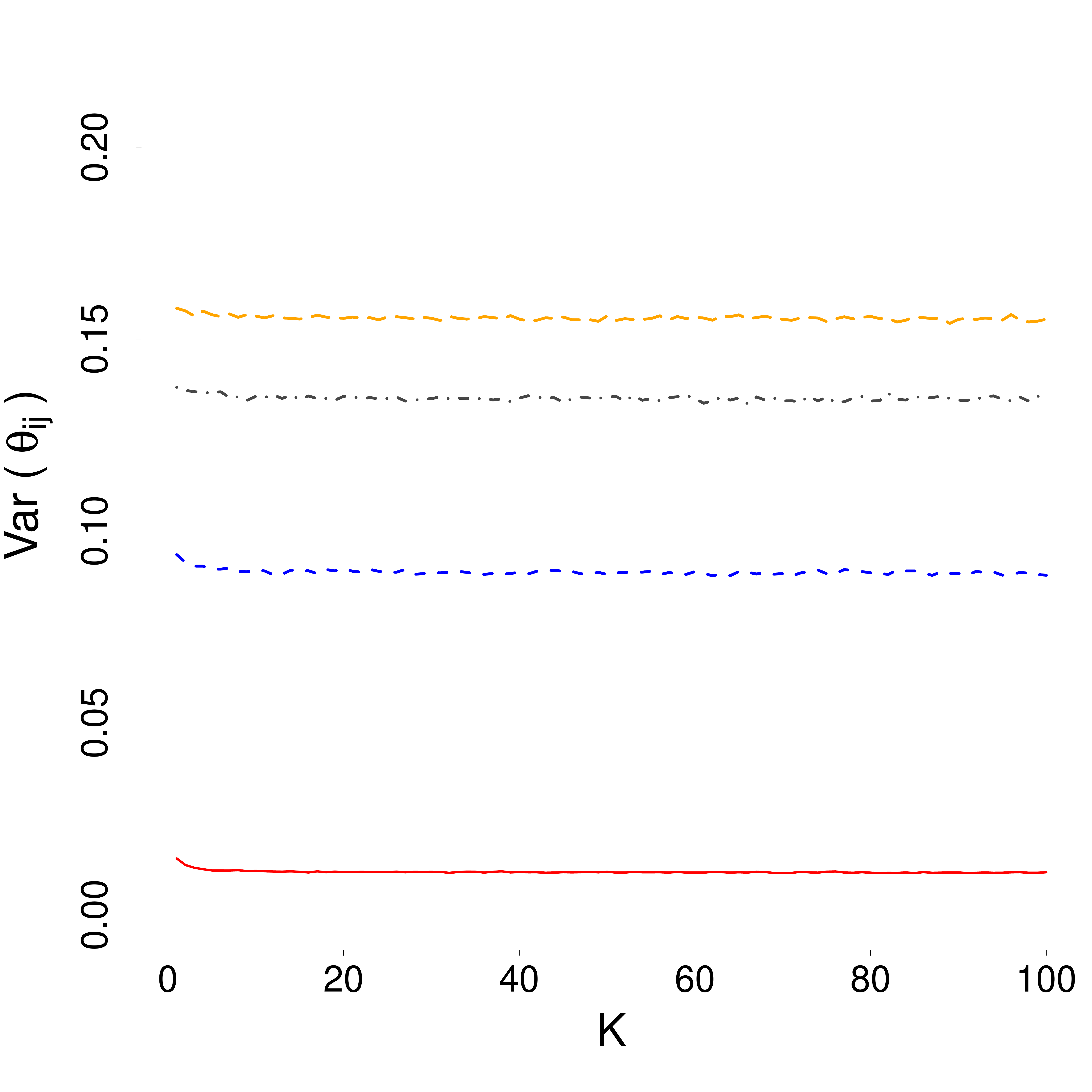} &
    \includegraphics[width=.3\textwidth]{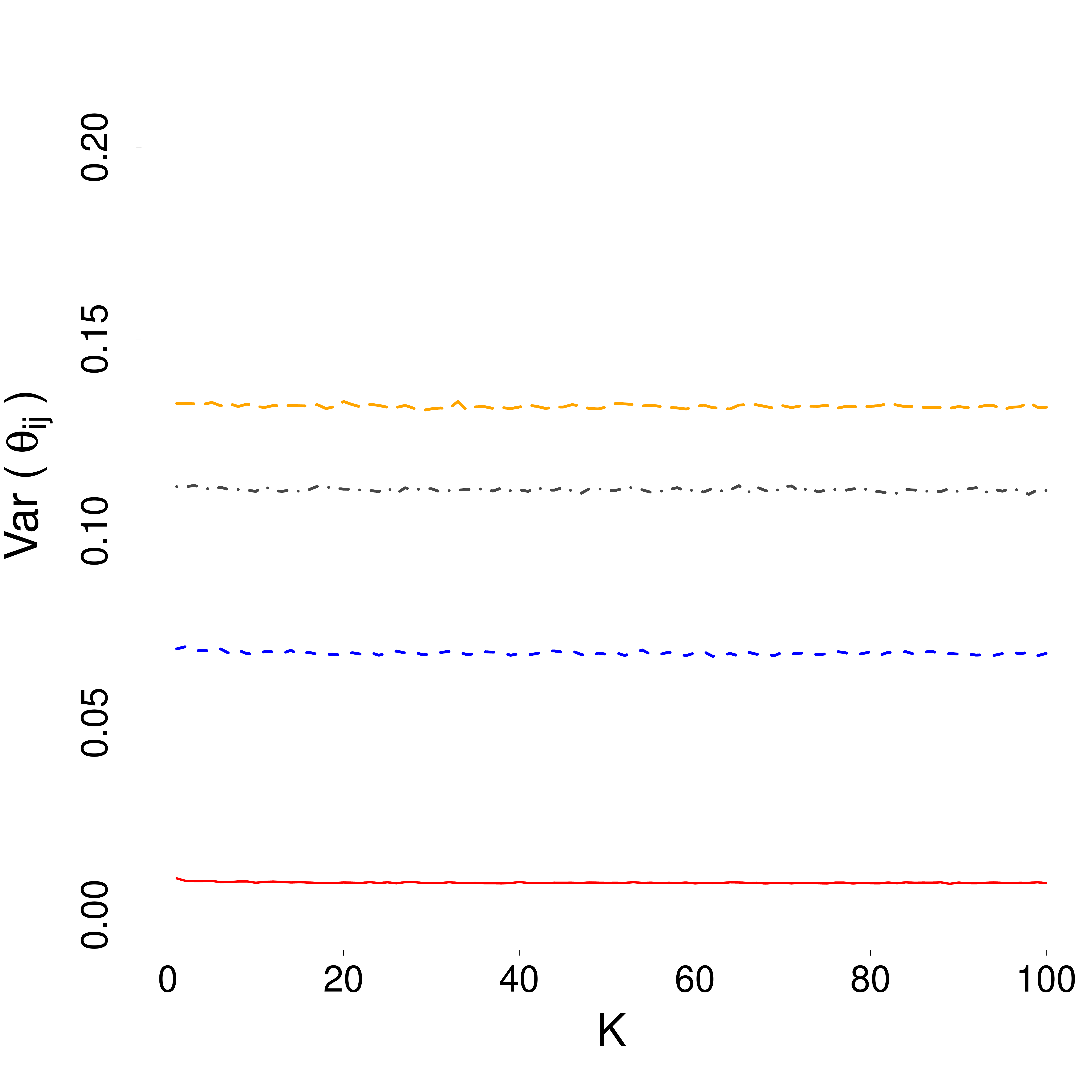} \\
    \begin{sideways} \;\;\;\;\;\;\;\;\;\;\;\;\;\;\;\;\;\;\; $\tau = 10$ \end{sideways} & 
    \includegraphics[width=.3\textwidth]{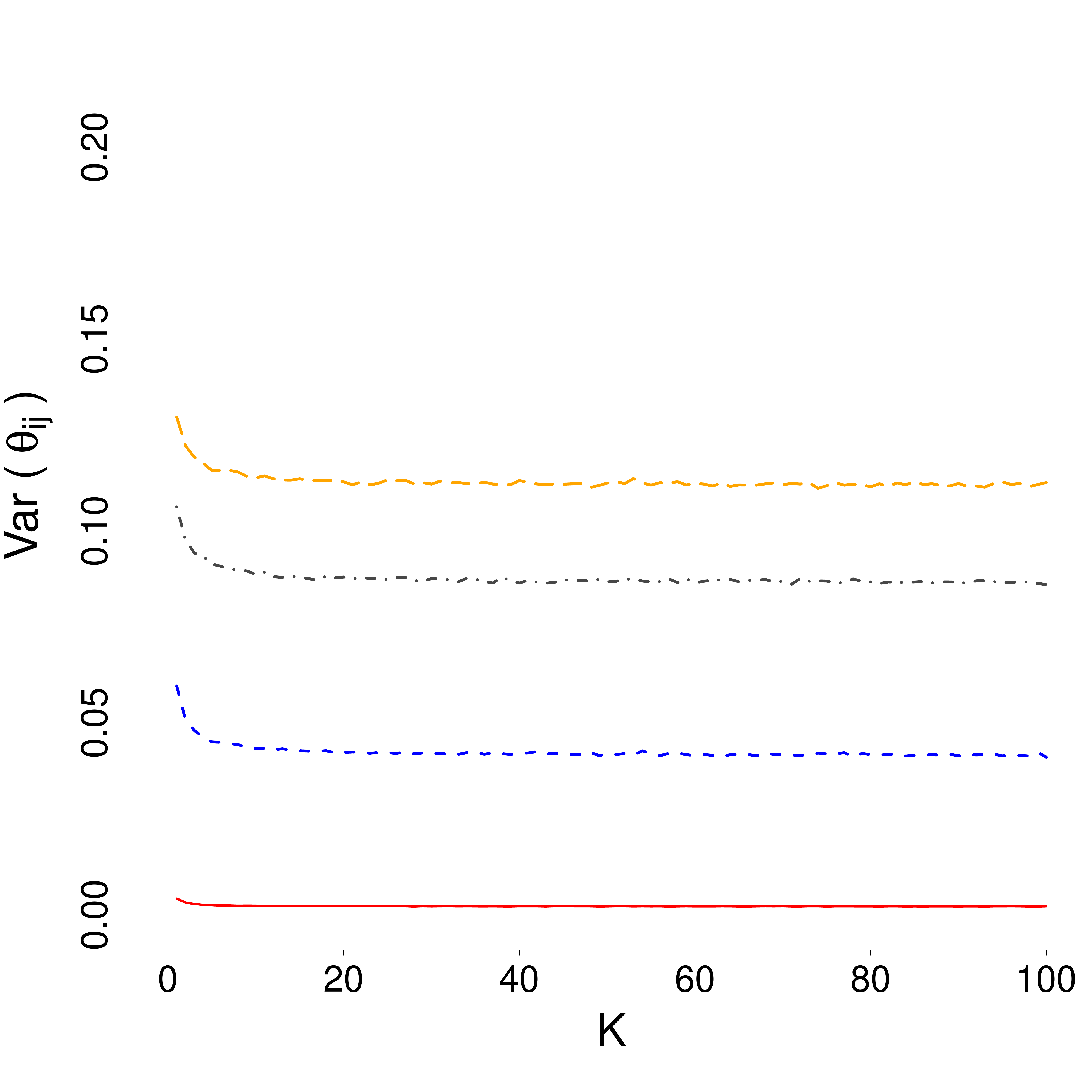} &
    \includegraphics[width=.3\textwidth]{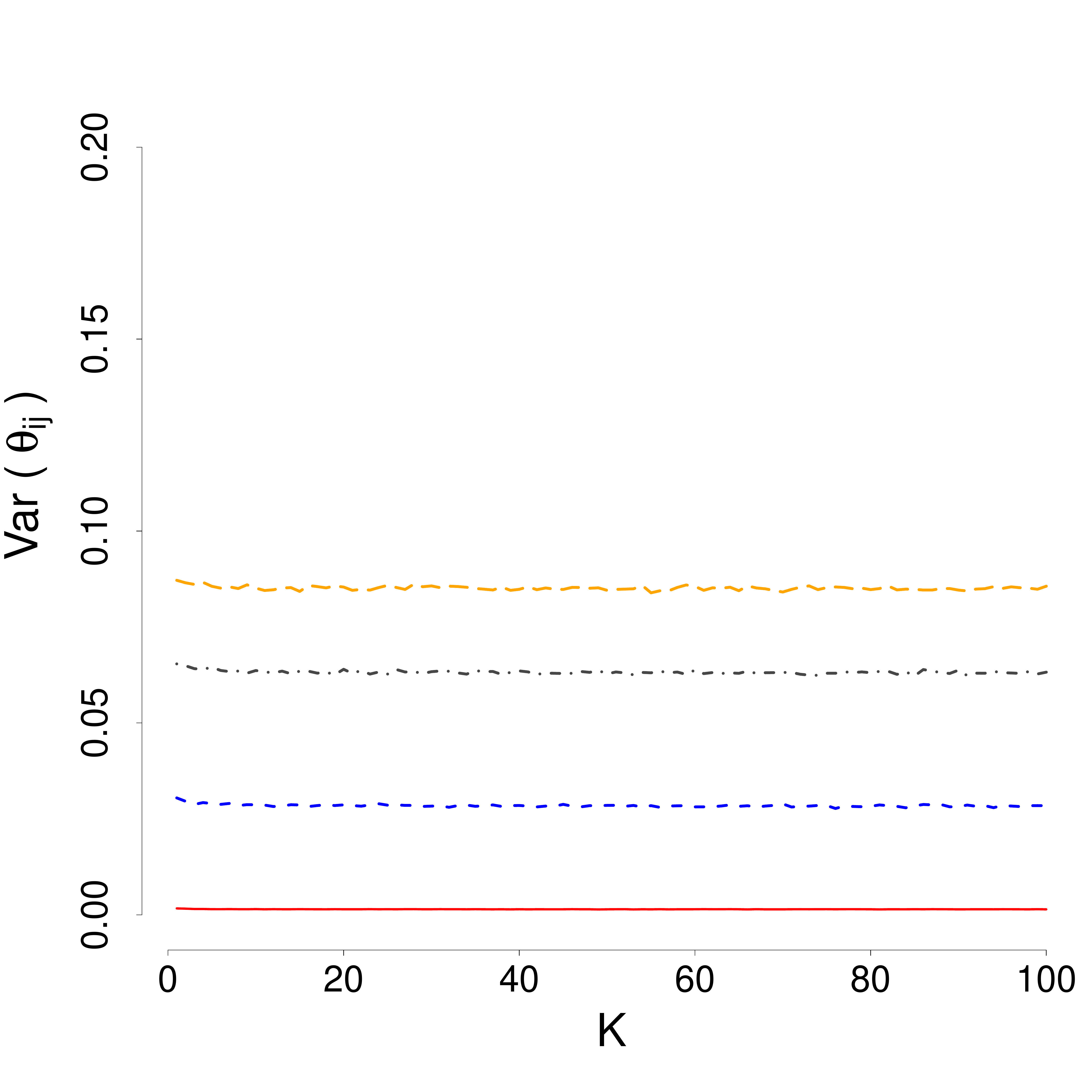} &
    \includegraphics[width=.3\textwidth]{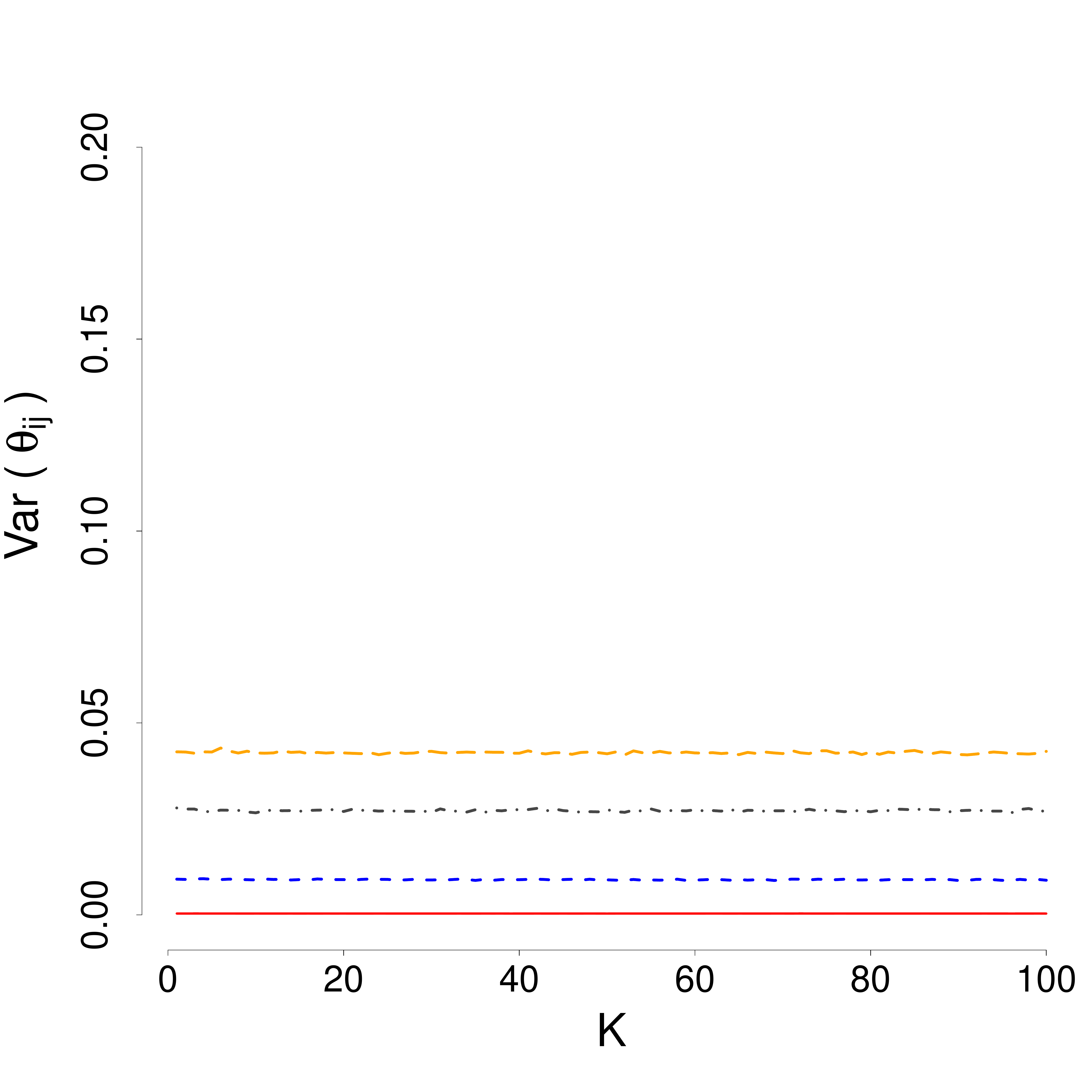} \\
\end{tabular}
\caption{Prior variance of $\theta_{i,j}$ induced by spherical von Mises priors with polynomially increasing marginal precision on the latent coordinates.}\label{fi:prioronthetanew}
\end{figure}

\begin{figure}
 \subfigure[$\kappa_j=200$, $\omega=20$, $\tau=20$]{\includegraphics[width=0.49\linewidth]{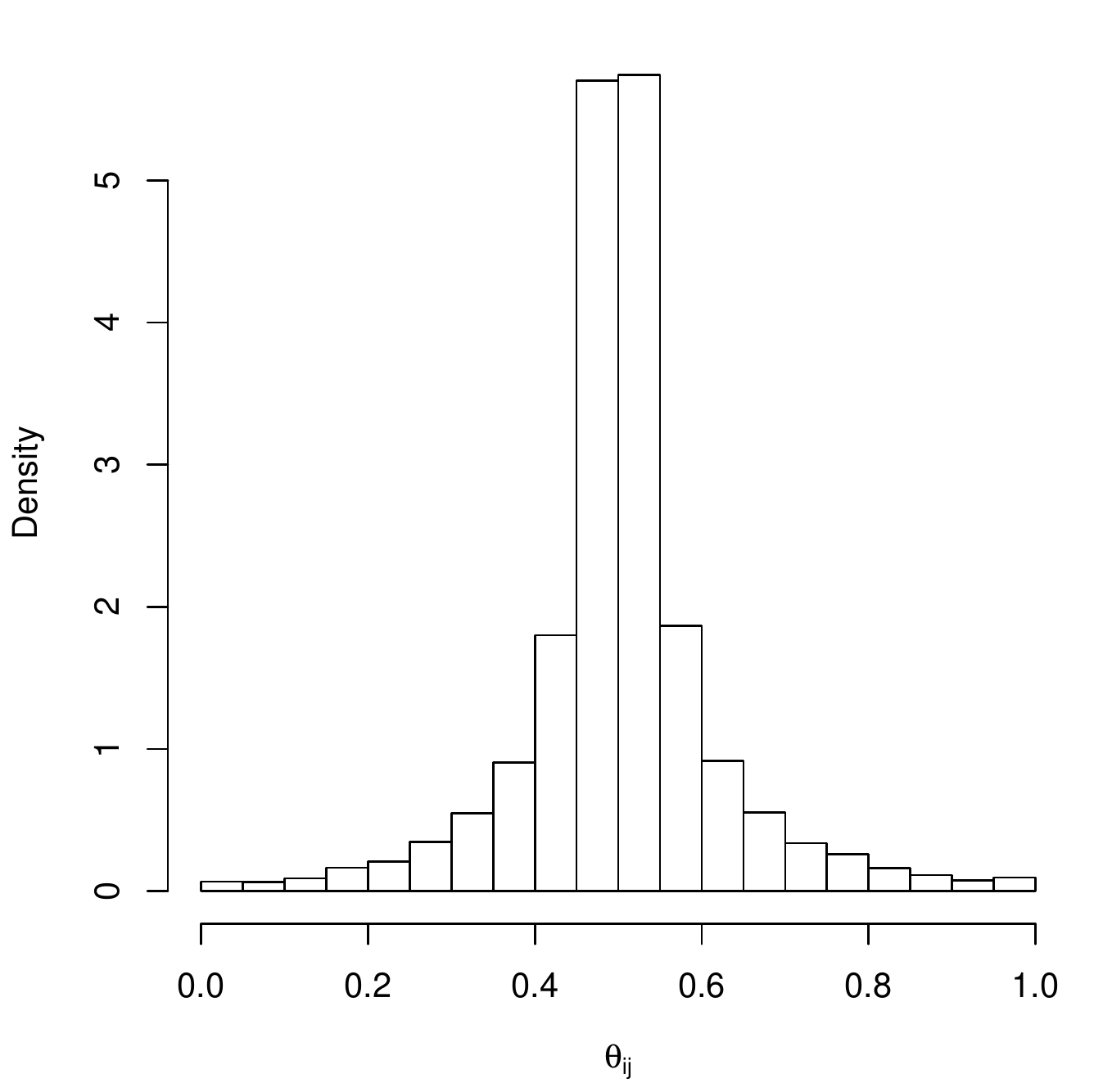}}
 \subfigure[$\kappa_j=50$, $\omega=1$, $\tau=1$]{\includegraphics[width=0.49\linewidth]{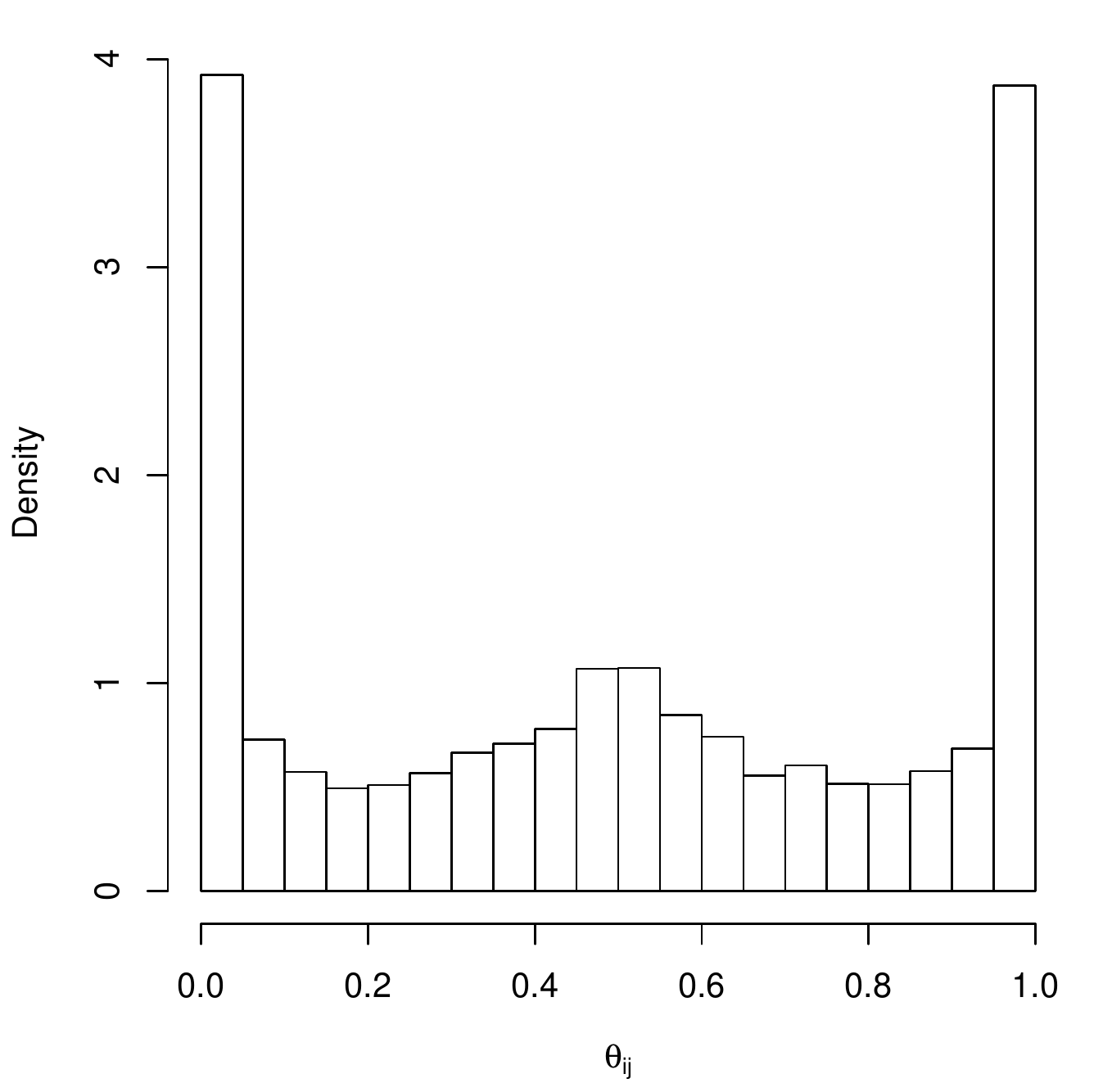}}
  \caption{Histogram of 10,000 draws from the prior distribution on $\theta_{i,j}$ implied by (\ref{eq:finalpriors1}-\ref{eq:finalpriors3}) for  $K=10$ and two different combinations of hyperparameters.}\label{fi:shapesdentheta}
\end{figure}

Coming back to our spherical factor model, we use the spherical von Mises distribution as priors for the $\bspsi_j$s, $\bszeta_j$s and $\bsbeta_i$s.  In particular, we set 
\begin{align}
\bspsi_i &\sim \mbox{SvM}(\tau, 2^2\tau, 3^2\tau, \ldots, K^2\tau)  \label{eq:finalpriors1}  \\
\bszeta_j &\sim \mbox{SvM}(\tau, 2^2\tau, 3^2\tau, \ldots, K^2\tau)  \label{eq:finalpriors2}  \\
\bsbeta_j &\sim \mbox{SvM}(\omega, 2^2\omega, 3^2\omega, \ldots, K^2\omega) \label{eq:finalpriors3}
\end{align}
independently across all $i=1,\ldots,I$ and $j=1,\ldots,J$.  
By setting up the priors on the latent positions to have increasing marginal precisions, we can avoid the degeneracy issues that arose in the case of the von Mises-Fisher priors.  This is demonstrated in Figure \ref{fi:prioronthetanew}, which shows the value of $\Var \left( \theta_{i,j} \right)$ as function of the latent dimension $K$ for various combinations of $\omega$, $\tau$ and $\kappa_j$ under (\ref{eq:finalpriors1}-\ref{eq:finalpriors3}).  Note that, in every case, the variance of the prior decreases slightly with the addition of the first few dimensions, but then seems to quickly stabilize around a constant that is determined by the value of the hyperparameters.  Figure \ref{fi:shapesdentheta} explores in more detail the shape of the implied prior distribution on $\theta_{i,j}$, as well as the effect of various hyperparameters.  Note that the implied prior can either be unimodal (which typically happens for relatively large values $\omega$ and $\tau$), or trimodal (which happens for low values of $\omega$ and $\tau$).  The intuition behind the formulation in Equations (\ref{eq:finalpriors1}-\ref{eq:finalpriors3}) comes from the fact that these polynomial rates ensures that $\Var \left( \theta_{i,j} \right)$ converges to a constant with the number of dimensions under the Gaussian approximation in Equation \eqref{eq:limitprior}.

In addition to avoiding degeneracy issues, using an increasing sequence for the marginal precision of all three sets of $\bspsi_j$s, $\bszeta_j$s and $\bsbeta_i$s means that the prior encourages the model to concentrate most of the mass on a embedded sub-sphere of $\mathcal{S}^K$ (recall the discussion at the end of Section \ref{se:likelihood}).  Hence, we can think of $K$ as representing the highest intrinsic dimension allowed by our prior, and as the combination of $\omega$ and $\tau$ as controlling the 
``effective'' prior dimension of the space, $K^{*} \le K$.  Another implication of this prior structure is that we can think of our prior as encouraging a decomposition of the (spherical) variance that is reminiscent of the principal nested sphere analysis introduced in \cite{jung2012analysis}.

\subsection{Connection to the Euclidean model}\label{se:coneceuclid}

In Section \ref{se:likelihood} we argued that the spherical model of dimension $K$ contains all other spherical models of lower dimension as special cases.  Similarly, the Euclidean space of dimension $K+1$ includes all spherical spaces of dimension $K$ or lower. It is also true that the $K$-dimensional Euclidean model discussed in Section \ref{se:euclideanmodel} (with a probit link) can be seen as a limit case of our spherical model on $\mathcal{S}^{K}$.

The argument relies on projecting the spherical model onto the tangent bundle of the manifold, and making
$\omega \to \infty$, $\tau \to \infty$ and $\kappa_j \to \infty$ while keeping $\omega/\tau$ and $\omega/\kappa_j$ constant for all $j$.  As mentioned in previous sections, under these circumstances, (i) the link function $G_{\kappa_j}$ defined in Equation \eqref{eq:linkcircular} projects onto the probit link function, (ii) the geodesic distance between the original latent positions converges to the Euclidean distance between their projections onto the tangent space, and (iii) the spherical von-Mises priors defined in Equations (\ref{eq:finalpriors1}-\ref{eq:finalpriors3}) project onto Gaussian priors.

This observation is important for at least three reasons.  Firstly, it can guide prior elicitation, specially for the precision parameters $\omega$ and $\tau$.  Secondly, spherical models can be seen as interpolating (in terms of complexity) between Euclidean models of adjacent dimensions.  In particular, note that the likelihood for Euclidean models involves $\mathcal{O}\left(\{I+2J\}K\right)$ parameters (since the $\sigma_j$s are not identifiable and typically fixed) while the likelihood of the spherical model relies on $\mathcal{O}\left( \{I+2J\}K + J\right)$ parameters, and finally, $\mathcal{O}\left(\{I+2J\}K\right) < \mathcal{O}\left( \{I+2J\}K + J\right) < \mathcal{O}\left(\{I+2J\}\{K+1\}\right)$.  Thirdly, it enhances the interpretability of the model.  In particular, the reciprocal of the precision, $\frac{1}{\omega}$ can be interpreted as a measure of sphericity of the latent space that can be directly contrasted across datasets.  

\subsection{Parameter identifiability}

The likelihood function for the spherical factor model discussed in Section \ref{se:likelihood} is invariant to simultaneous rotations of all latent positions.  However, the structure of the priors in (\ref{eq:finalpriors1}-\ref{eq:finalpriors3}) induces weak identifiability for the $\bfpsi_j$s, $\bfzeta_j$s, and $\bsbeta_i$s in the posterior distribution.  Reflections, which are not addressed by our prior formulation, can be accounted for by fixing the octant of the hypersphere to which a particular $\bsbeta_i$ belongs.  In our implementation, this constraint is enforced a posteriori by post-processing the samples generated by the Markov chain Monte Carlo algorithm described in Section \ref{se:comput}.  On the other hand, while the scale parameters $\sigma_1, \ldots, \sigma_J$ are not identifiable in the Euclidean model, the analogous precisions $\kappa_1, \ldots, \kappa_J$ are identifiable in the spherical model.  This is because $\mathcal{S}^K$ has finite measure, and therefore the likelihood is not invariant to rescalings of the latent positions.  More generally, partial Procrustes analysis (which accounts for invariance to translations and rotations, but retains the scale) can be used to postprocess the samples to ensure identifiability.

\subsection{Hyperpriors}

\begin{figure}
 \centering
 \includegraphics[width=0.5\linewidth]{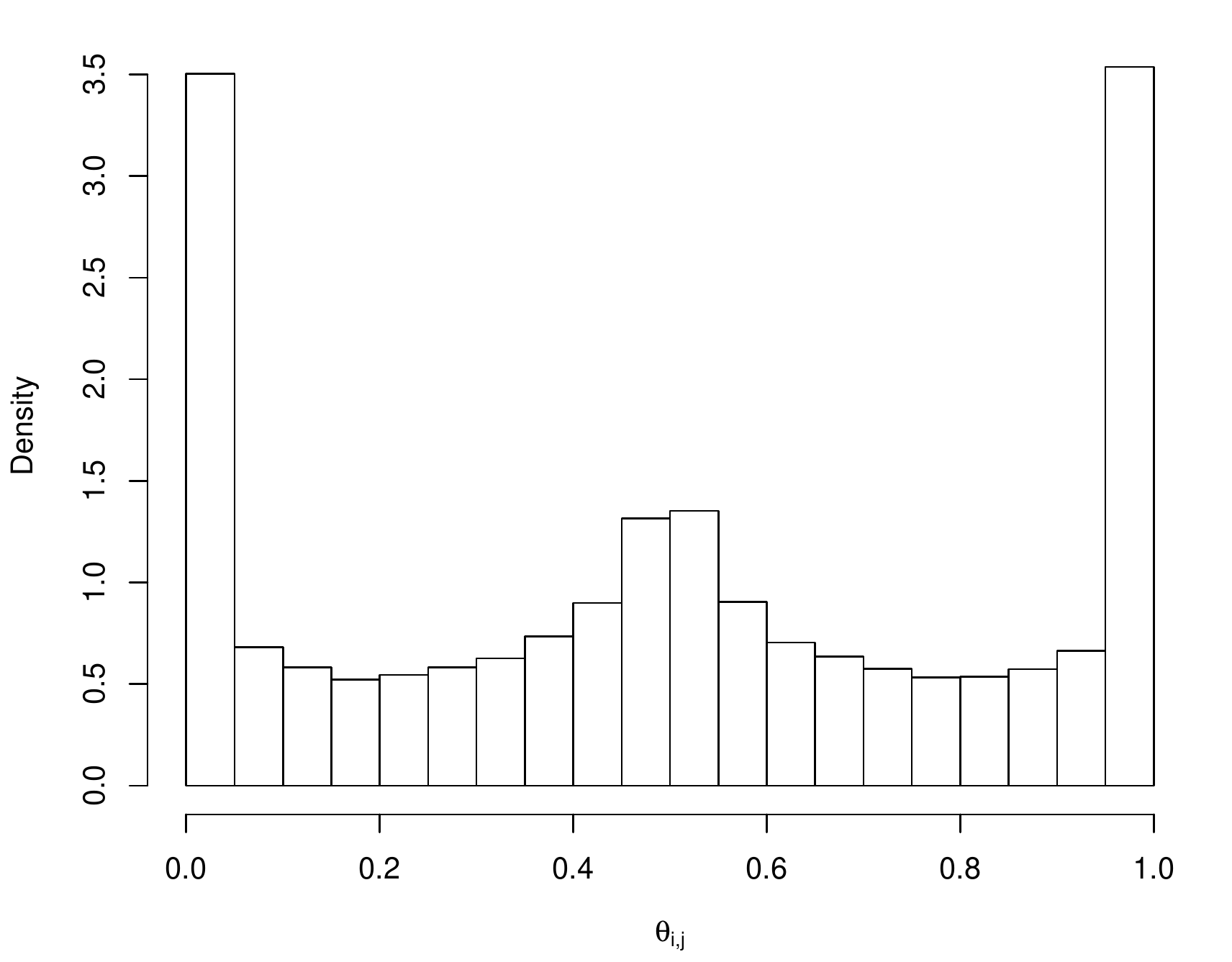}
 \caption{Histogram of 10,000 samples from the prior on $\theta_{i,j}$ implied by the hyperparameters $a_{\omega} = a_{\tau} = 1$, $b_{\omega} = 1/10$, $b_{\tau} = 5$, $a_{\lambda} = 2$, $b_{\lambda}=150$, and $c=1$ for $K=10$.  Unless otherwise noted, these are the parameter settings we use to carry out all the data analyses in this paper.}\label{fi:finaldenthetaprior}
\end{figure}

Completing the specification of the model requires that we assign hyperpriors to $\omega$, $\tau$ and $\kappa_1, \ldots, \kappa_J$.  The precisions $\omega$ and $\tau$ are assigned independent Gamma distributions, $\omega \sim \mbox{Gam}(a_{\omega}, b_{\omega})$ and $\tau \sim \mbox{Gam}(a_{\tau}, b_{\tau})$.  Similarly, the concentration parameters associated with the link function are assumed to be conditionally independent and given a common prior, $\kappa_j \sim \mbox{Gam}(c, \lambda)$, where $\lambda$ is in turn given a conditionally conjugate Gamma hyperprior, $\lambda \sim \mbox{Gam}(a_{\lambda}, b_{\lambda})$.  The parameters $a_{\omega}$, $b_{\omega}$, $a_{\tau}$, $b_{\tau}$, $a_{\lambda}$, $b_{\lambda}$ and $c$ for these hyperpriors are assigned to  strongly favor configurations in which $\beta_{i,1} \in [-\pi/2, \pi/2]$ (which is consistent with the assumption that a Euclidean model is approximately correct), and so that the induced prior distribution on $\theta_{i,j}$ is, for large $K$,  close to a $\mbox{Beta}(1/2,1/2)$ distribution (the proper, reference prior for the probability of a Bernoulli distribution).  Various combinations of hyperparameters satisfy these requirements, most of which lead to priors on $\theta_{i,j}$ that are trimodal.  We recommend picking one in which the hyperpriors are not very concentrated (e.g., see Figure \ref{fi:finaldenthetaprior}) and studying the sensitivity of the analysis to the prior choice.  In our experience, inferences tend to be robust to reasonable changes in the hyperparameters (see Section \ref{se:sensitivity}).

\section{Computation}\label{se:comput}

The posterior distribution for the spherical factor model
is analytically intractable. Hence, inference for the model parameters is carried out using Markov chain Monte Carlo (MCMC) algorithms.  The algorithm we propose is a hybrid that combines Gibbs sampling, random walk Metropolis-Hastings and Hamiltonian Monte Carlo steps to generate samples from the full conditional distributions of each parameter.  The simplest steps correspond to sampling the parameters $\omega$, $\tau$, $\lambda$ and $\kappa_1, \ldots, \kappa_J$.  
More specifically, we sample $\lambda$ from its inverse-Gamma full conditional posterior distribution, and sample $\omega$, $\tau$ and each of the $\kappa_j$s using random walk Metropolis Hastings with log-Gaussian proposals.  The variance of the proposals for these steps are tuned so that the acceptance rate is roughly 40\%.  On the other hand, for sampling the latent positions we employ the Geodesic Hamiltonian Monte Carlo (GHMC) algorithm described in \cite{byrne2013geodesic}, which provides a scalable and efficient way to obtain samples from target distributions defined on manifolds that can be embedded in a Euclidean space.

As an example, consider the step associated with updating $\bsbeta_i$, the latent factor for individual $i$. Denoting the associated coordinates in $\reals^{K+1}$ by $\bfx_{\beta_i}$ (recall Equation \eqref{eq:sph_coord}), the density of the Hausdorff measure associated with the full conditional posterior is given by 
\begin{multline*}
p_{\mathcal{H}}(\bfx_{\beta_i} \mid \cdots ) \propto 
\left[\prod_{j=1}^{J}   G_{\kappa_j}(e_{ij})^{y_{ij}} \left ( 1 -G_{\kappa_j}(e_{ij})\right)^{1-y_{ij}}  \right] \\ \left[\exp\left\{\omega\frac{x_{\beta_i,1}}{\sqrt{x_{\beta_i,1}^2+x_{\beta_i,2}^2}}\right\}\right]  
\left[\exp\left\{ -\sum_{k=2}^{K} k^2\omega \Bigg(2 \frac{x_{\beta_i,k+1}^2}{ \sum_{t=1}^{k+1} x_{\beta_i,t}^2} -1 \Bigg)\right\}\right] \\ \left[\frac{1}{\prod_{k=1}^{K} \left(\sum_{t=1}^{k+1} x_{\beta_i,t}^2  \right)^{\frac{1}{2}}}\right], \quad\quad\quad \bfx_{\beta_i}^T\bfx_{\beta_i} = 1 ,
\end{multline*}
where $\bfx_{\beta_i} = \left(x_{\beta_i,1},x_{\beta_i,2},\dots,x_{\beta_i,{K+1}} \right)$.  Then, given tuning parameters $L$ and $\epsilon$, the GHMC step takes the form: 

\begin{enumerate}
\item Initialize $\bfx_{\beta_i} = \bfx^{(c)}_{\beta_i}$, as well as the auxiliary momentum variable $\bsgamma$ by sampling $\bsgamma \sim N(0, \bfI_d)$.

\item Project the momentum onto the tangent space at $\bfx_{\beta_i}$ by setting $\bsgamma \leftarrow \left(\bfI_d-\bfx_{\beta_i}\bfx_{\beta_i}^T \right)\bsgamma$, and then set $\bsgamma^{(c)} = \bsgamma$.

\item For each of the $L$ leap steps:
\begin{enumerate}
\item Update the momentum by setting $\bsgamma  \leftarrow \bsgamma + \frac{\epsilon}{2} \nabla  \log p_{\mathcal{H}}(\bfx_{\beta_i} \mid \cdots)$.

\item Project the momentum onto the tangent space at $\bfx_{\beta_i}$ by setting $\bsgamma \leftarrow \left(\bfI_d-\bfx_{\beta_i}\bfx_{\beta_i}^T \right)\bsgamma$, and then set the angular velocity of the geodesic flow $\nu=\norm{\phi}$.

\item  Update $\bfx_{\beta_i}$ and $\bsgamma$ jointly according to the geodesic flow with step size of $\epsilon$,
\begin{align*}
\bfx_{\beta_i} \leftarrow&\bfx_{\beta_i} \,\cos(\nu \epsilon)+\frac{\phi}{\nu}\,\sin(\nu \epsilon), &
\phi\leftarrow&\phi \,\cos(\nu \epsilon)-\nu \, \bfx_{\beta_i} \,\sin(\nu \epsilon).
\end{align*}

\item Update $\bsgamma  \leftarrow \bsgamma+\frac{\epsilon}{2} \nabla \, \log p_{\mathcal{H}}(\bfx_{\beta_i} \mid \cdots)$.

\item Project the momentum onto the tangent space at $\bfx_{\beta_i}$ by setting $\bsgamma \leftarrow \left(\bfI_d-\bfx_{\beta_i}\bfx_{\beta_i}^T \right)\bsgamma$.
\end{enumerate}

\item Set the proposed values as $\bfx_{\beta_i}^{(p)} = \bfx_{\beta_i}$ and the proposed value $\bfx_{\beta_i}^{{(p)}}$ is accepted with probability
$$
\min \left\{ 1 , \frac{p_{\mathcal{H}}\left(\bfx^{(p)}_{\beta_i} \mid \cdots\right) \exp \left\{ -\frac{1}{2} \left[ \bsgamma^{(p)} \right]^T \bsgamma^{(p)} \right\}}{p_{\mathcal{H}}\left(\bfx^{(c)}_{\beta_i} \mid \cdots \right)\exp \left\{ -\frac{1}{2} \left[ \bsgamma^{(c)} \right]^T \bsgamma^{(c)} \right\}}  \right\}
$$
\end{enumerate}

The gradient of the logarithm of the Hausdorff measure may appear forbidding to derive.  One practical difficulty is dealing with the spherical constraint $\bfx_{\beta_i}^T\bfx_{\beta_i} = 1$.  We discuss the calculation of the gradient in the online supplementary materials, where we also derive a recursive formula linking the gradient of the prior density in dimensions $K$ to that of the gradient in dimension $K-1$.  Detailed expressions for the Hausdorff measures associated with the full conditional posteriors of the $\bfx_{\beta_i}$s, $\bfx_{\psi_j}$s and $\bfx_{\zeta_j}$s, as well as their corresponding gradients, can be found in the online supplementary materials. In our experiments, we periodically ``jitter'' the step sizes and the number of leap steps (e.g., see \citealp[pg.~306]{gelman2013bayesian}).  The specific range in which $\epsilon$ and $L$ move for each (group of) parameter and each dataset is selected to target an average acceptance probability between 60\% and 90\% \citep{beskos2013optimal,betancourt2014optimizing}.

Our implementation of the algorithm is available at \url{https://github.com/forreviewonly/review/}.

\section{Illustrations}\label{se:illustration}

In this section, we illustrate the performance of the proposed model on both simulated and real data sets. In all of these analyses, the number of leaps used in the HMC steps is randomly selected from a discrete uniform distribution between 1 and 10 every 50 samples.  Similarly, the leap sizes are drawn from uniform distribution on $(0.01,0.03)$ or $(0.01,0.05)$ for each $\bsbeta_i$, and from a uniform distribution on $(0.01,0.07)$ or $(0.01,0.105)$ for each $\bszeta_j$ and $\bspsi_j$.  All inference presented in this Section are based on 20,000 samples obtained after convergence of the Markov chain Monte Carlo algorithm.  The length of the burn in period varied between 10,000 and 20,000 iterations depending on the dataset, with a median around 10,000.  Convergence was checked by monitoring the value of the log-likelihood function, both through visual inspection of the trace plot, and by comparing multiple chains using the procedure in \cite{GeRu92}.

\subsection{Simulation Study}\label{se:simulation}

We conducted a simulation study involving four distinct scenarios to evaluate our spherical model.  For each scenario, we generate one data set consisting of $J=700$ items and $I=100$ subjects (similar in size to the roll call data from the U.S.\ Senate presented in Section \ref{se:senate_102_115}).  In our first three scenarios, the data is simulated from spherical factor models on $\mathcal{S}^2$, $\mathcal{S}^3$ and $\mathcal{S}^5$, respectively.  In all cases, the subject-specific latent positions, as well as 
the item-specific latent positions, are sampled from spherical von-Mises distributions where all component-wise precisions are equal to 2.  Note that this data generation mechanism is slightly different from the model we fit to the data (in which the precision of the components increase with the index of the dimension).  For the fourth scenario, data was simulated from a Euclidean probit model (recall Section \ref{se:euclideanmodel}) in which the intercepts $\mu_1, \ldots, \mu_J$, the  discrimination parameters $\bsalpha_1, \ldots, \bsalpha_J$, and latent traits $\bsbeta_1, \ldots, \bsbeta_I$  are sampled from standard Gaussian distributions. 

\begin{figure}
  \centering
\begin{tabular}{cccc}
    & \multirow{2}{1cm}{$DIC$} & In-sample  & Principal nested \\ 
    & & accuracy & sphere decomposition \\ 
    \begin{sideways} \;\;\;\;\;\;\; Scenario 1 ($\mathcal{S}^2$) \end{sideways}&
    \includegraphics[width=.25\textwidth]{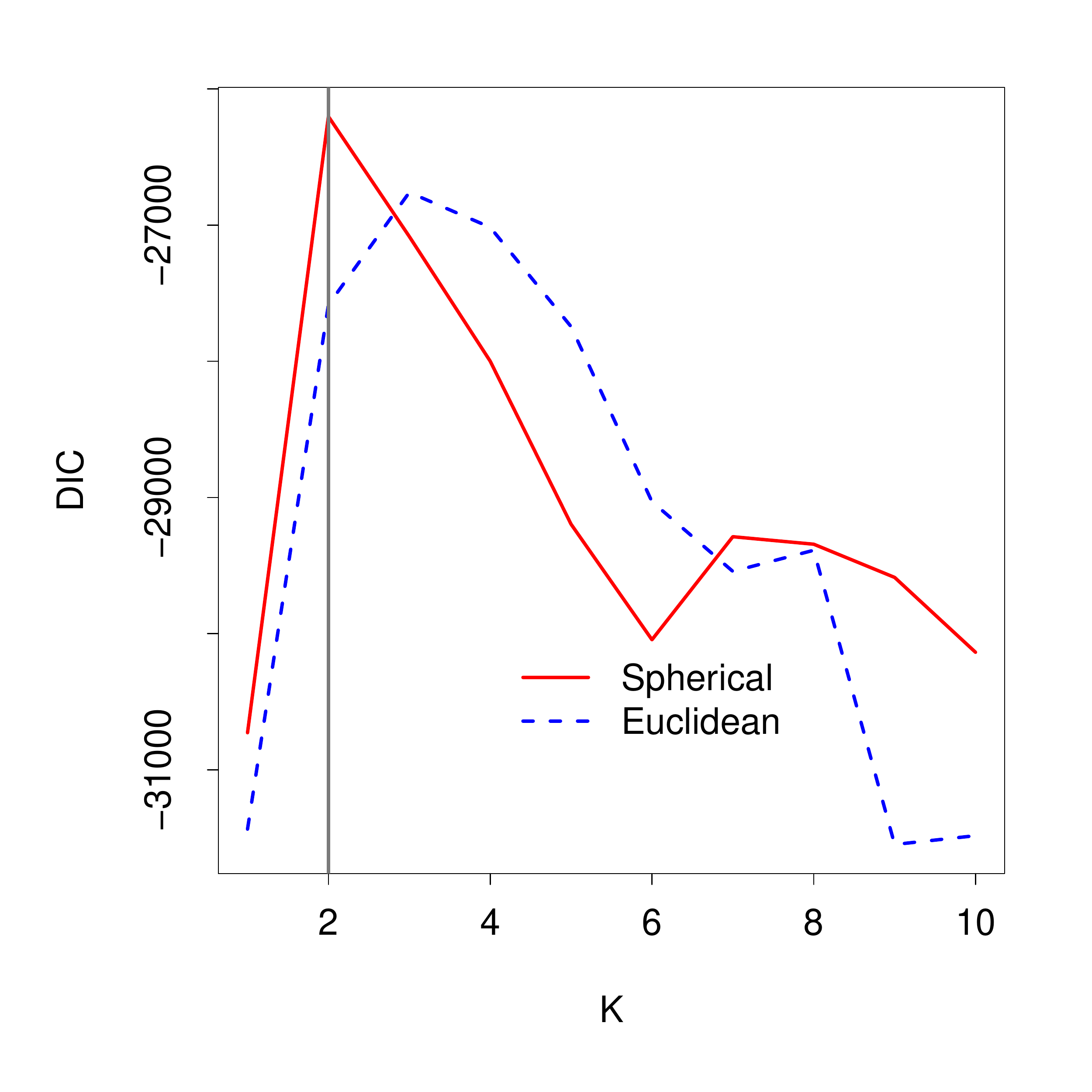} &
    \includegraphics[width=.25\textwidth]{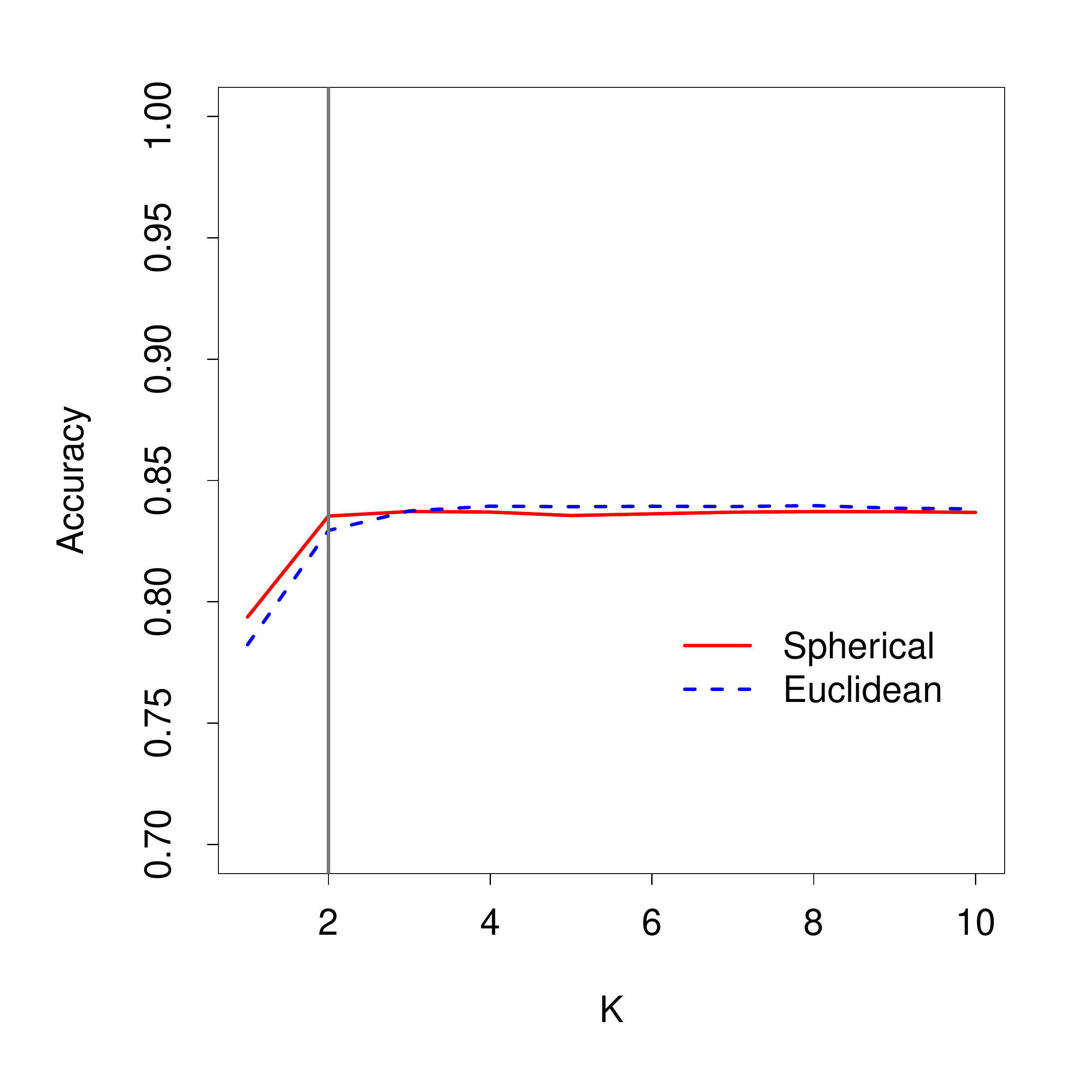} &
    \includegraphics[width=.25\textwidth]{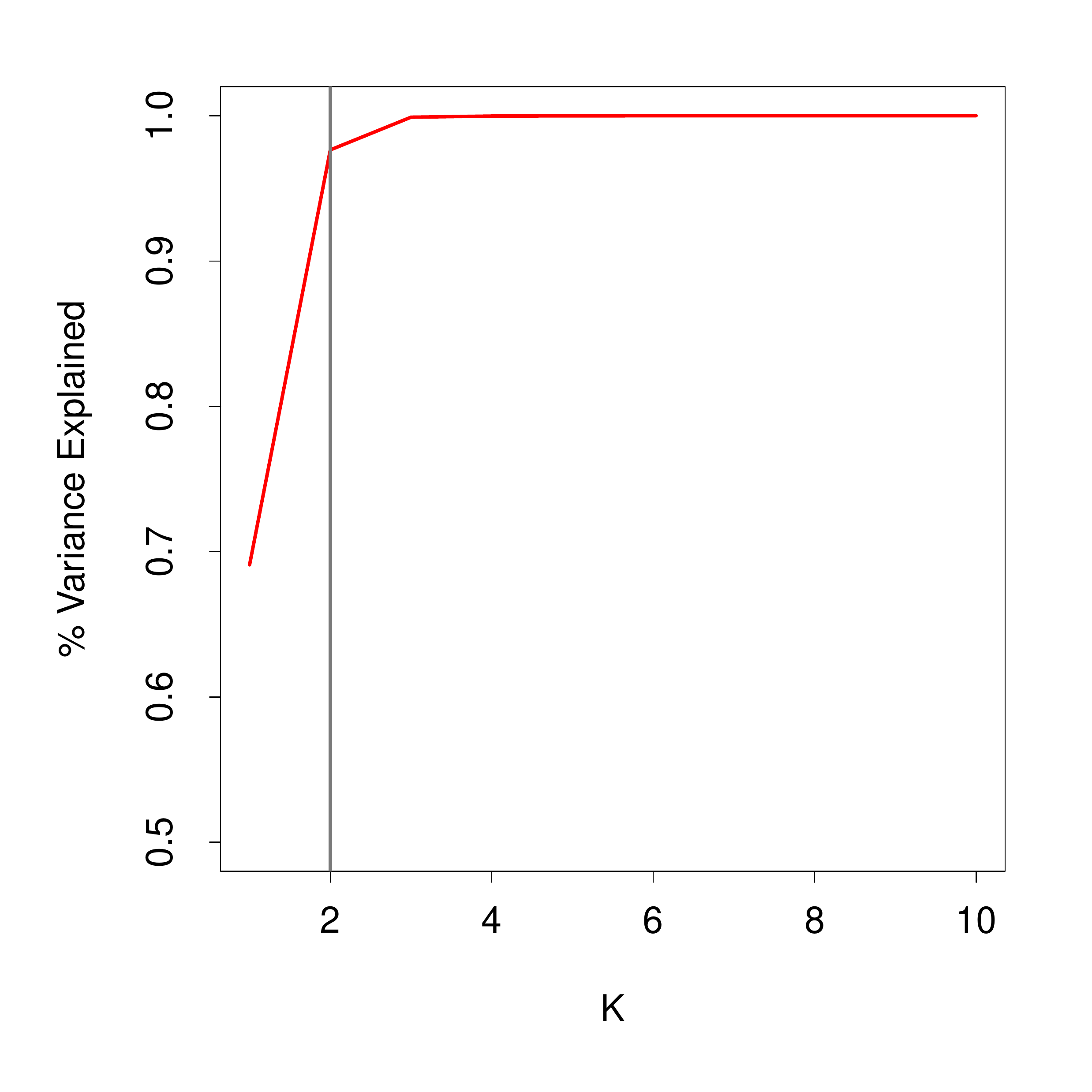} \\
    \begin{sideways} \;\;\;\;\;\;\; Scenario 2 ($\mathcal{S}^3$) \end{sideways} & 
    \includegraphics[width=.25\textwidth]{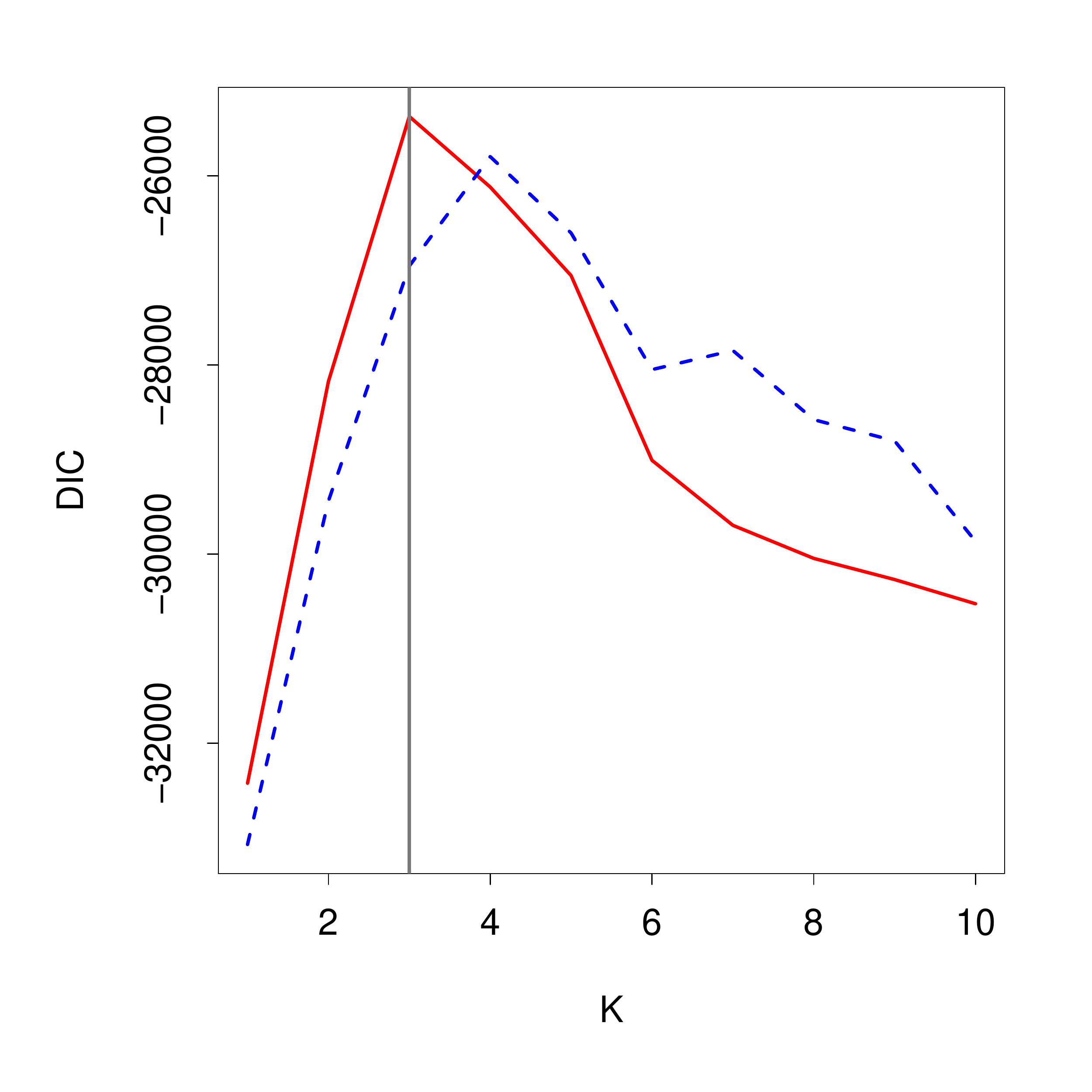} &
    \includegraphics[width=.25\textwidth]{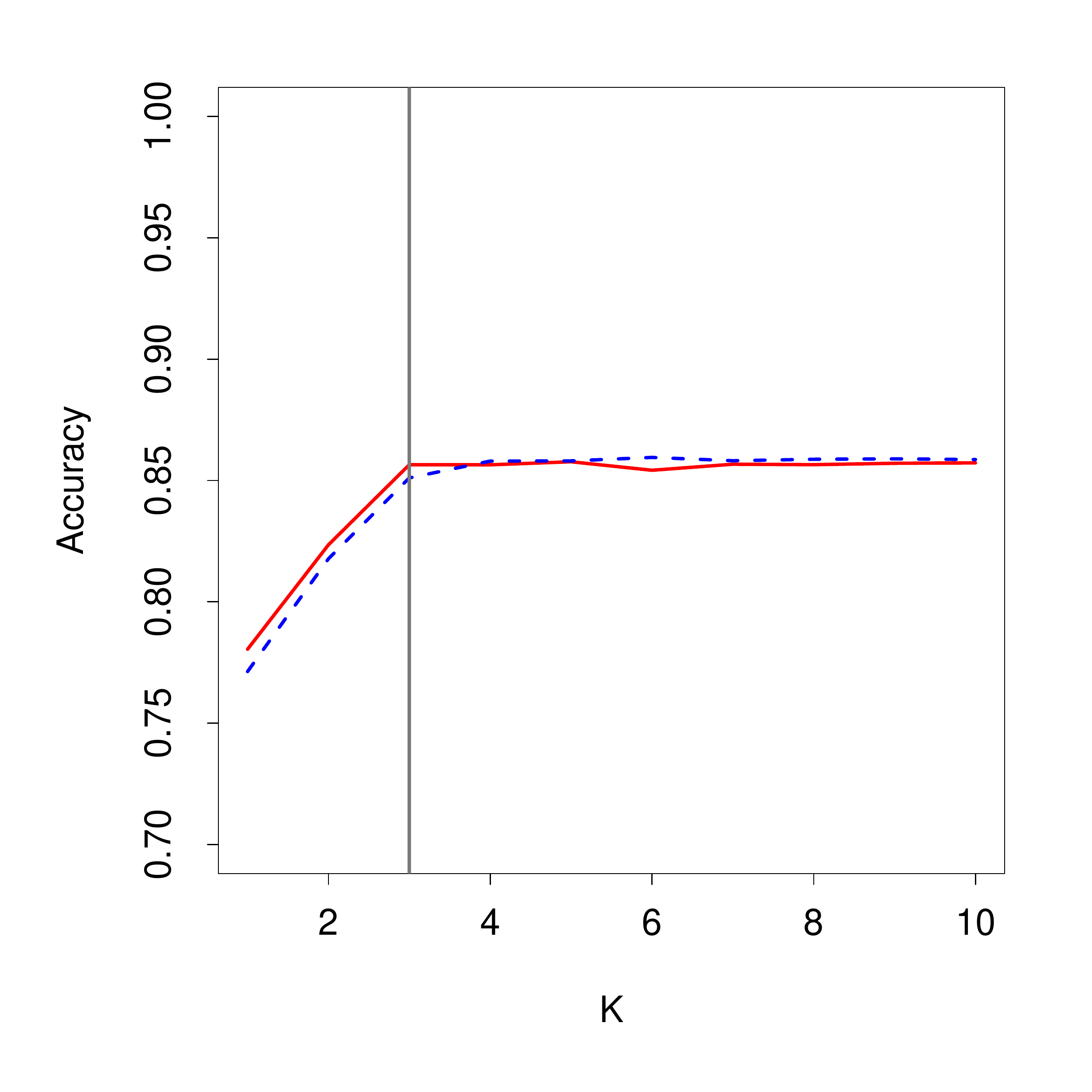} &
    \includegraphics[width=.25\textwidth]{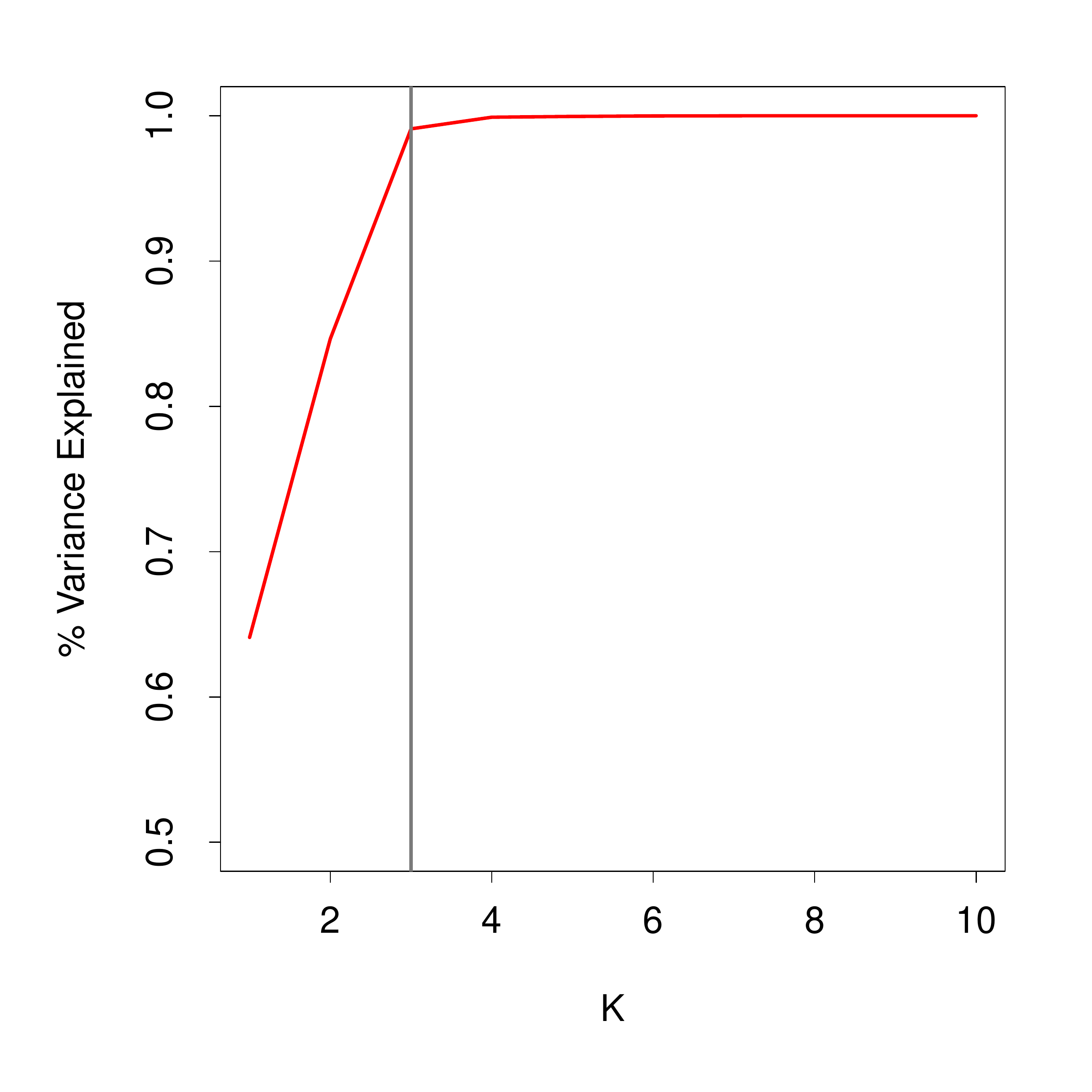} \\
    \begin{sideways} \;\;\;\;\;\;\; Scenario 3 ($\mathcal{S}^5)$ \end{sideways} & 
    \includegraphics[width=.25\textwidth]{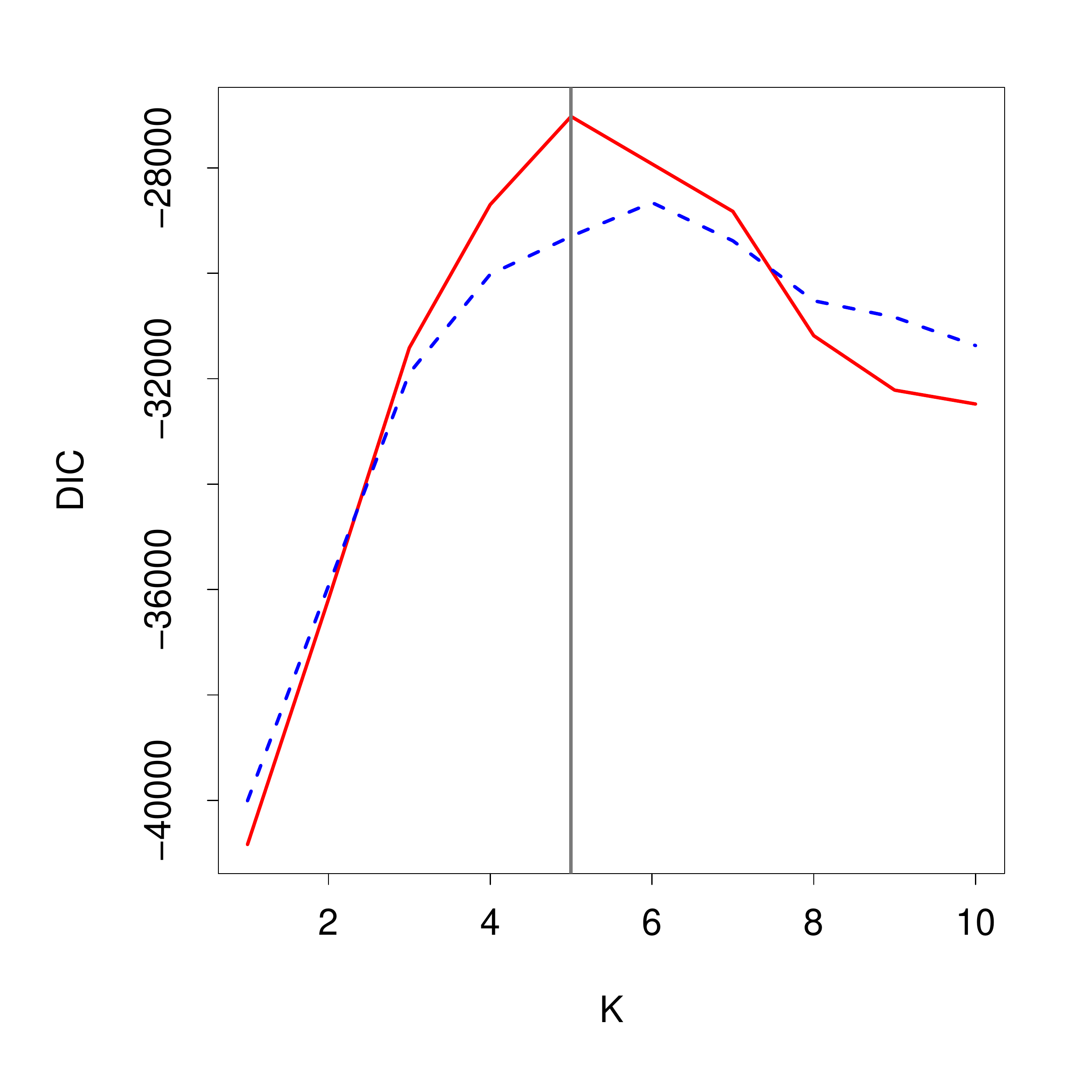} &
    \includegraphics[width=.25\textwidth]{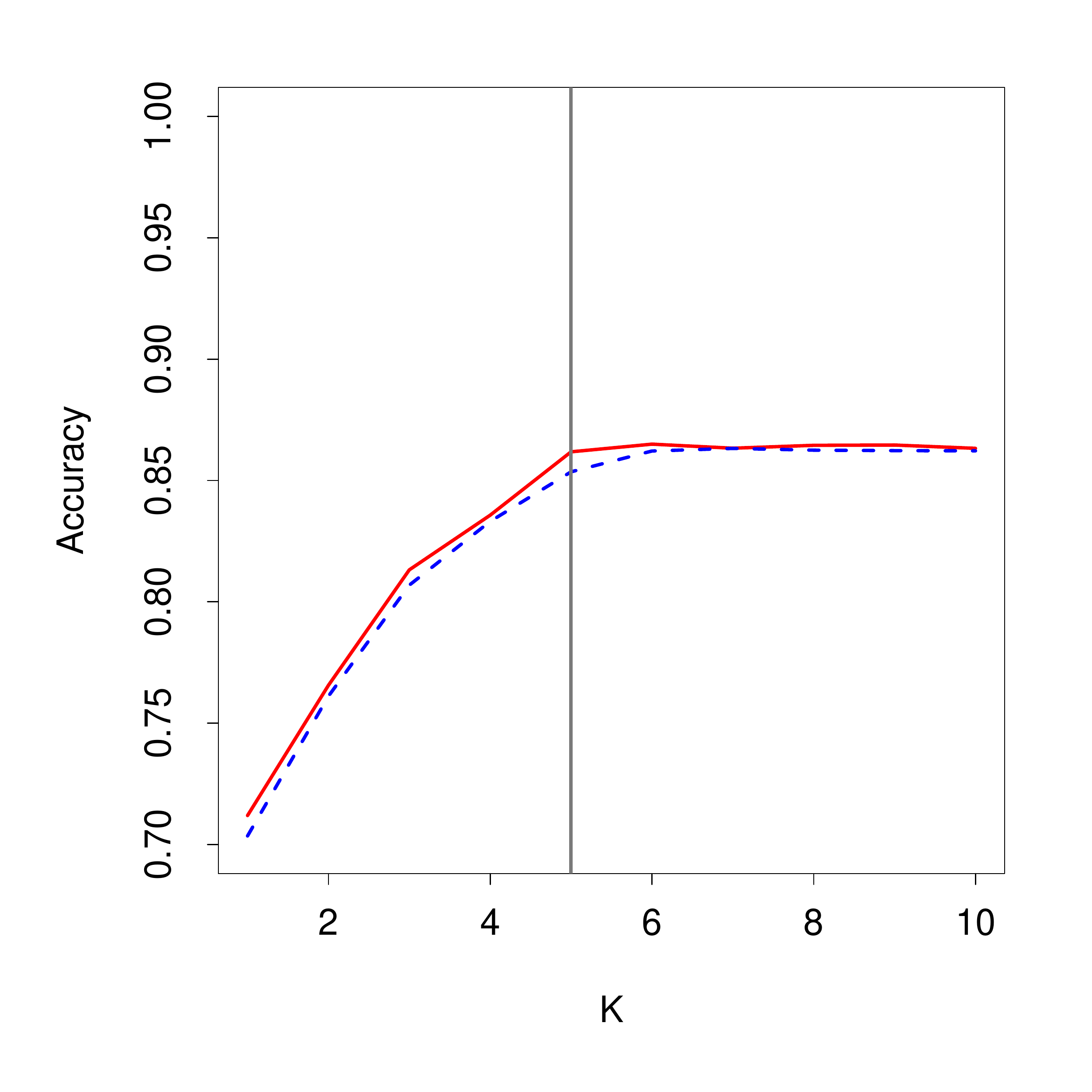} &
    \includegraphics[width=.25\textwidth]{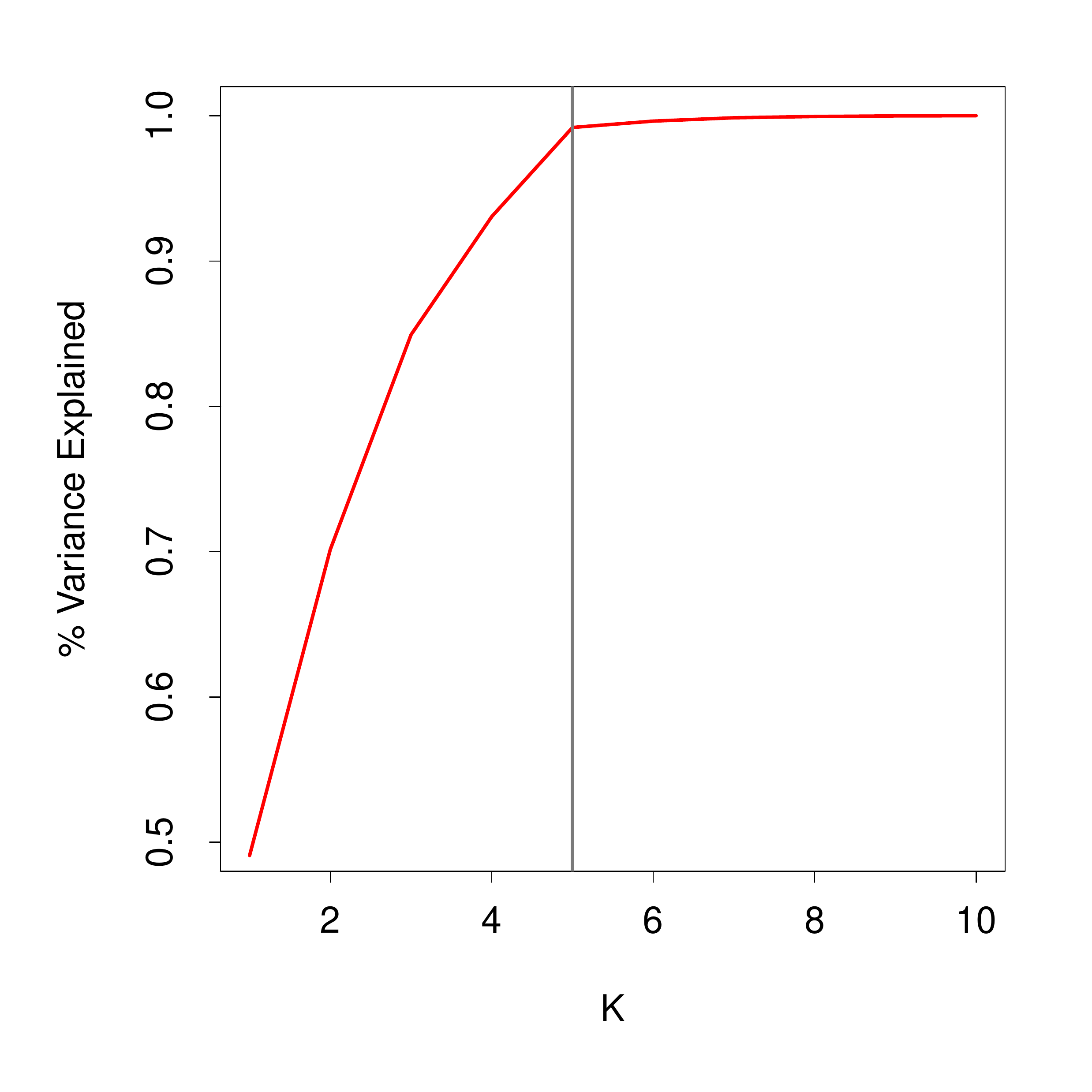} \\
    \begin{sideways} \;\;\;\;\;\;\; Scenario 4 ($\reals^3$) \end{sideways} & 
    \includegraphics[width=.25\textwidth]{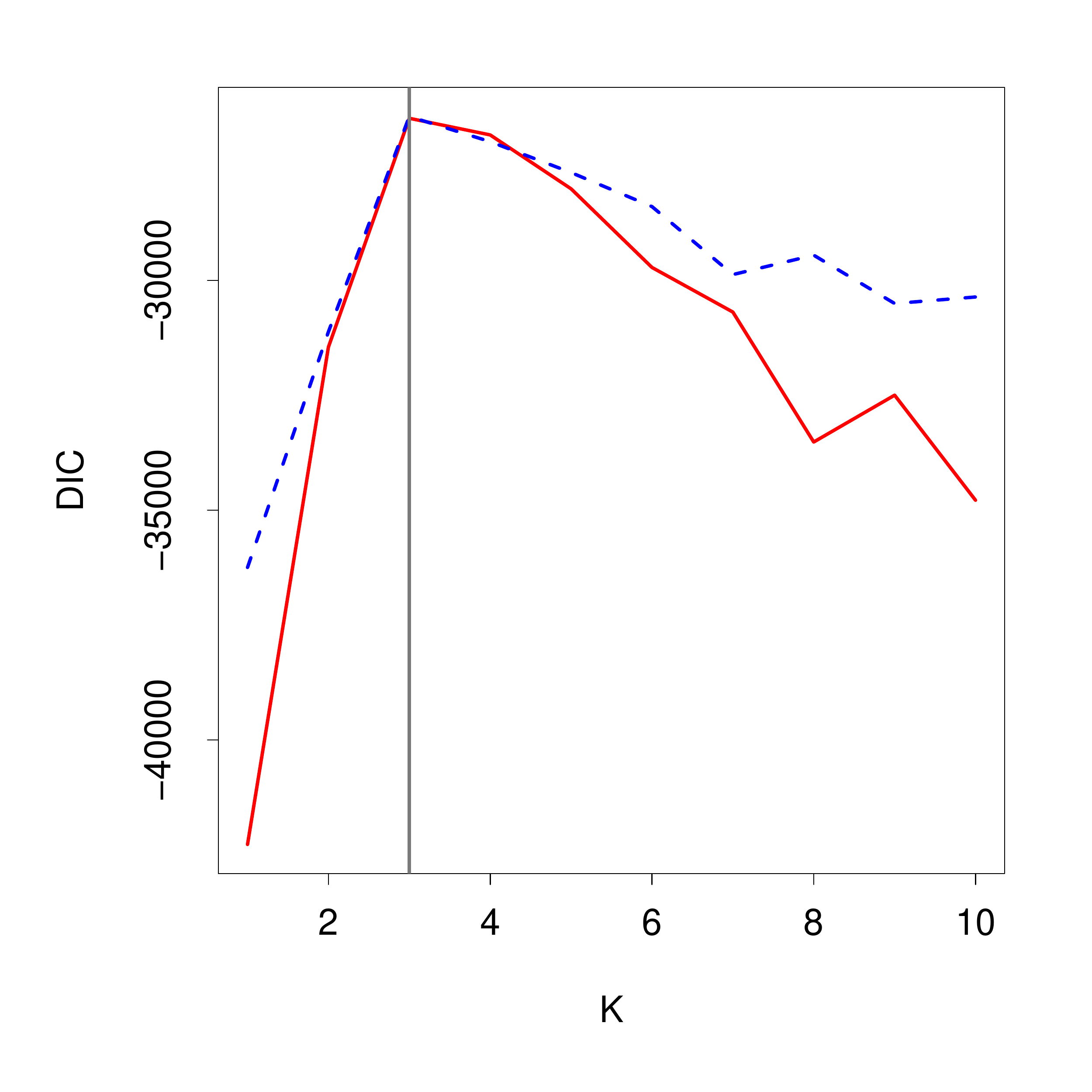} &
    \includegraphics[width=.25\textwidth]{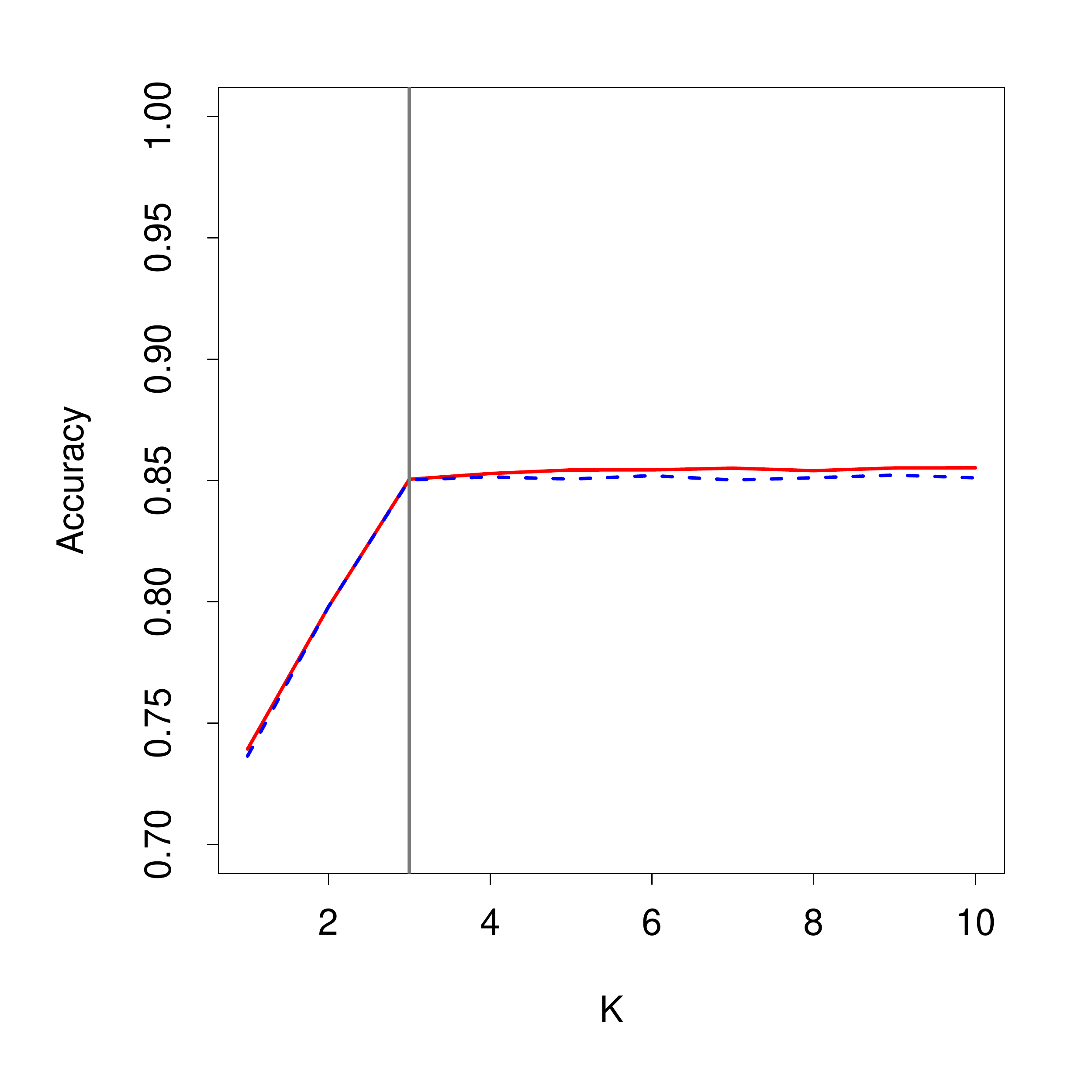} &
    \includegraphics[width=.25\textwidth]{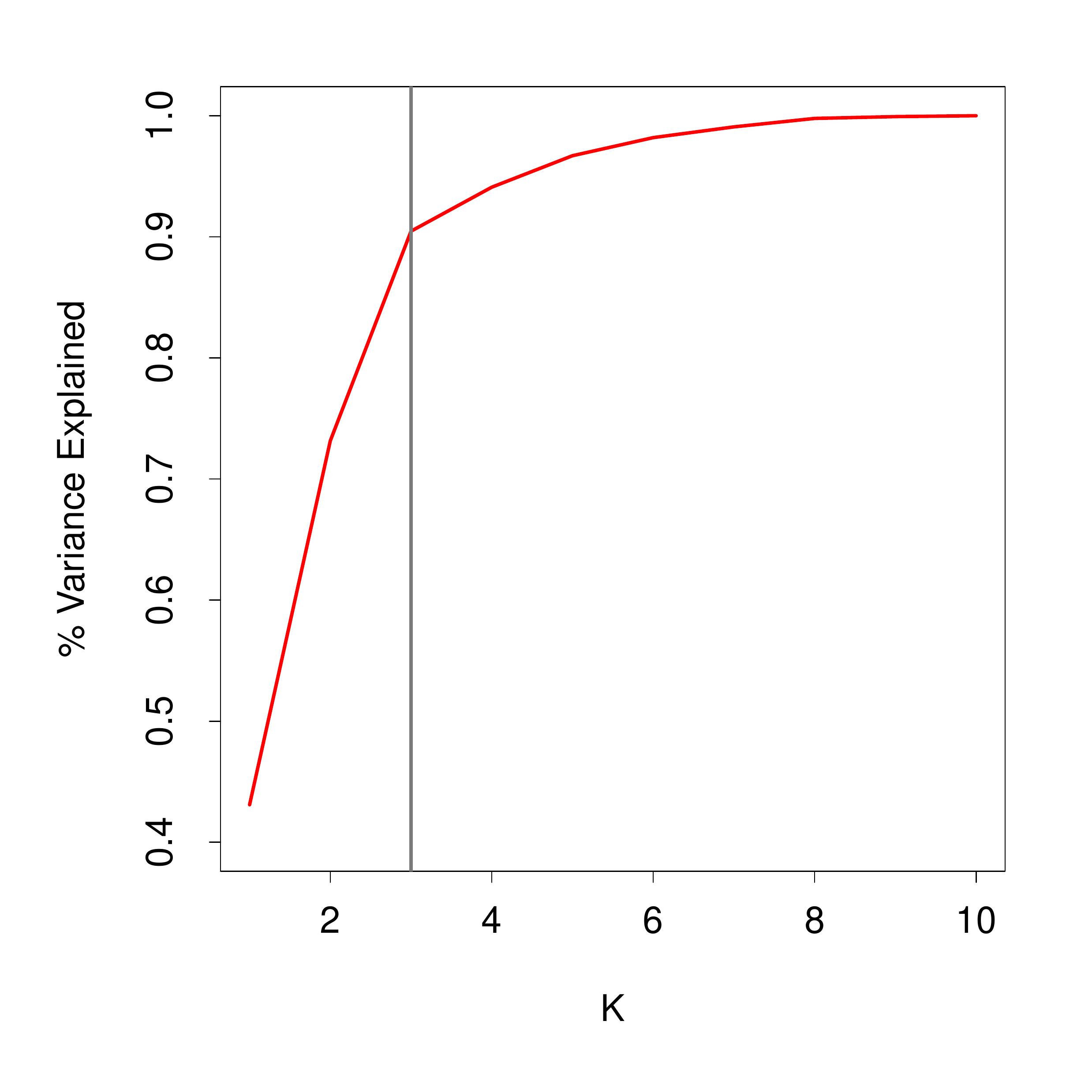} \\\end{tabular}
\caption{Deviance information criteria, in-sample predictive accuracy and principal nested sphere decomposition of the fitted models in each of the four simulation scenarios.}\label{fi:simulationdimres}
\end{figure}

In each scenario, both spherical and Euclidean probit factor models of varying dimension are fitted to the simulated datasets.  The left column of Figure \ref{fi:simulationdimres} shows the value of the Deviance Information Criteria (DIC) \citep{spiegelhalter2002bayesian,spiegelhalter2014deviance,gelman-2014-information} as a function of the dimension $K$ of the fitted model's latent space for each of the four scenarios in our simulation.  In our models, the DIC is defined as:
$$
DIC =  \ell \left(\E\{\bsTheta \mid \mbox{data}\}\right) - 2\Var \left( \ell \left(\bsTheta\right) \mid \mbox{data} \right) ,
$$
where $\bsTheta = [\theta_{i,j}]$ is the matrix of probabilities given by
$
\theta_{i,j} = G_{\kappa_j} \left( \left\{ \rho(\bszeta_j, \bsbeta_i) \right\}^2 - \left\{ \rho(\bspsi_j, \bsbeta_i) \right\}^2 \right)
$
for the spherical model and
$
\theta_{i,j} = \Phi \left(  \mu_j + \alpha_j \beta_i \right)
$
for the Euclidean probit model, $\ell \left( \bsTheta \right)= \sum_{i=1}^{I}\sum_{j=1}^{J} \left\{y_{i,j} \log \theta_{i,j} + (1 - y_{i,j}) \log (1 - \theta_{i,j}) \right\}$, and the expectations and variances are computed with respect to the posterior distribution of the parameters (which are in turn approximated from the samples generated by our Markov chain Monte Carlo algorithm).  The first term in the DIC can be understood as a goodness-of-fit measure, while the second one can be interpreted as a measure of model complexity.

From Figure \ref{fi:simulationdimres}, we can see that DIC is capable of identifying the presence of sphericity in the underlying latent space, as well as its correct dimension.  In Scenarios 1 to 3, the optimal model under DIC is correctly identified as a spherical model with the right dimension.  Furthermore, as we would expect, the optimal Euclidean model in each case has one additional dimension (i.e., the dimension of the space corresponds to the lowest-dimensional Euclidean space in which the true spherical latent space can be embedded).  For Scenario 4, DIC correctly selects a Euclidean model in three dimensions, followed very closely by a spherical space of the same dimension. Again, these results match our previous discussion about the ability of a spherical model to approximate a Euclidean model.

The second column of Figure \ref{fi:simulationdimres} shows the in-sample predictive accuracy of the models in each scenario.  This accuracy is closely linked to the goodness-of-fit component of the DIC.  Note that, from the point of view of this metric, both the spherical and Euclidean models have about the same performance.  Furthermore, in both cases, as the dimension increases, the predictive accuracy increases sharply at first, but quickly plateaus once we reach the right model dimension, generating a clear elbow in the graph.  Similar elbows can be seen in the third column of Figure \ref{fi:simulationdimres}, which shows the amount of variance associated with each component of a principal nested spheres decomposition \citep{jung2012analysis} of the subject's latent positions for the $\mathcal{S}^{10}$.  This decomposition can be interpreted as a version of principal component analysis for data on an spherical manifold.  Note, however, that the elbow in Scenario 4 is much less sharp than the elbows we observed in Scenarios 1, 2 and 3.

\begin{figure}
    \centering
\begin{tabular}{cccc}
    & Scenario 2 ($K=3$) & Scenario 4 ($K=3$) \\
    \begin{sideways} \;\;\;\;\;\;\;\;\;\;\;\;\;\;\;\;\;\;\;\;\;\;\;\;\;      $\omega$  \end{sideways}&
    \includegraphics[width=.33\textwidth]{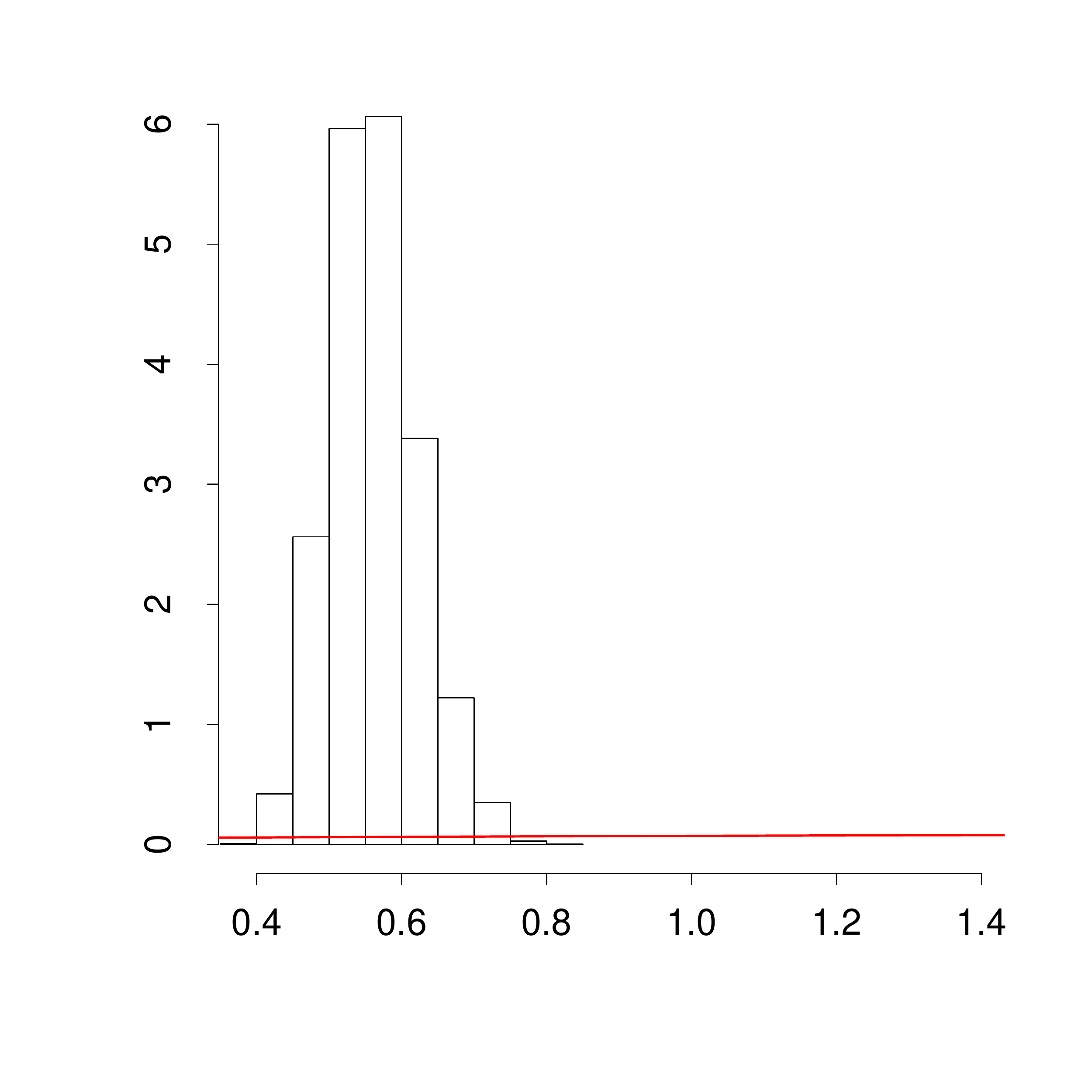} &
    \includegraphics[width=.33\textwidth]{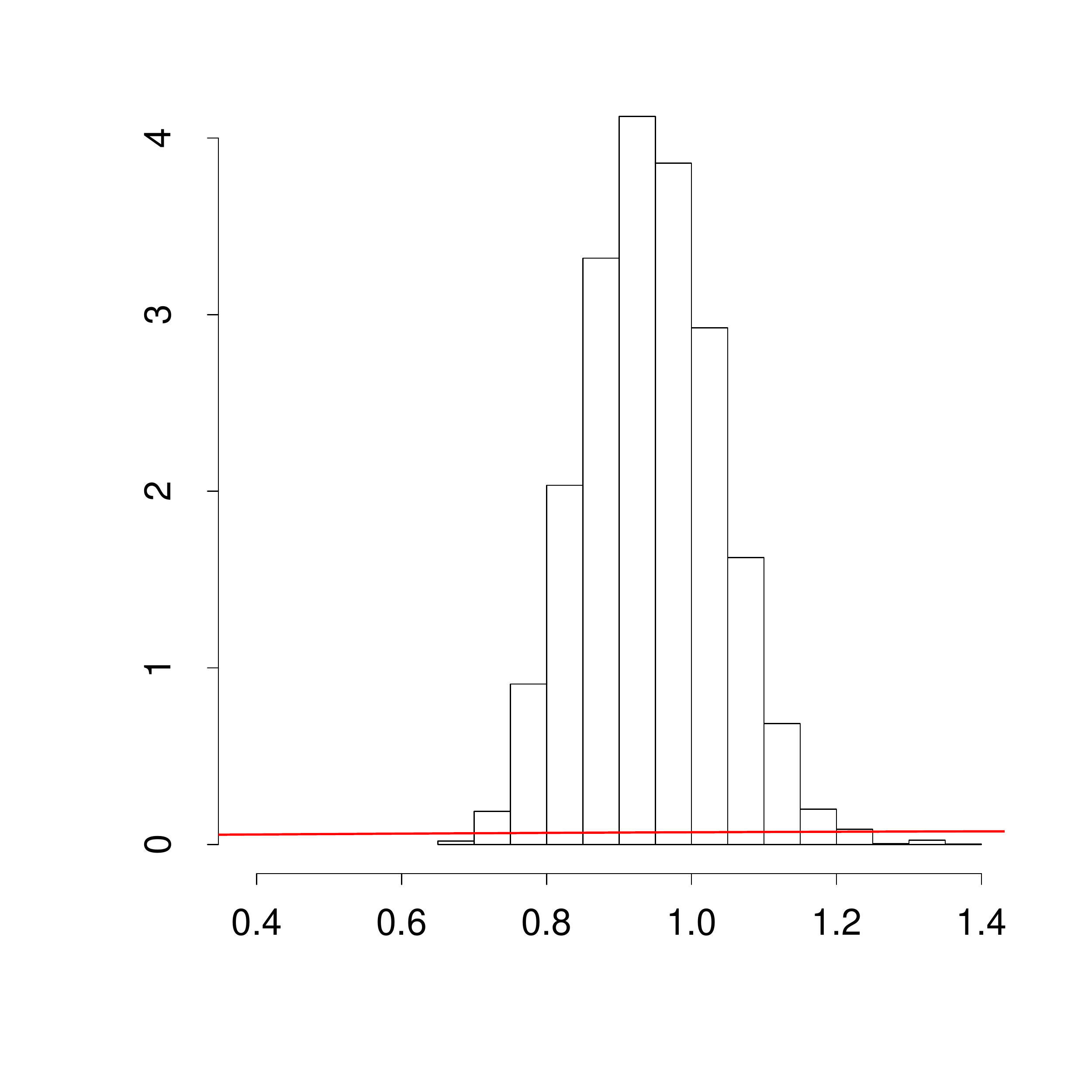} \\
    \begin{sideways} \;\;\;\;\;\;\;\;\;\;\;\;\;\;\;\;\;\;\;\;\;\;\;\;\;       $\tau$  \end{sideways}&
    \includegraphics[width=.33\textwidth]{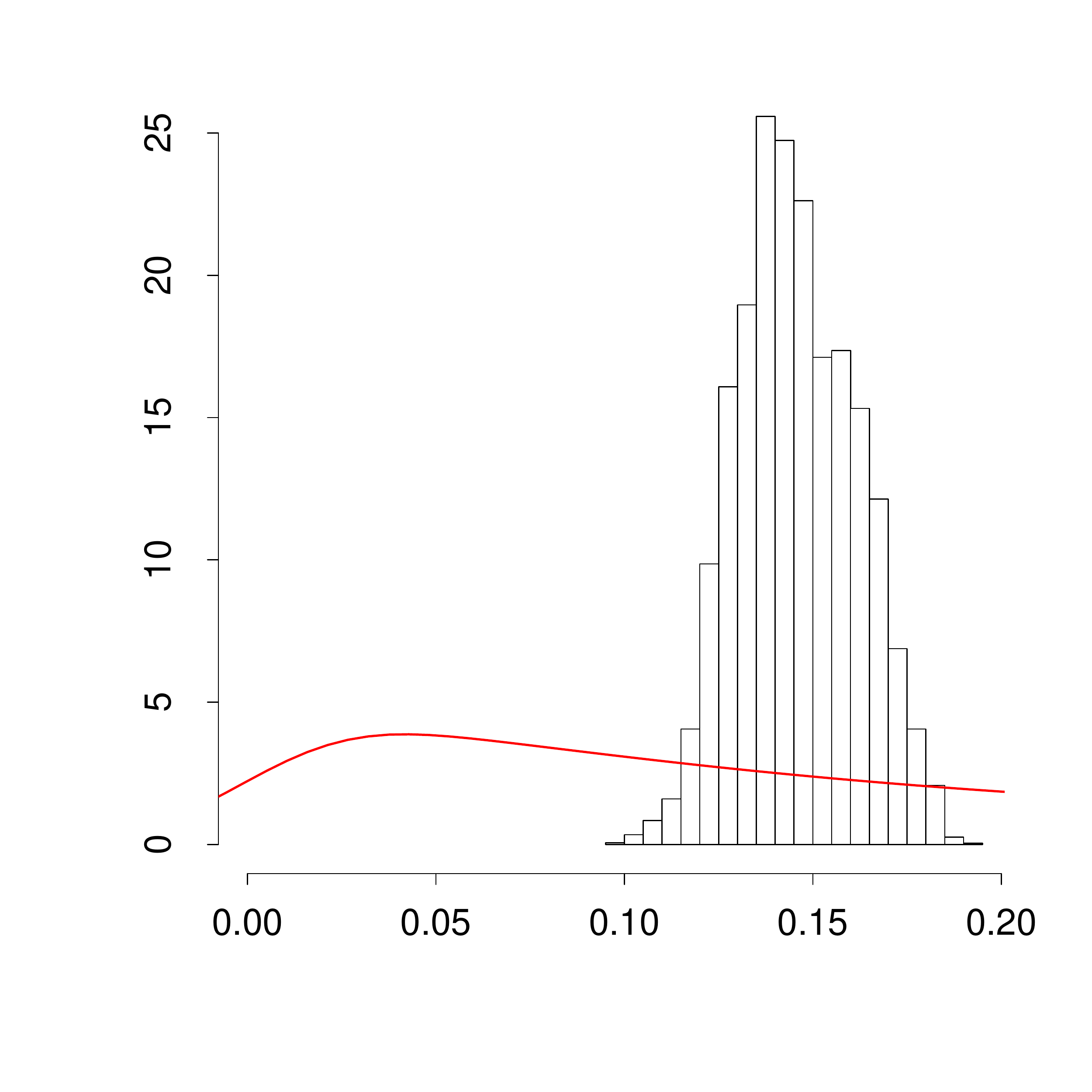} &
    \includegraphics[width=.33\textwidth]{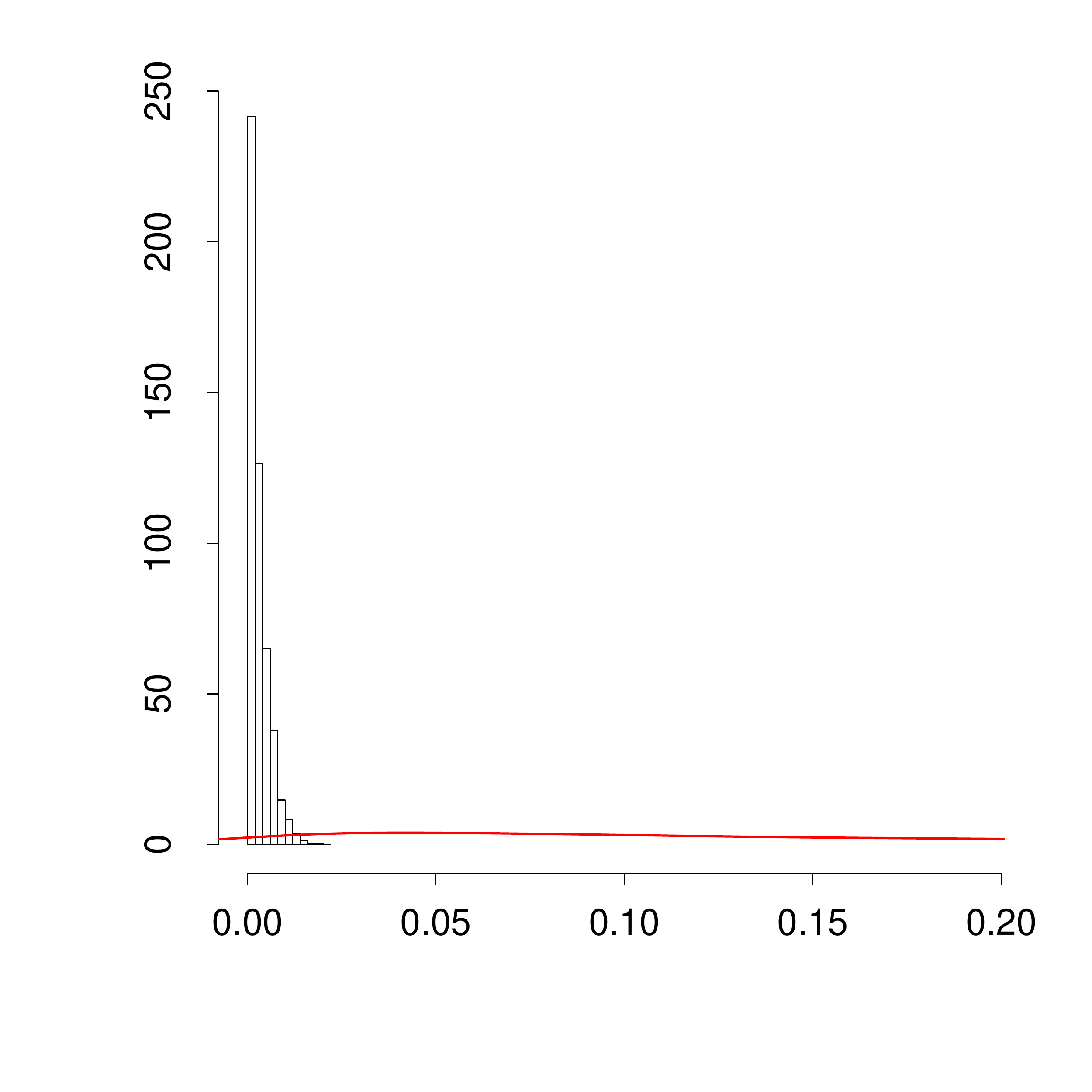} \\
    \begin{sideways} \;\;\;\;\;\;\;\;\;\;\;\;\;\;\;\;\;\;\;\;\;\;   $1/\lambda$  \end{sideways} & 
\includegraphics[width=.33\textwidth]{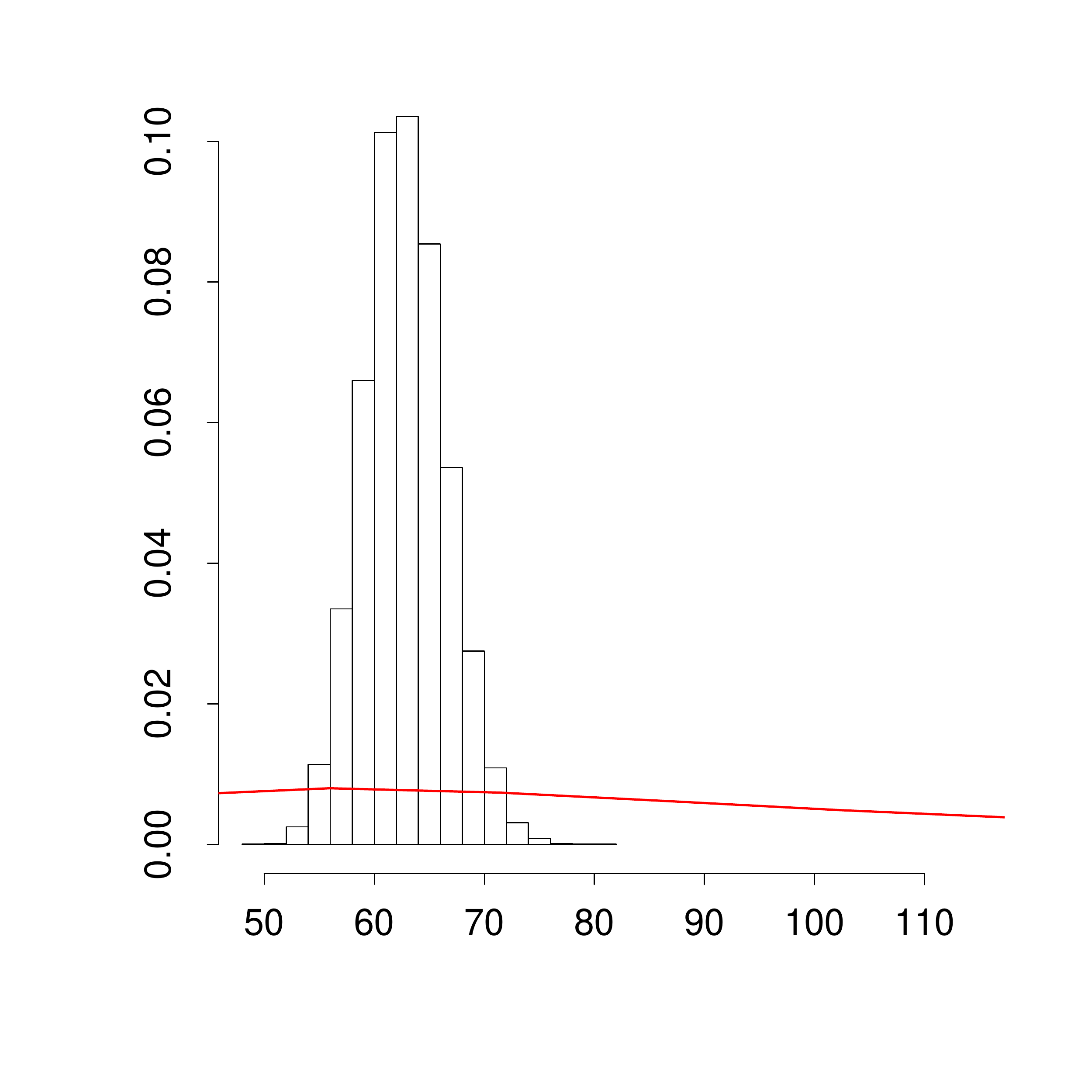}&
    \includegraphics[width=.33\textwidth]{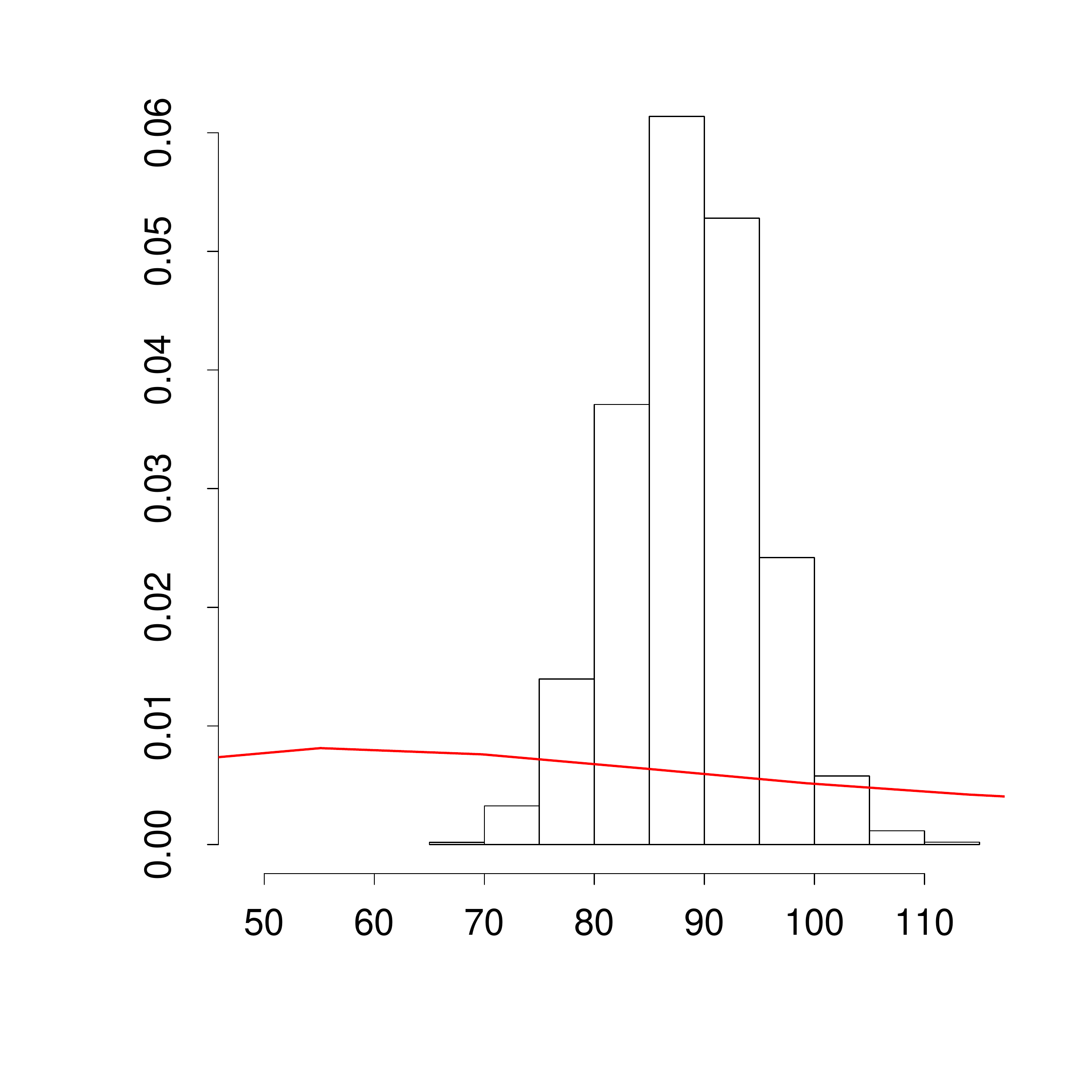} \\
    \end{tabular}
    \caption{Histograms of the posterior samples for the hyperparameters $\omega$, $\tau$ and $1/\lambda$ in scenarios 2 and 4 for $K=3$.  The continuous line corresponds to the prior distribution used in the analysis.}
    \label{fi:omegataulambda_sim}
\end{figure}

To get a better sense of the relationship between the spherical and Euclidean models, Figure \ref{fi:omegataulambda_sim} presents histograms of the posterior samples for the hyperparameters $\omega$, $\tau$ and $1/\lambda$ of our spherical model for Scenario 2 (left column) and Scenario 4 (right column).  We show these two scenarios because the true dimension of the latent space is 3 in both cases.  Note that the posterior distributions of these hyperparameters concentrate on very different values depending on whether the truth corresponds to a Euclidean or a spherical model.  Furthermore, in all cases the posterior distributions are clearly different from the priors.

\begin{figure}
    \centering
    \subfigure[Dimension 1]{\includegraphics[width=0.45\textwidth]{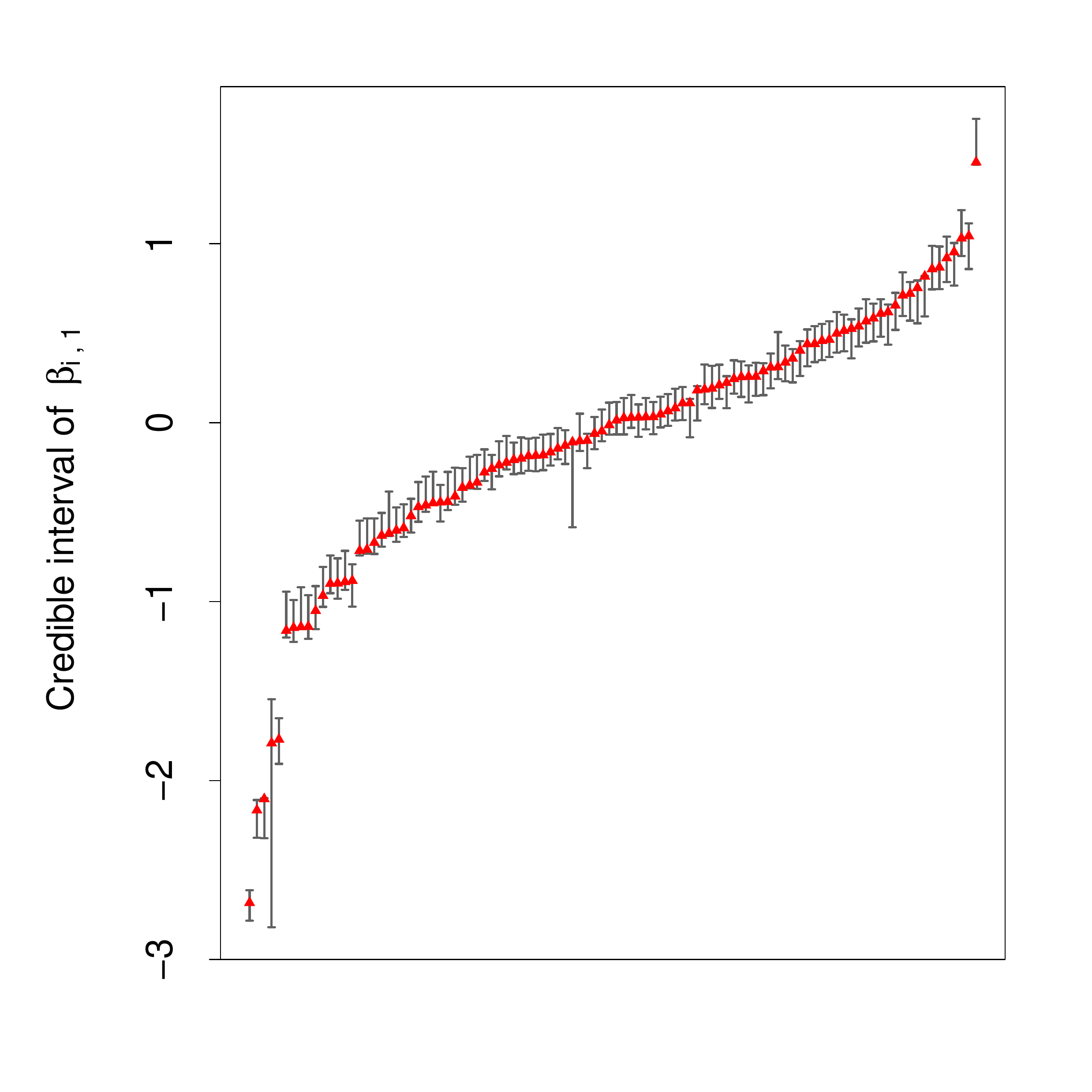}}
    \subfigure[Dimension 2]{\includegraphics[width=0.45\textwidth]{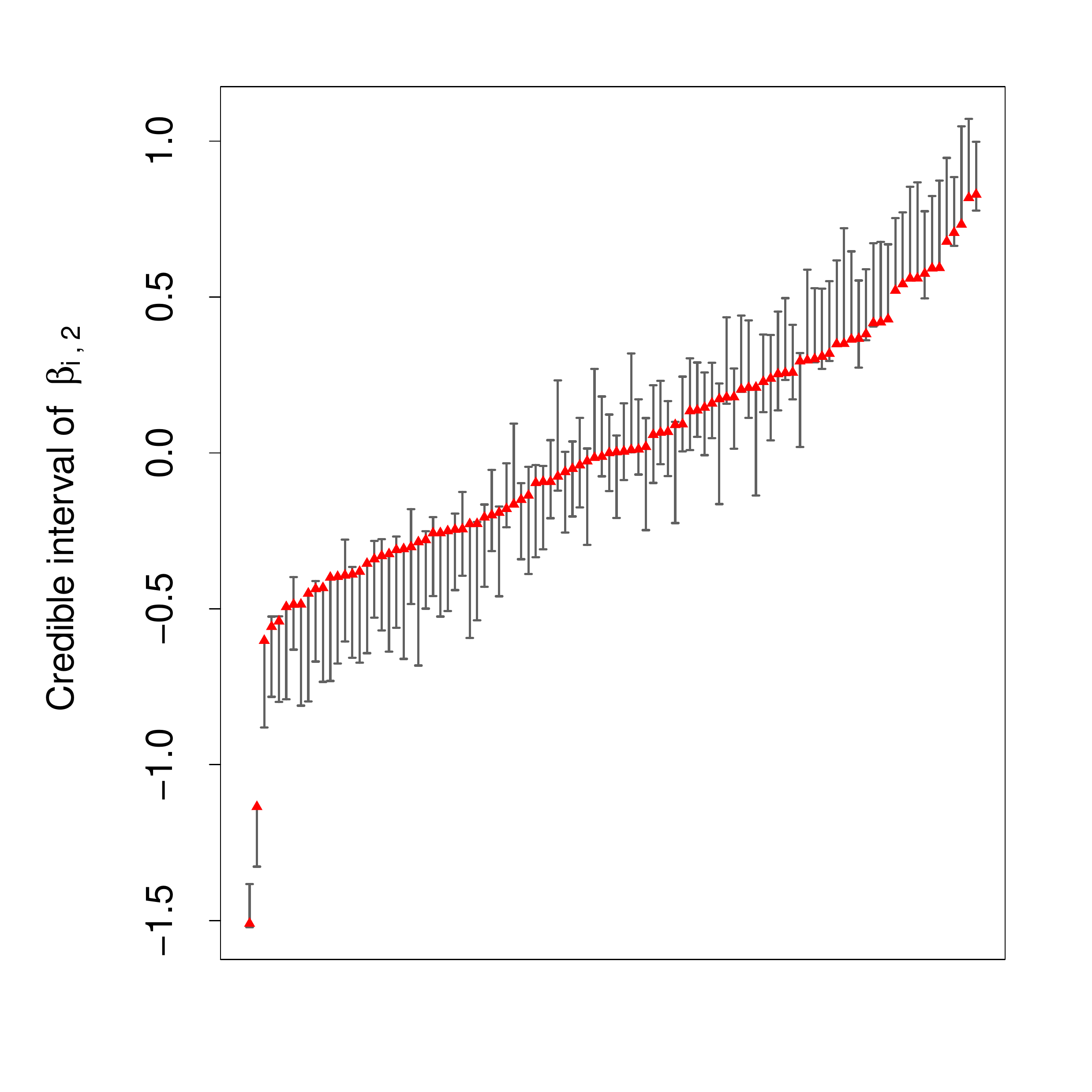}}
    \caption{Scenario 1, 95\% credible interval for the latent traits in each of the two dimensions.  The true value of the traits is represented using a  red triangle.  Coverage rates are 99\% and 68\%, respectively.}
    \label{fi:simulationlatentresor}
\end{figure}

Finally, Figure \ref{fi:simulationlatentresor} presents posterior estimates for the subject's latent positions for Scenario 1. To facilitate comparisons between the truth and the estimates, we plot each dimension separately.  While the first dimension seems to be reconstructed quite accurately (the coverage rate for the 95\% credible intervals shown in the left panel is 98\%, with an approximate 95\% confidence interval of $(0.953, 1.000)$), the reconstruction of the second component is less so (the coverage in this case is 68\%, with an approximate 95\% confidence interval of $(0.578,0.782)$).

\subsubsection{Sensitivity analysis}\label{se:sensitivity}

\begin{figure}
 \centering
 \includegraphics[width=0.5\linewidth]{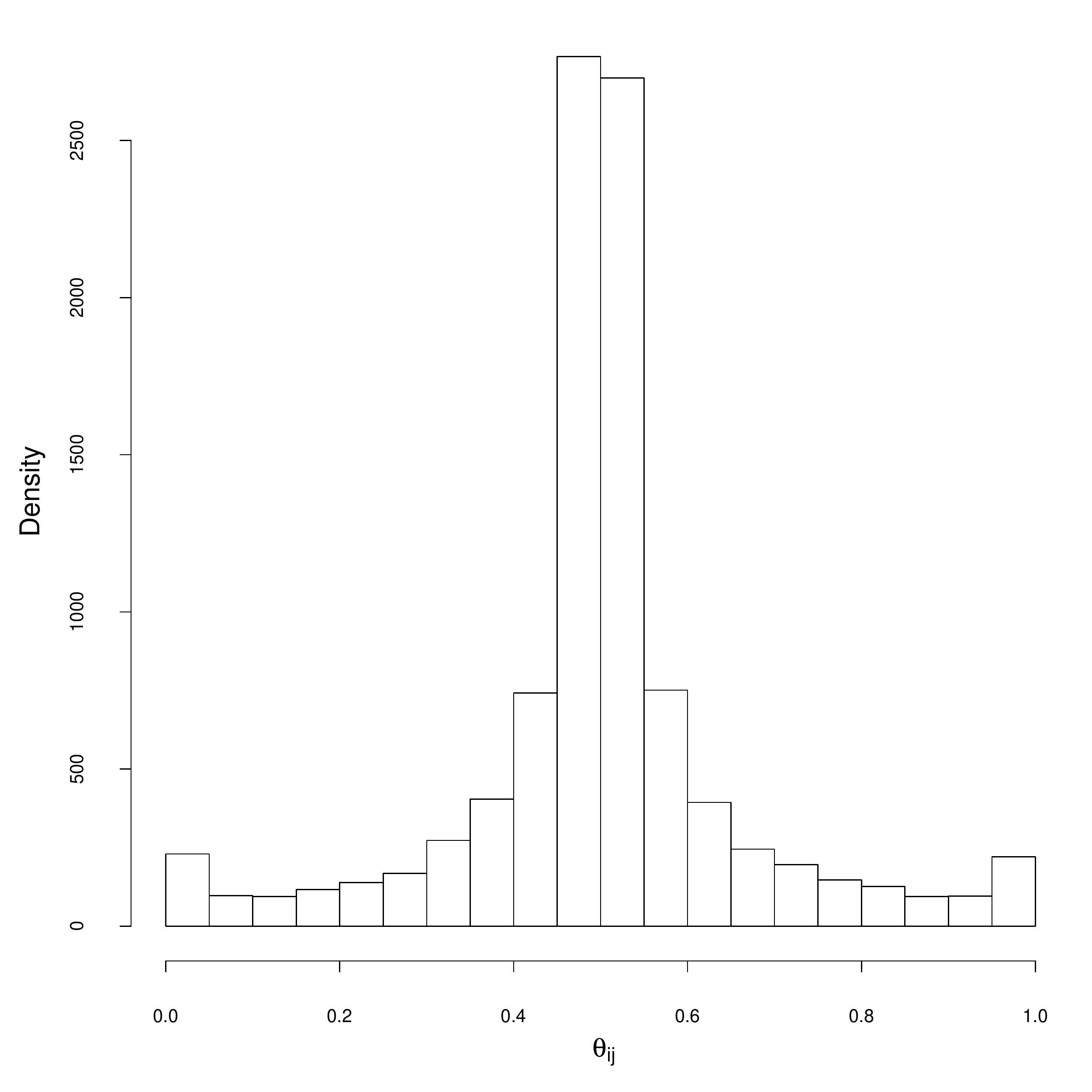}
 \caption{Histogram of 10,000 samples from the prior on $\theta_{i,j}$ implied by our alternative set of hyperparameters: $a_{\omega} = a_{\tau} = 1$, $b_{\omega} = b_{\tau} = 1/10$, $a_{\lambda} = 2$, $b_{\lambda}=25$, and $c=1$ for $K=10$.}\label{fi:finaldenthetaprioralt}
\end{figure}

To assess the sensitivity of our estimates  to the values of the hyperparameters, we repeated our analysis using a different prior specification in which $\omega \sim \mbox{Gam}(1, 1/10)$, $\tau \sim \mbox{Gam}(1, 1/10)$ and $\lambda \sim \mbox{Gam}(2, 25)$.  The induced prior on $\theta_{i,j}$ for $K=10$ can be seen in Figure \ref{fi:finaldenthetaprioralt}. (Note the very different shape when compared with Figure \ref{fi:finaldenthetaprior}.)  

\begin{figure}
    \centering
    \subfigure[$DIC$]{\includegraphics[width=.32\textwidth]{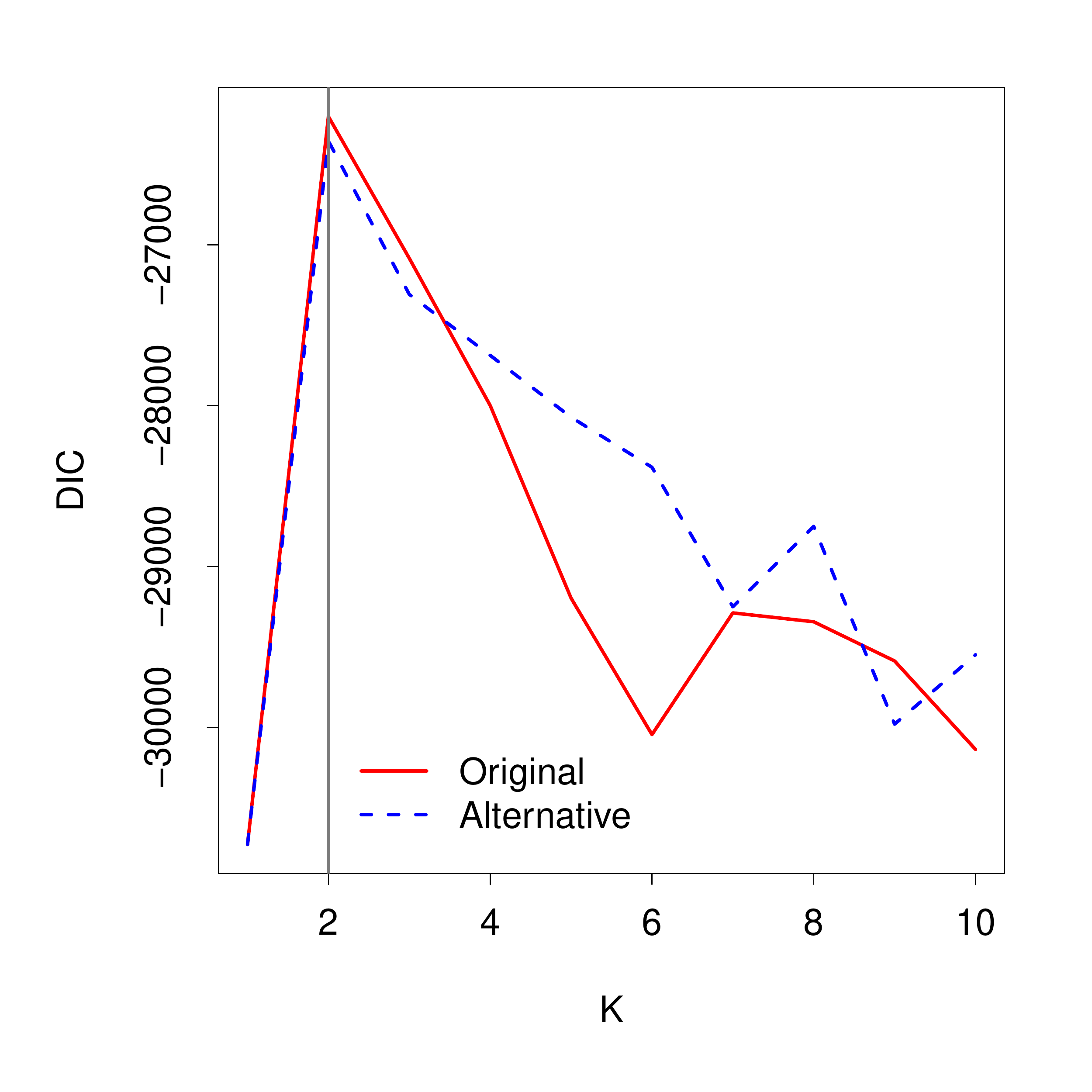}}
    \subfigure[In-sample predictive accuracy]{\includegraphics[width=.32\textwidth]{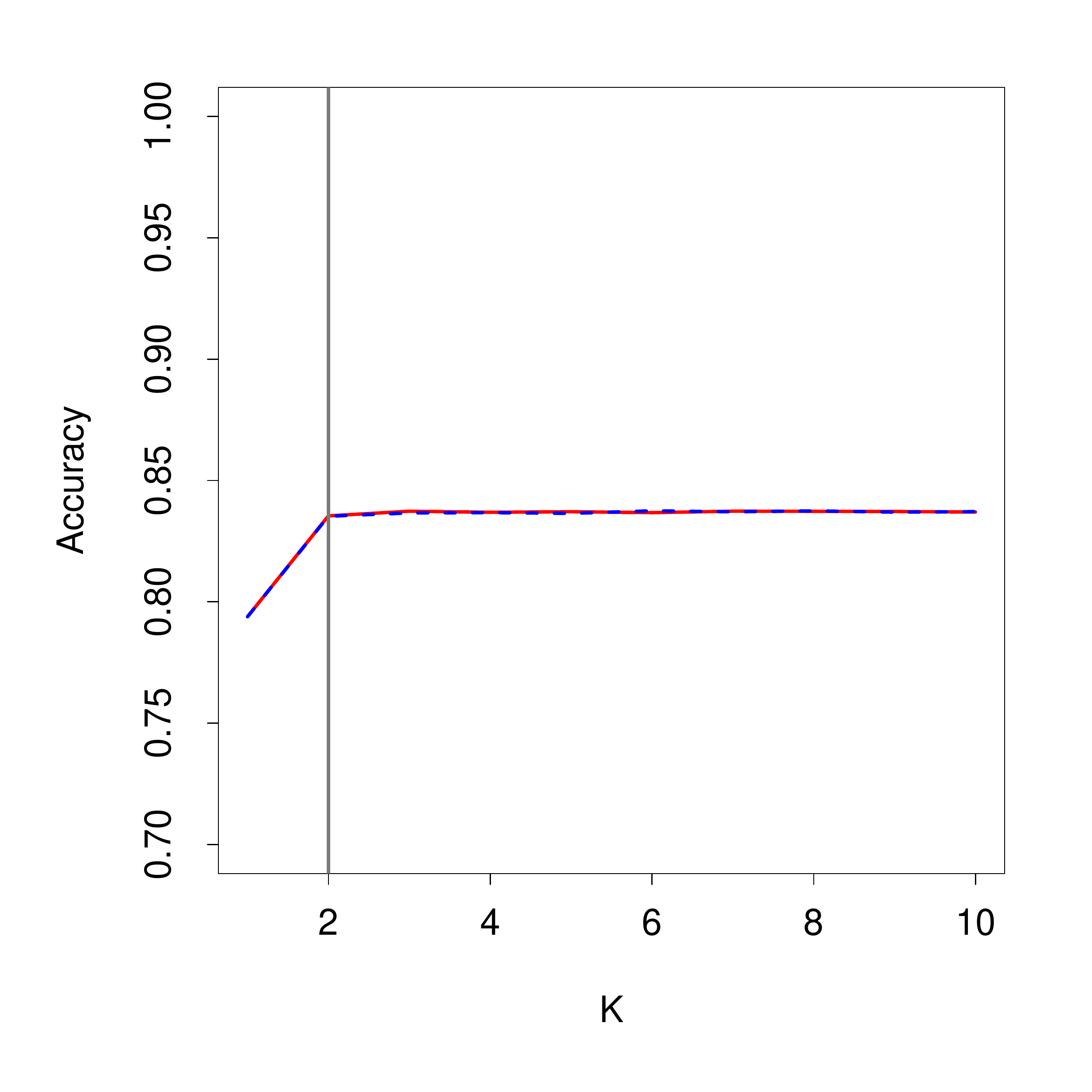}}
    \subfigure[Principal nested sphere decomposition]{\includegraphics[width=.32\textwidth]{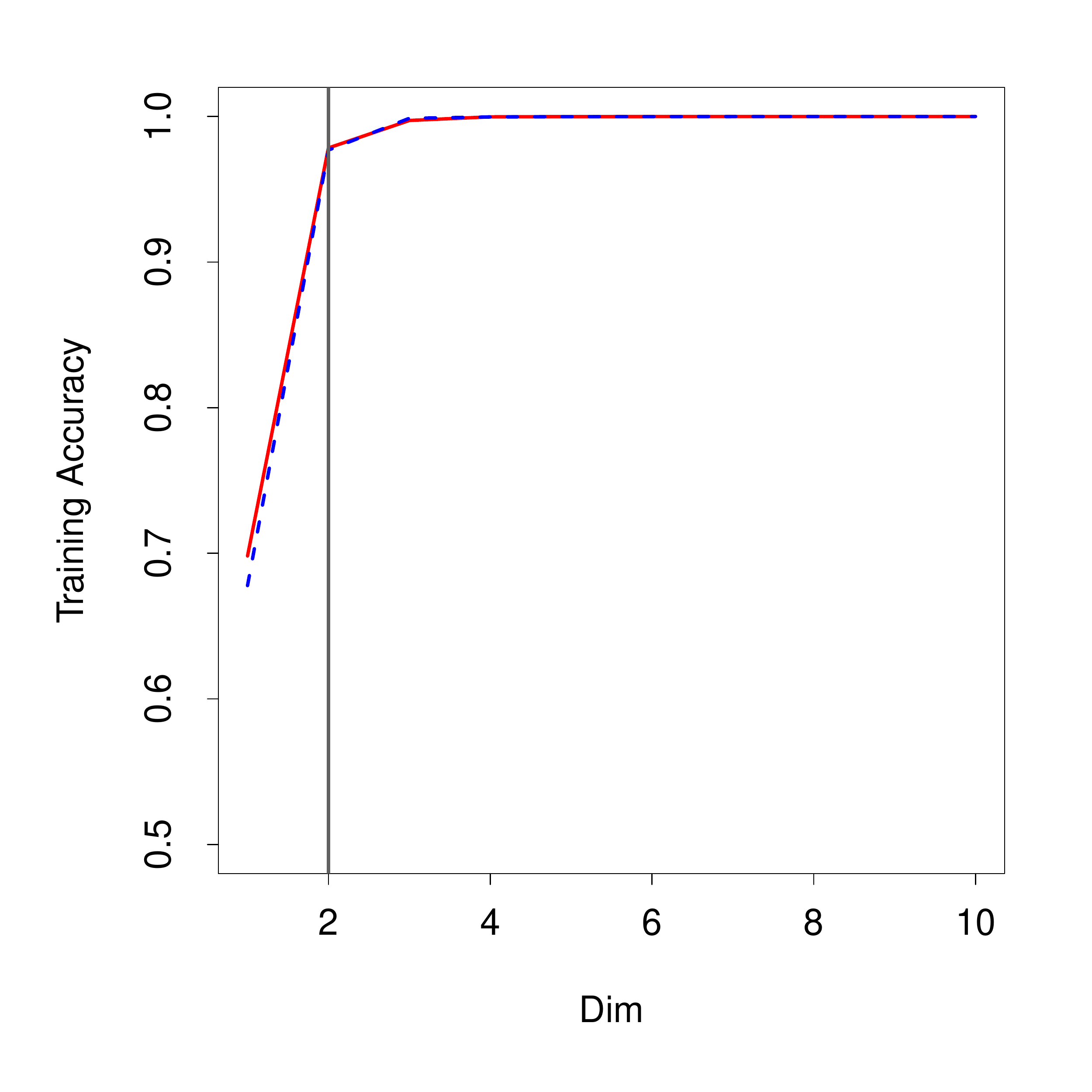}}
    \caption{Deviance information criteria, in-sample predictive accuracy and principal nested sphere decomposition of the fitted models for Scenario 1 under our original and alternative priors.}
    \label{fi:simulationdimressensitivity}
\end{figure}

\begin{figure}
    \centering
    \begin{tabular}{cccc}
    & Original prior & Alternative prior \\
    \begin{sideways} \;\;\;\;\;\;\;\;\;\;\;\;\;\;\;\;\;\;\;\;\;\;\;\;\;      $\omega$  \end{sideways}&
    \includegraphics[width=.33\textwidth]{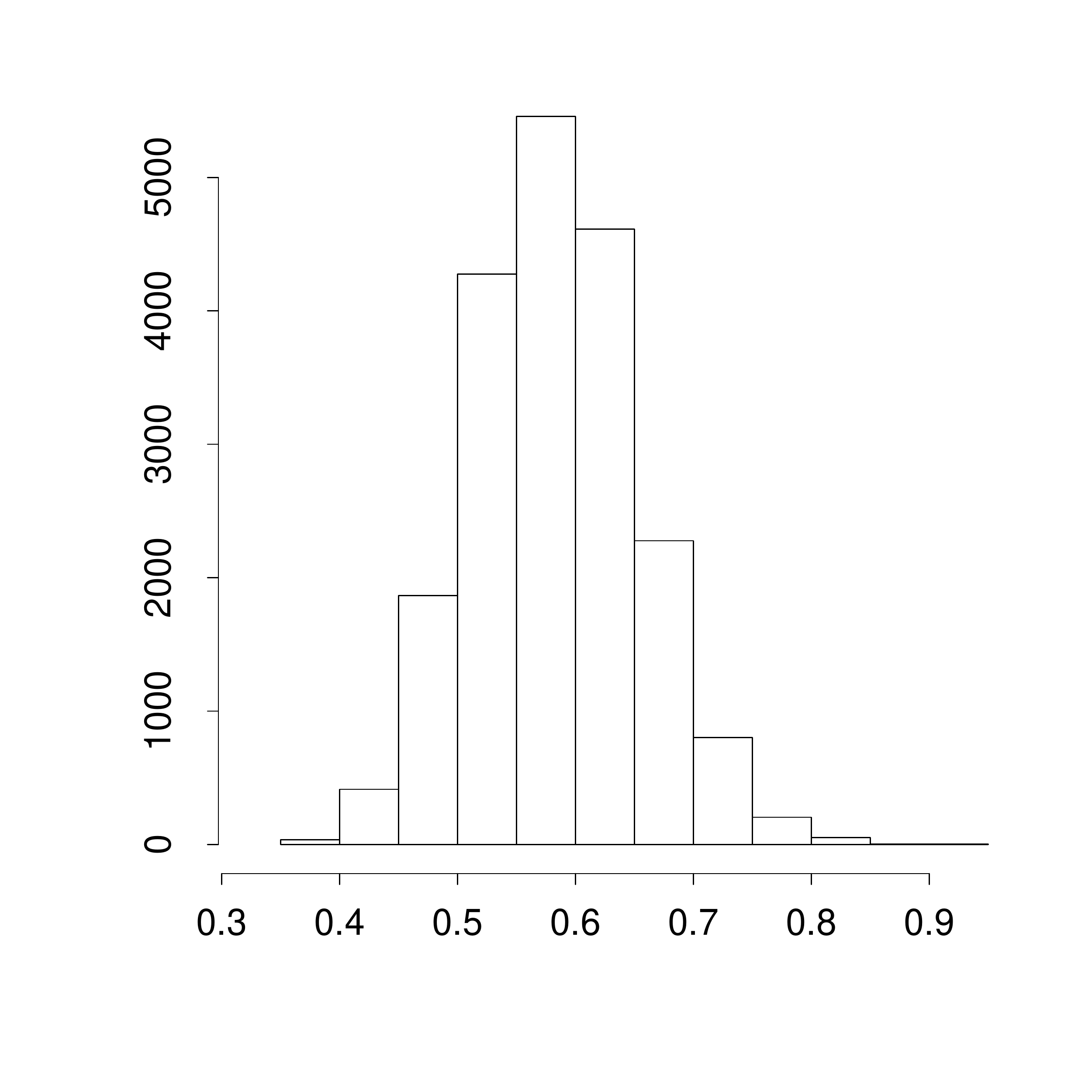} &
    \includegraphics[width=.33\textwidth]{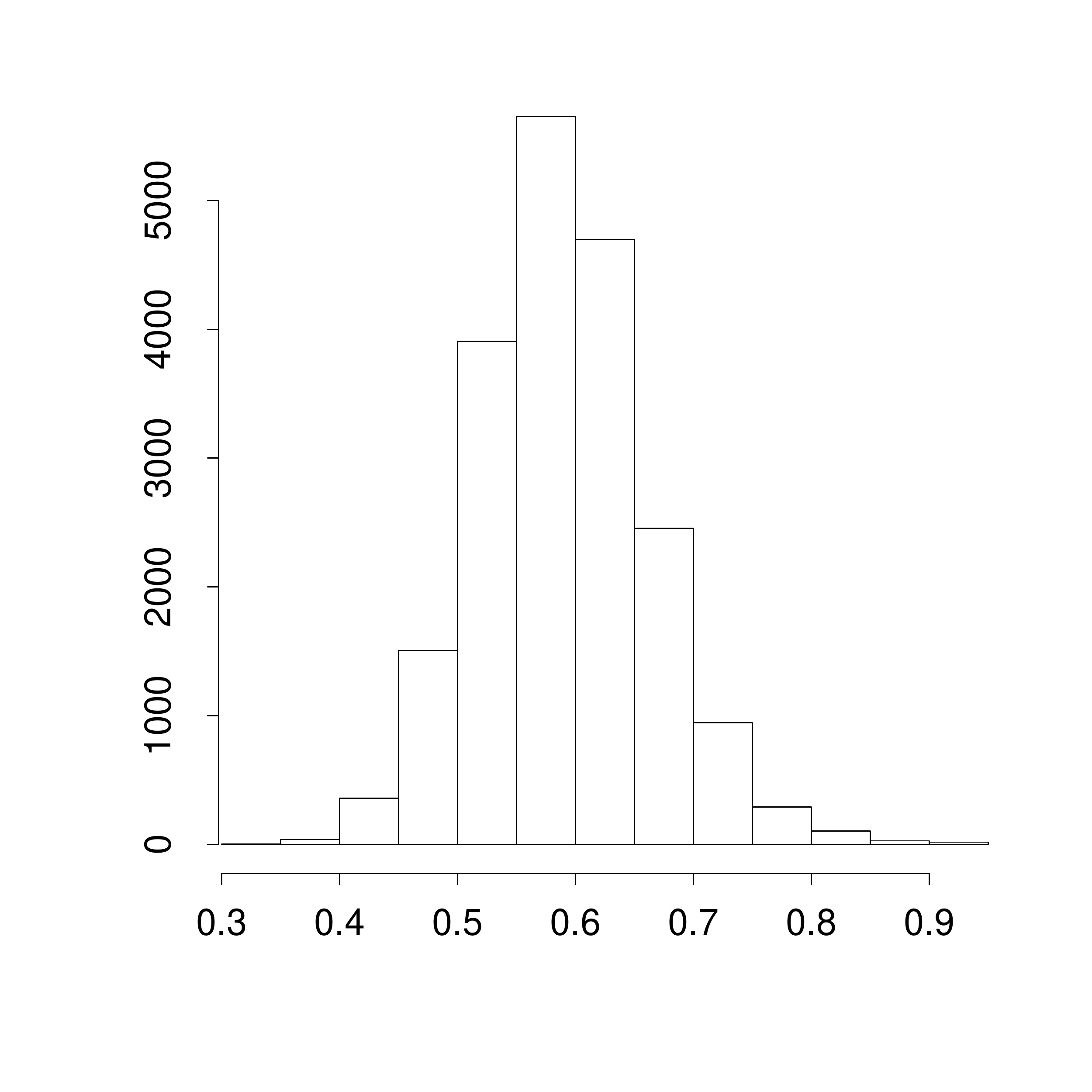} \\
    \begin{sideways} \;\;\;\;\;\;\;\;\;\;\;\;\;\;\;\;\;\;\;\;\;\;\;\;\;       $\tau$  \end{sideways}&
    \includegraphics[width=.33\textwidth]{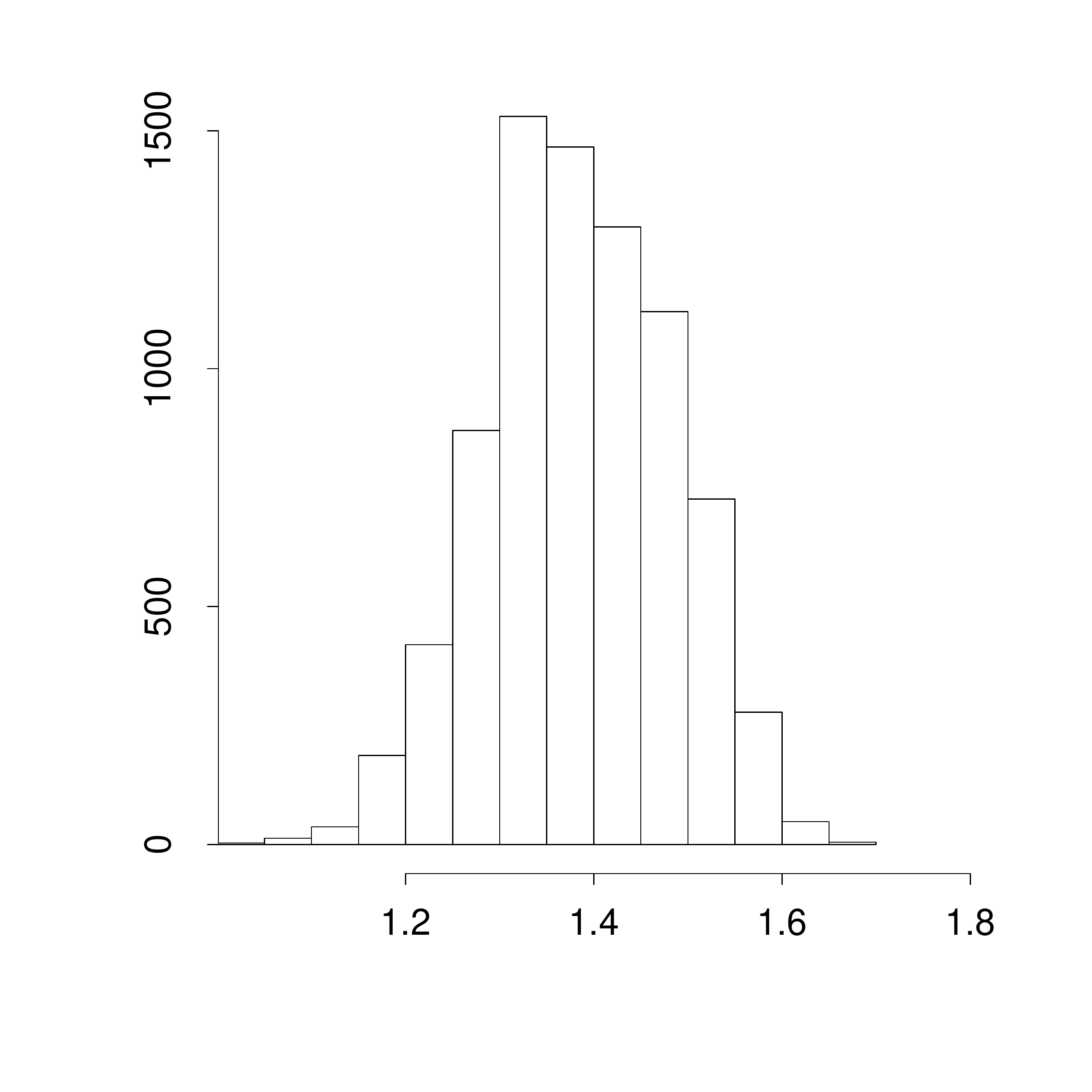} &
    \includegraphics[width=.33\textwidth]{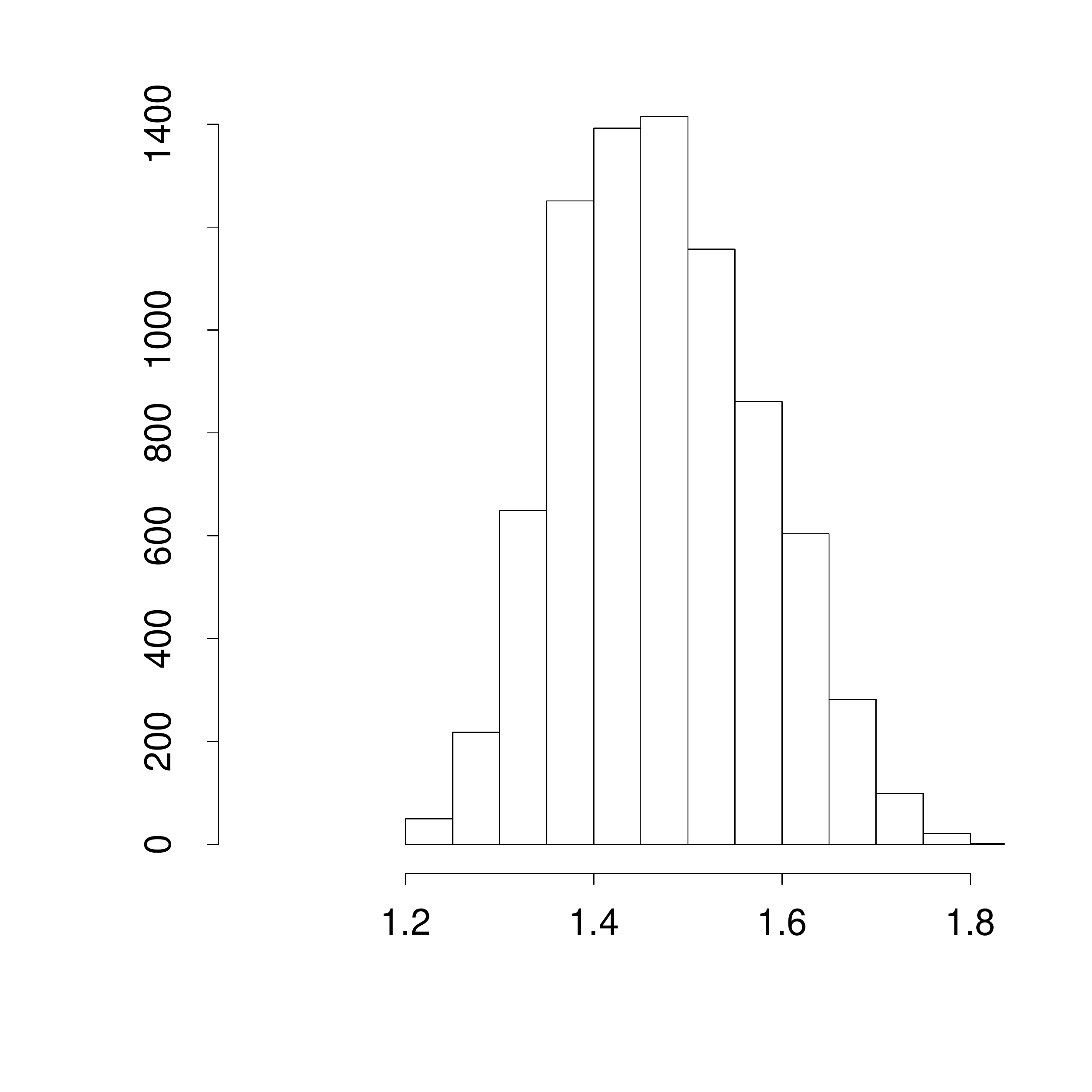} \\
    \begin{sideways} \;\;\;\;\;\;\;\;\;\;\;\;\;\;\;\;\;\;\;\;\;\;   $1/\lambda$  \end{sideways} & 
\includegraphics[width=.33\textwidth]{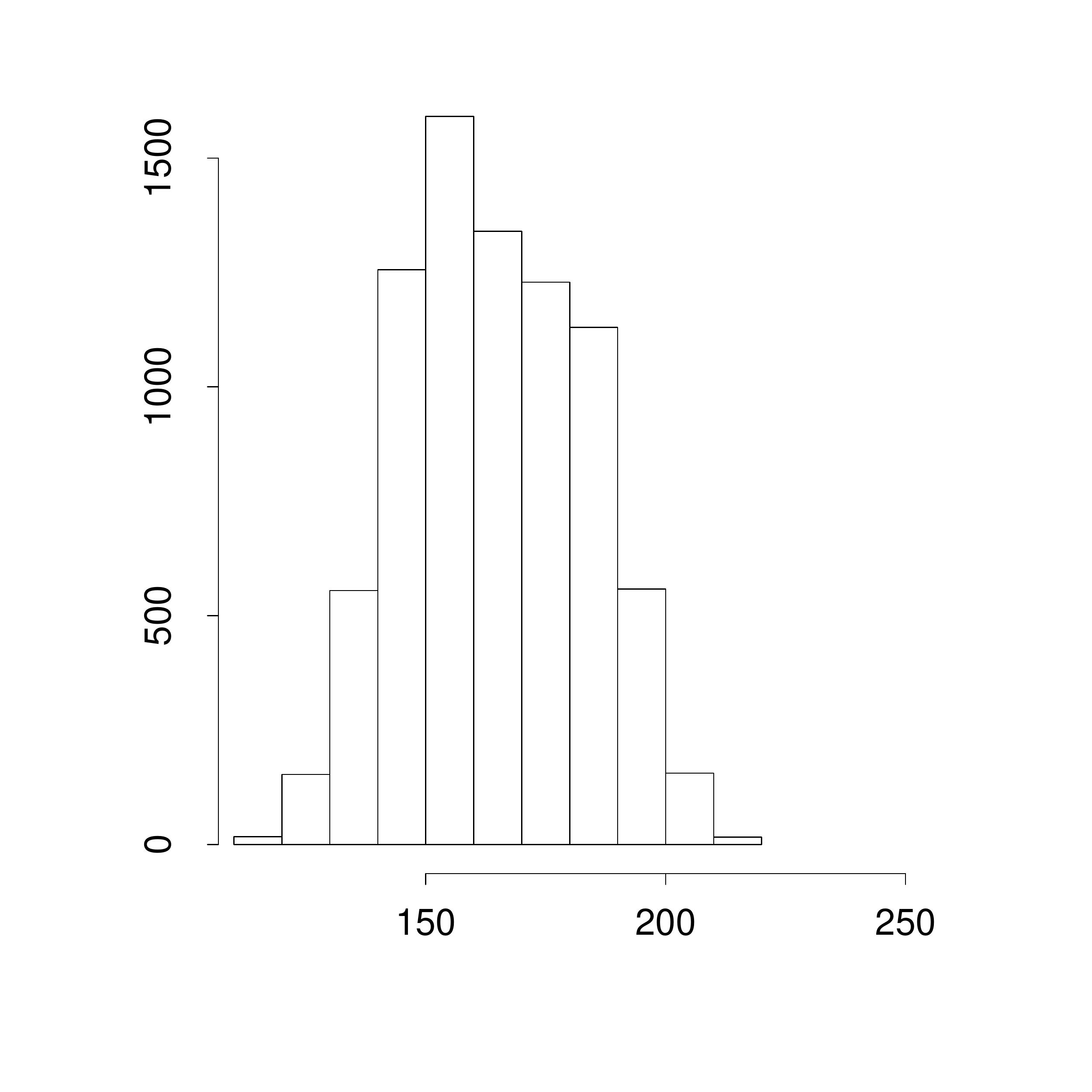}&
    \includegraphics[width=.33\textwidth]{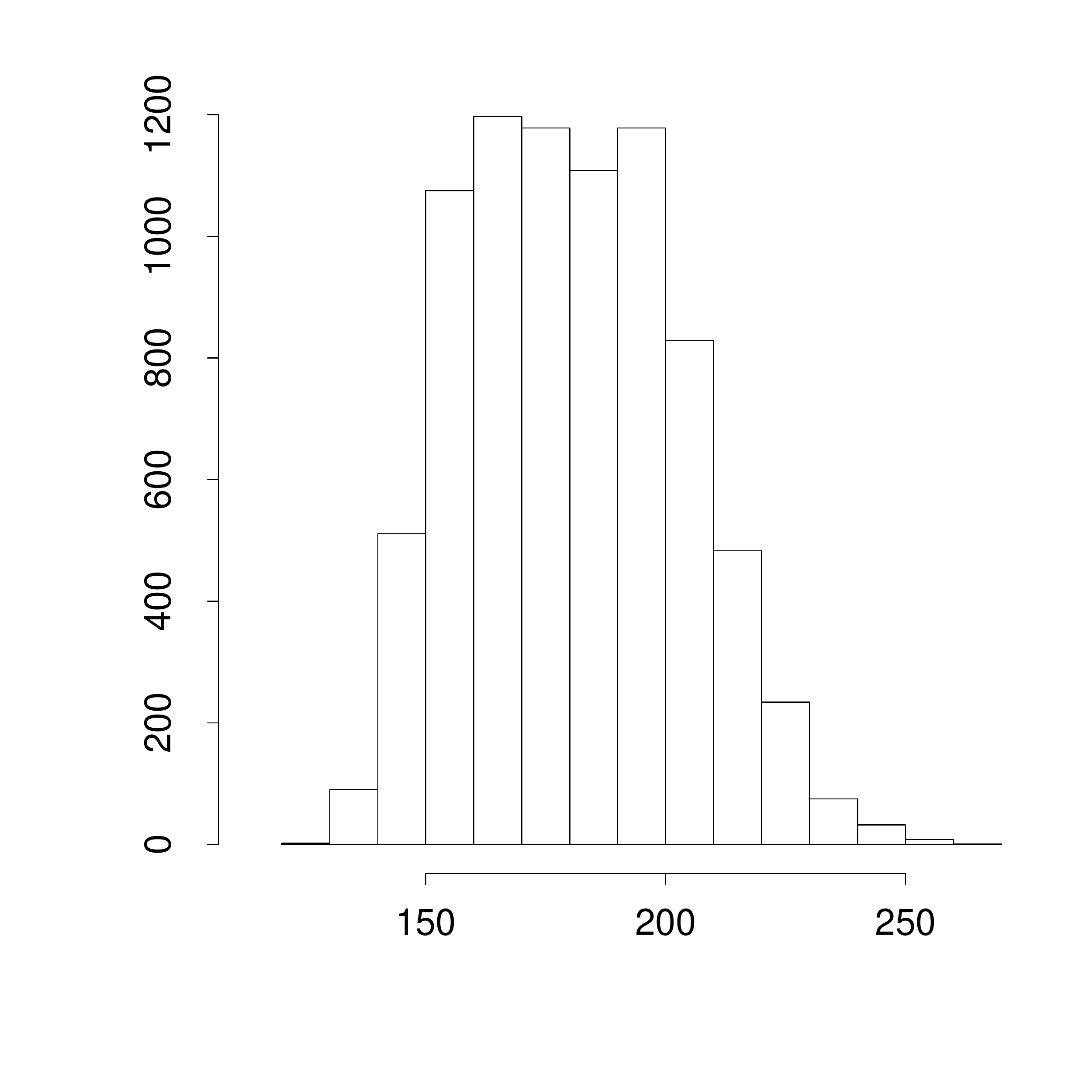} \\
    \end{tabular}
    \caption{Histograms of the posterior samples for the hyperparameters $\omega$, $\tau$ and $1/\lambda$ in Scenario 1 under the original (Figure \ref{fi:finaldenthetaprior}) and the alternative (Figure \ref{fi:finaldenthetaprioralt}) prior specifications.}
    \label{fi:omegataulambda_sim_alt}
\end{figure}

\begin{figure}
    \centering
    \subfigure[Dimension 1]{\includegraphics[width=0.45\textwidth]{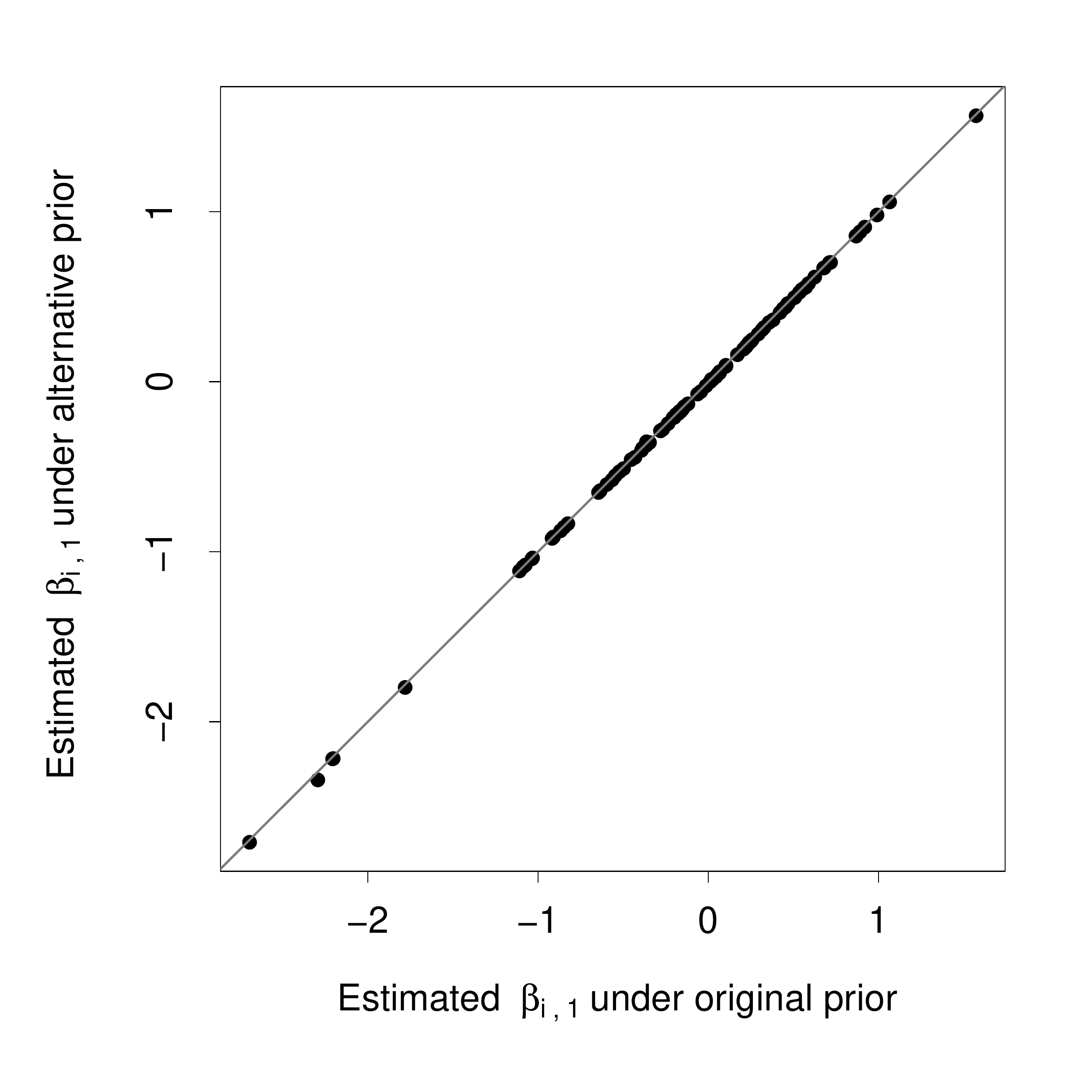}}
    \subfigure[Dimension 2]{\includegraphics[width=0.45\textwidth]{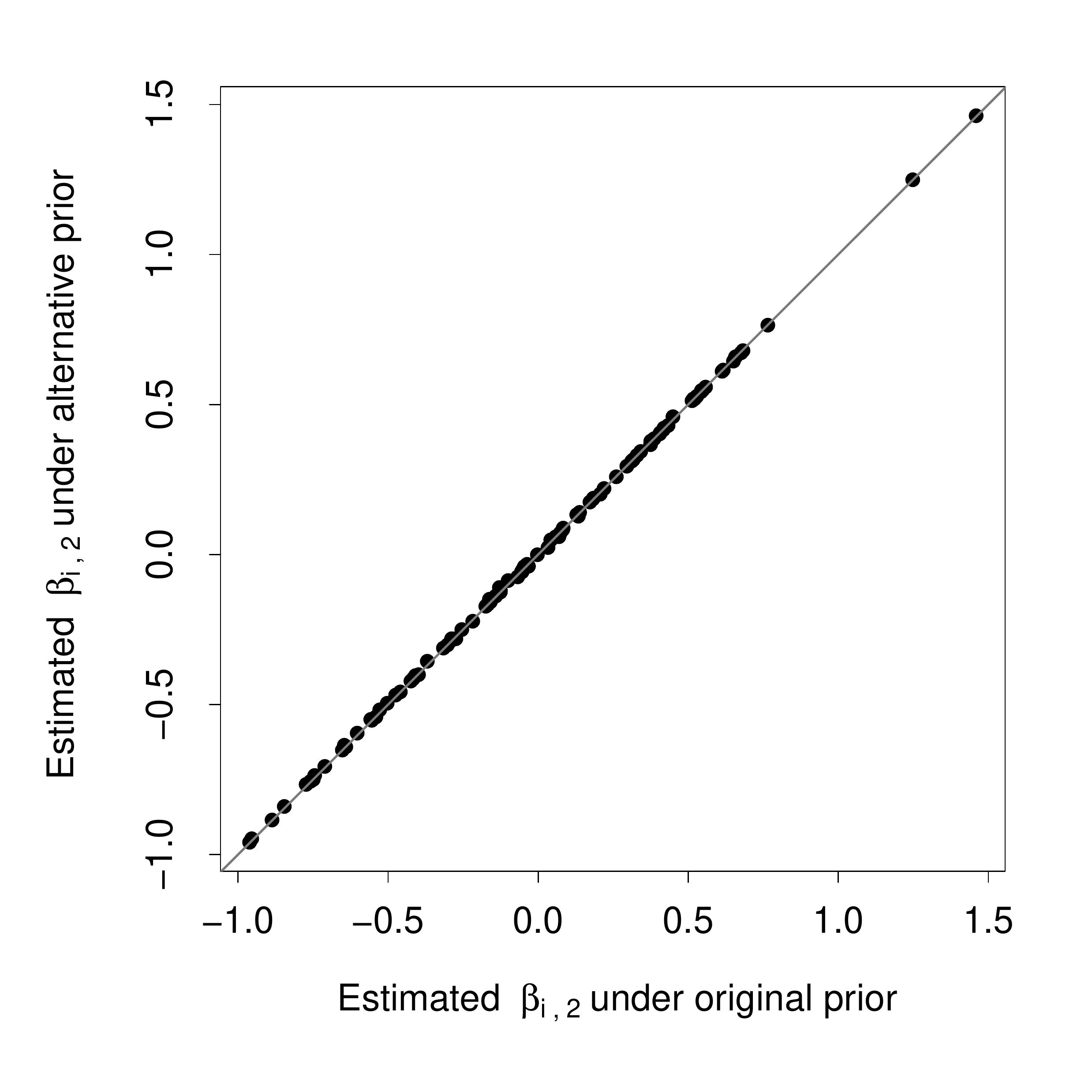}}
    \caption{Comparison of the posterior means of the locations $\bsbeta_i$ across two different prior for Scenario 1.  Left panel compares the first dimension of the latent traits, while the right panel compares the values along the second dimension.}
    \label{fi:simulationlatentres_alt}
\end{figure}

We present here details only for Scenario 1; the results for the other 3 scenarios are very similar.  Similarly to Figure \ref{fi:simulationdimres}, Figure \ref{fi:simulationdimressensitivity} presents the value of the DIC, the in-sample accuracy and the principal nested sphere decomposition associated under both the original and the alternative priors.  The main difference we observe is in the values of the DIC, with the alternative prior tending to give higher plausibility to models dimension 5 and above.
Furthermore, Figure \ref{fi:omegataulambda_sim_alt} shows histograms of the posterior distributions for the hyperparameters $\omega$, $\tau$ and $1/\lambda$ under our two priors.  The posterior distribution of these parameters appear to be very similar, suggesting robustness to this prior change.  
Finally, Figure \ref{fi:simulationlatentres_alt} presents the posterior means of the subject's latent positions under our alternative prior.  Note that the point estimates are nearly identical under both priors.

\subsection{Roll call voting in the U.S.\ Senate}\label{se:senate_102_115}

Factor models are widely used for the analysis of voting records from deliberative bodies (e.g., see \citealp{jackman2001multidimensional}, \citealp{clinton2004statistical} and \citealp{bafumi13andrew}).  In this context, the embedding space provides an abstract representation for the different policy positions, and legislator-specific latent traits are interpreted as their revealed preferences over one or more policy dimensions (e.g., economic vs.\ social issues in a two dimensional setting, see \citealp{moser2020multiple}) or as their ideological preferences in a liberal-conservative scale (in one-dimensional settings, see for example \citealp{jessee2012ideology}).

While Euclidean policy spaces often provide an appropriate description of legislator's decision-making process, they might not be appropriate in all settings.  The idea that spherical policy spaces might be appealing alternatives to Euclidean ones dates back at least to \cite{weisberg1974dimensionland}, who provides a number of examples and notes that ``circular shapes may be expected for alliance structures and for vote coalitions where extremists of the left and right coalesce for particular purposes''.  Spherical policy spaces also provide a mechanism to operationalize the so-called  ``horseshoe theory'' (\citealp{pierre2002siecle}; 
\citealp{taylor2006did}, pg.\ 118), which asserts that the far-left and the far-right are closer to each other than they are to the political center, in an analogous way to how the opposite ends of a horseshoe are close to each other. 

In this Section we use our spherical factor models to investigate the geometry of policy spaces in two different U.S.\ Senates, the 102\textsuperscript{nd} (which met between January 3, 1991 and January 3, 1993, during the last two years of George H.\ W.\ Bush's presidency) and the 115\textsuperscript{th} (which met between January 3, 2017 and January 3, 2019, during the first two years of Donald Trump's presidency).  The data we analyze consists of the record of roll call votes, and it is publicly available from \url{https://voteview.com/}.  
Before presenting our results, we provide some additional background into the data and the application.  Roll call votes are those in which each legislators votes "yea" or "nay" as their names are called by the clerk, so that the names of senators voting on each side are recorded.  It is important to note that not all votes fall into this category. Voice votes (in which the position of individual legislator are not recorded) are fairly common.  This means that roll call votes represent a biased sample from which to infer legislator's preferences.  Nonetheless, roll call data is widely used in political science.  The U.S.\ Senate is composed of two representatives from each of the 50 States in the Union.  The total number of Senators included in each raw dataset might be slightly higher because of vacancies, which are typically filled as they arise.  While turnover in membership is typically quite low,  these changes lead to voting records with (ignorable) missing values on votes that came up during the period in which a Senator was not part of the chamber.  Additionally, missing values can also occur because of temporary absences from the chamber, or because of explicit or implicit abstentions.  As is common practice, we exclude from our analysis any senators that missed more than 40\% of the votes cast during a given session, which substantially reduces the number missing values.  The remaining ones, which are a small percentage of the total number of votes, were treated in our analysis as if missing completely at random.  While this assumption is not fully supported by empirical evidence (e.g., see \citealp{rodriguez2015measuring}), it is extremely common in applications.  Furthermore, the relatively low frequency of missing values in these datasets suggests that deviations from it will likely have a limited impact in our results.  See table \ref{tab:summarySenate} for a summary of the features of the two post processed datasets.
\begin{table}
    \centering
    \begin{tabular}{cccc} \hline
        Session & Senators ($I$) & Measures ($J$) & Missing votes \\ \hline
        102\textsuperscript{nd} &100 & 550 & 2222 (4.04\%) \\
        115\textsuperscript{th} &96 & 599 & 1236 (2.15\%)\\ \hline
    \end{tabular}
    \caption{Summary information for the two roll call datasets analyzed in this paper.}
    \label{tab:summarySenate}
\end{table}

\begin{figure}
  \centering
    \begin{tabular}{ccc}
    & \multirow{2}{1cm}{$DIC$}   & Principal nested \\ 
    &  & sphere decomposition \\ 
    \begin{sideways} \;\;\;\;\;\;\;\;\;\;\;\;\;\;\;\;\;\;\;\;\;\;\;\; 102\textsuperscript{nd} Senate \end{sideways}&
    \includegraphics[width=.4\textwidth]{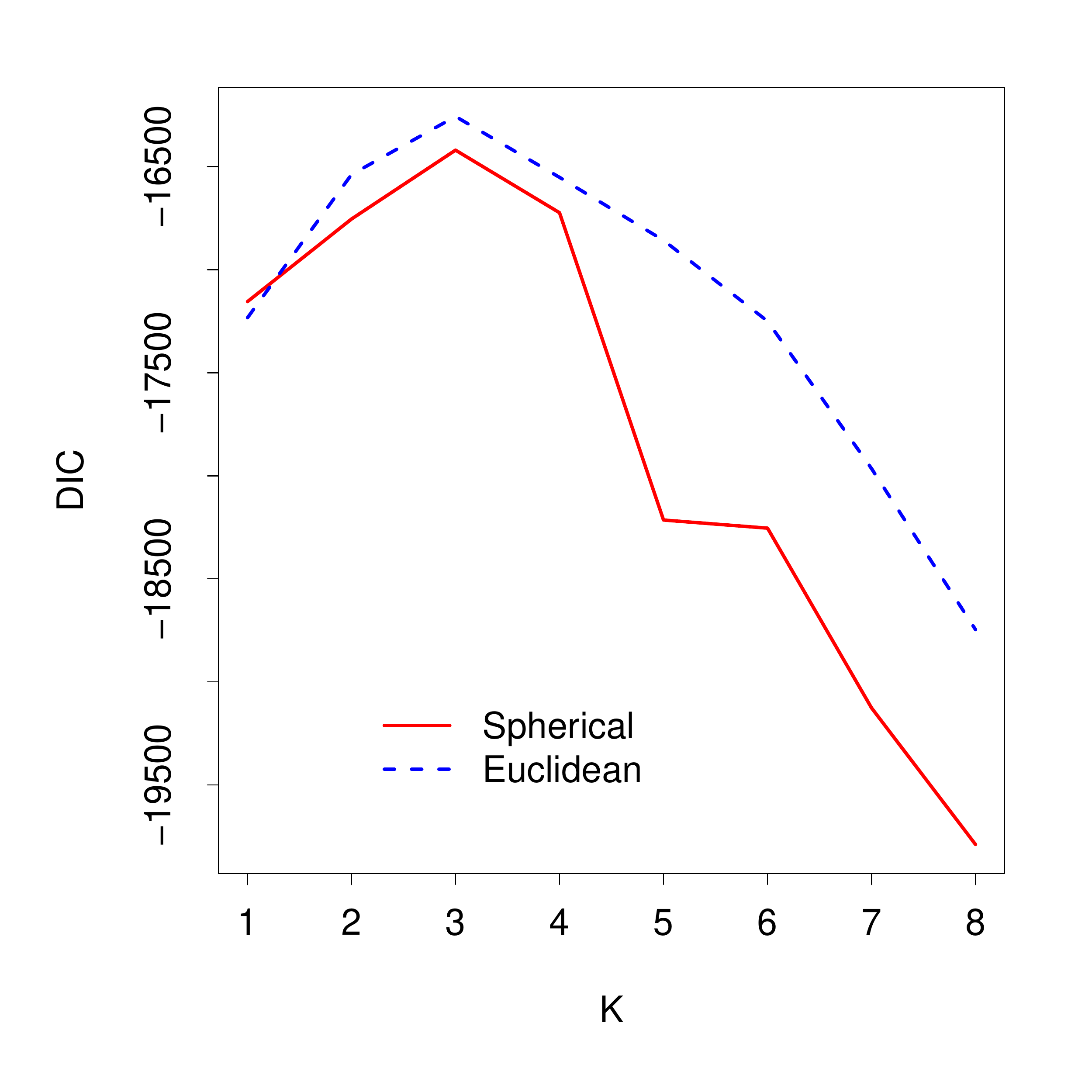} &
    \includegraphics[width=.4\textwidth]{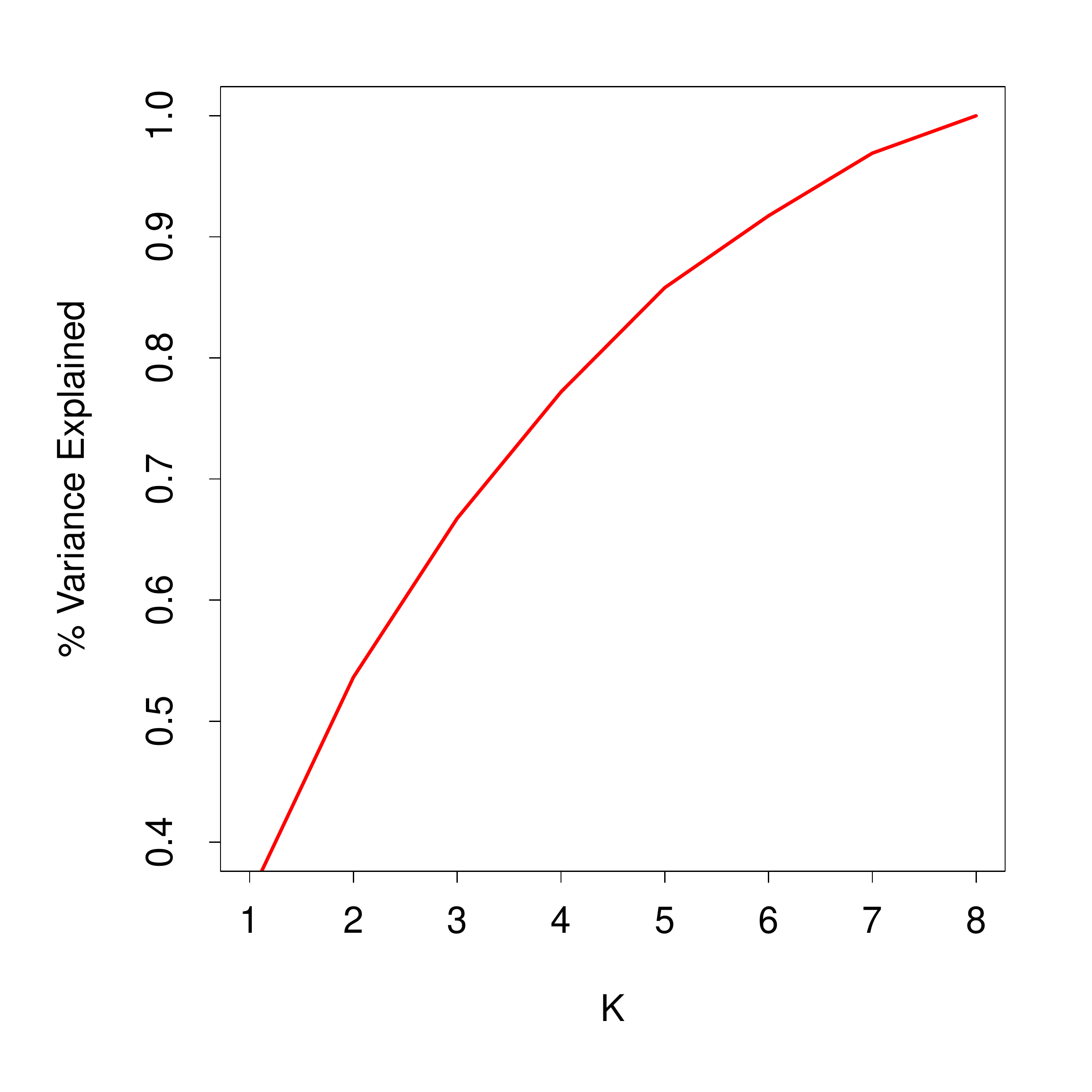} \\
    \begin{sideways} \;\;\;\;\;\;\;\;\;\;\;\;\;\;\;\;\;\;\;\;\;\;\;\; 115\textsuperscript{th} Senate \end{sideways} & 
    \includegraphics[width=.4\textwidth]{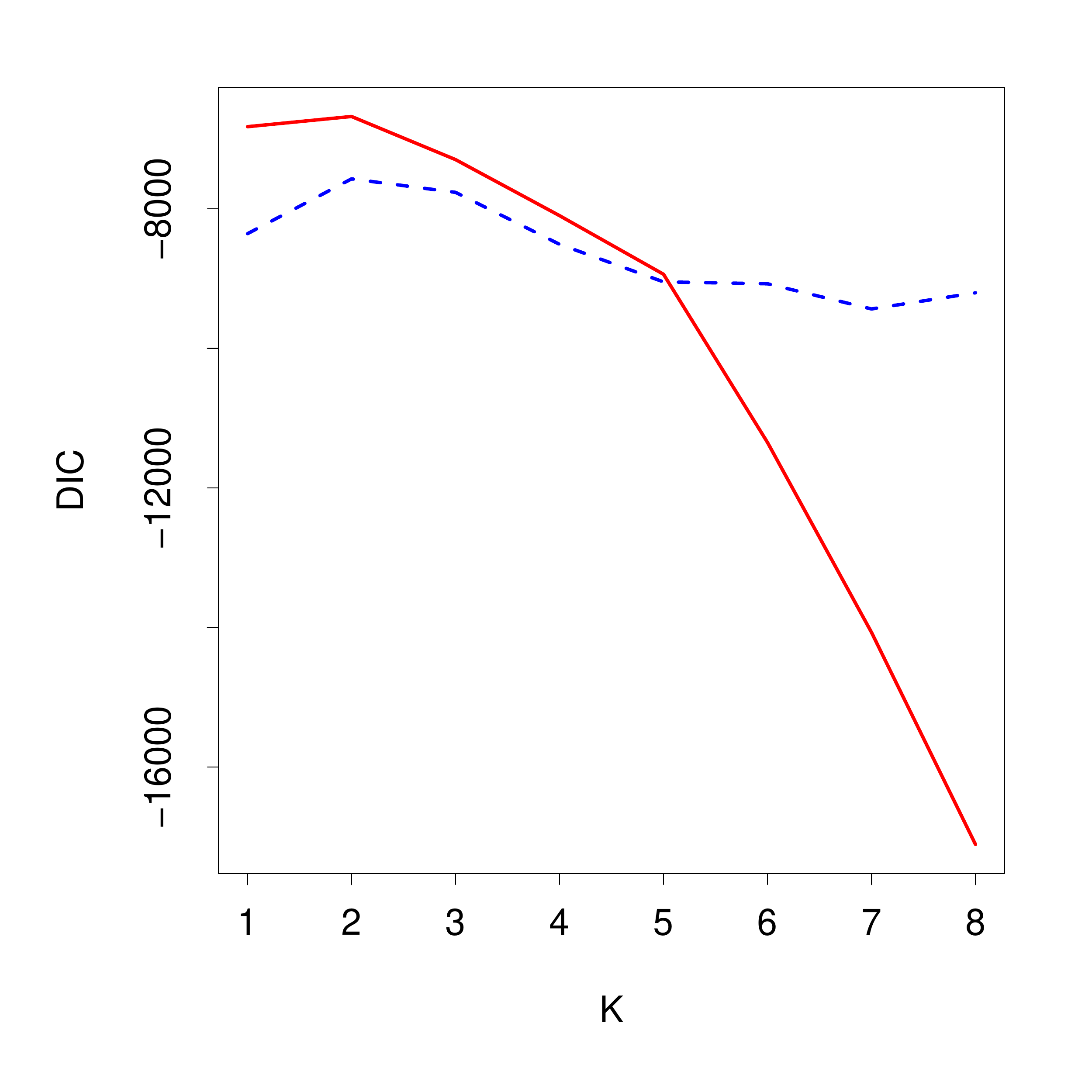} &
    \includegraphics[width=.4\textwidth]{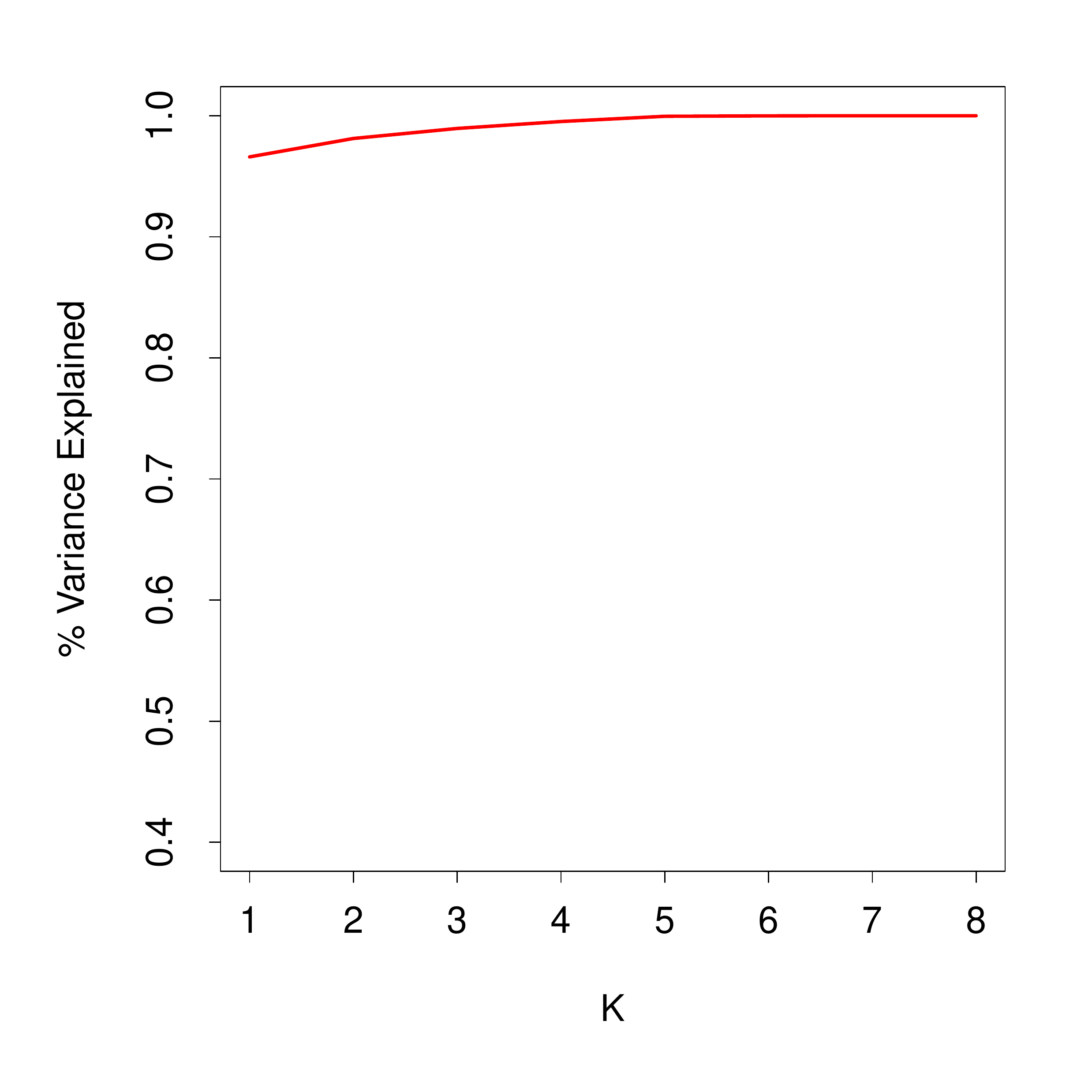} \\
    \end{tabular}
\caption{Deviance information criteria as a function of the embedding space's dimension $K$ (left column) and principal nested sphere decomposition associated with the $\mathcal{S}^{8}$ model for the 102\textsuperscript{nd} (top row) and the 115\textsuperscript{th} (bottom row) U.S.\ Senates.}\label{fi:DICSenate}
\end{figure}

As in Section \ref{se:simulation}, we fit both Euclidean and spherical factor models of varying dimensions to each of these two datasets.  The left column of Figure \ref{fi:DICSenate} shows the DIC values associated with these models.  For the 102\textsuperscript{nd} Senate, we can see that a Euclidean model of dimension 3 seems to provide the best fit overall while, among the spherical models, a model of dimension 3 also seems to outperform.  
On the other hand, for the 115\textsuperscript{th} Senate, a two-dimensional spherical model seems to provide the best fit to the data, closely followed by a circular (one-dimensional spherical) model.  The right column of Figure \ref{fi:DICSenate} presents the principal nested sphere decomposition associated with the posterior mean configuration estimated by an eight-dimensional spherical model on each of the two Senates.  The differences are substantial. In the case of the 115\textsuperscript{th} Senate (for which DIC suggests that the geometry of the latent space is indeed spherical), the first component of the decomposition explains more than 95\% of the variability in the latent space, and the first three components explain close to 99\%.  On the other hand, for the 102\textsuperscript{nd} Senate, the first three spherical dimensions explain less than 70\% of the total variability.  


At a high level, these results are consistent with the understanding that scholars of American politics have of contemporaneous congressional voting patterns.  While congressional voting in the U.S.\ has tended to be at least two-dimensional for most of its history (with one dimension roughly aligning with economic issues, and the other(s) corresponding to a combination of various social and cultural issues), there is clear evidence that it has become more and more unidimensional over the last 40 years (e.g., see \citealp{poole2000congress}).  Increasing polarization has also been well documented in the literature (e.g., see \citealp{jessee2012ideology} and \citealp{hare2014polarization}).  The rise of extreme factions within both the Republican and Democratic parties willing to vote against their more  mainstream colleagues, however, is a relatively new phenomenon that is just starting to be documented (see \citealp{ragusa2016s}, \citealp{Lewis2019b}, \citealp{Lewis2019a}) and can explain the dominance of the spherical model for the 115\textsuperscript{th} Senate voting data.

\begin{figure}
    \centering
    \begin{tabular}{cccc}
    & 102\textsuperscript{nd} Senate & 115\textsuperscript{th} Senate \\
    \begin{sideways} \;\;\;\;\;\;\;\;\;\;\;\;\;\;\;\;\;\;\;\;\;\;\;\;\;      $\omega$  \end{sideways}&
    \includegraphics[width=.33\textwidth]{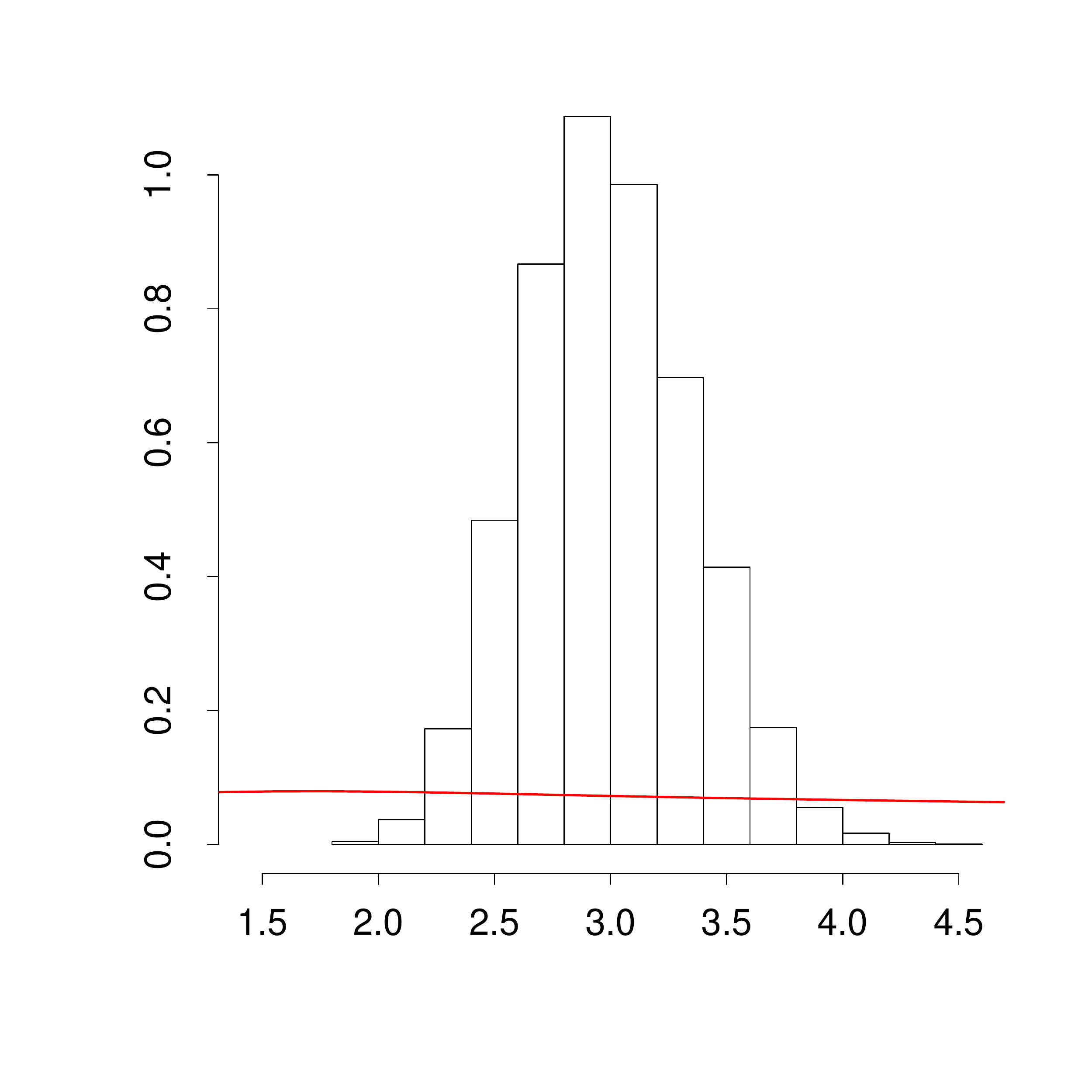} &
    \includegraphics[width=.33\textwidth]{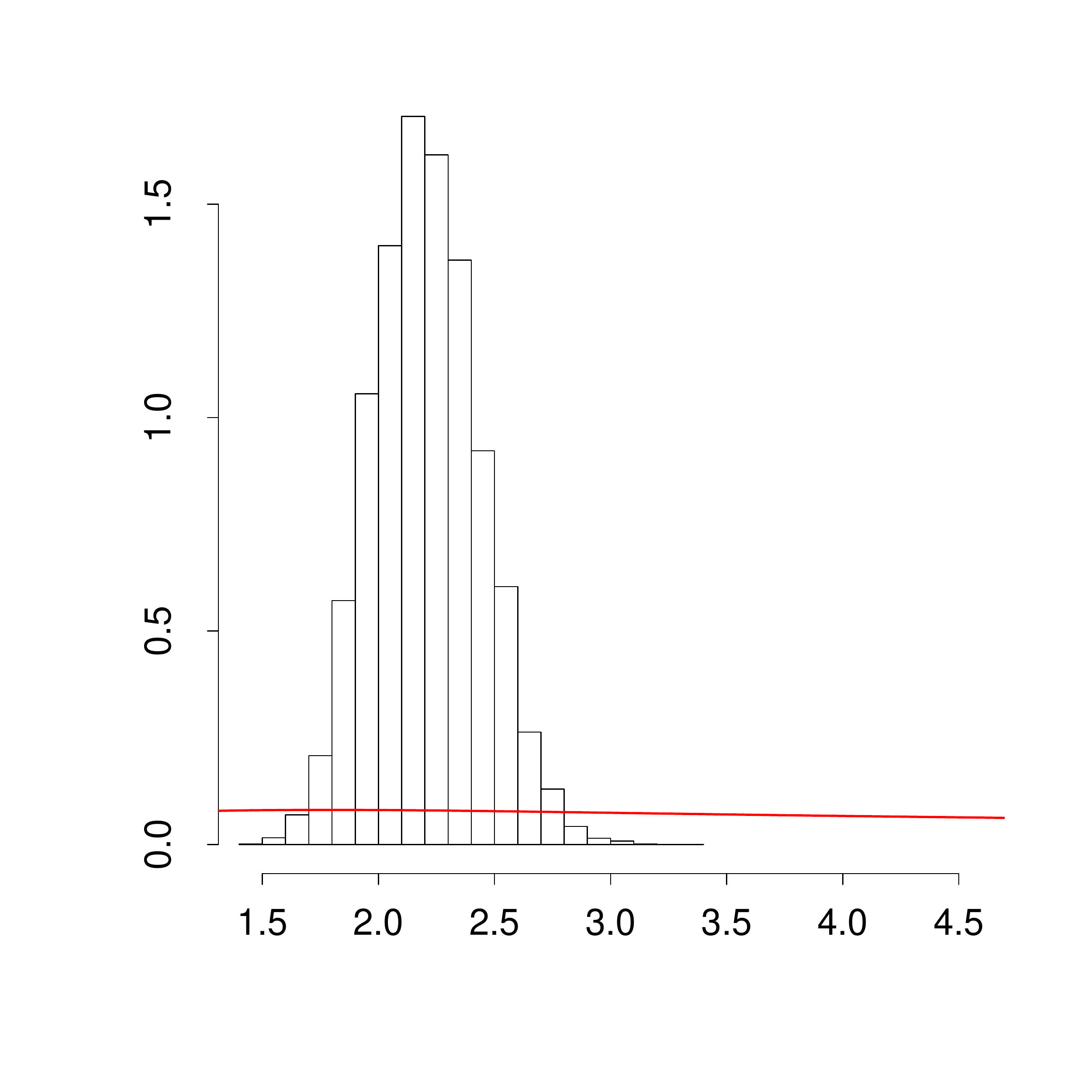} \\
    \begin{sideways} \;\;\;\;\;\;\;\;\;\;\;\;\;\;\;\;\;\;\;\;\;\;\;\;\;       $\tau$  \end{sideways}&
    \includegraphics[width=.33\textwidth]{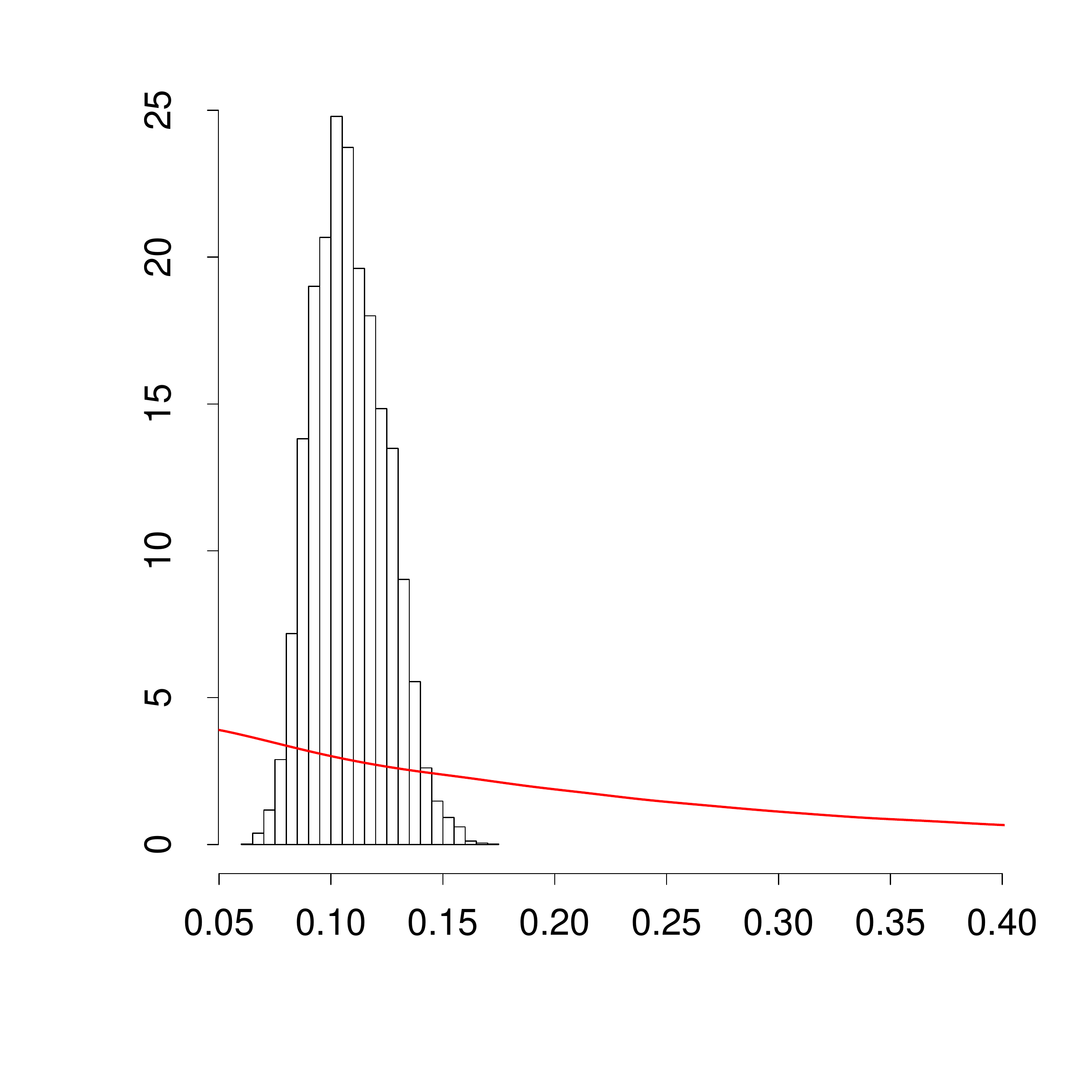} &
    \includegraphics[width=.33\textwidth]{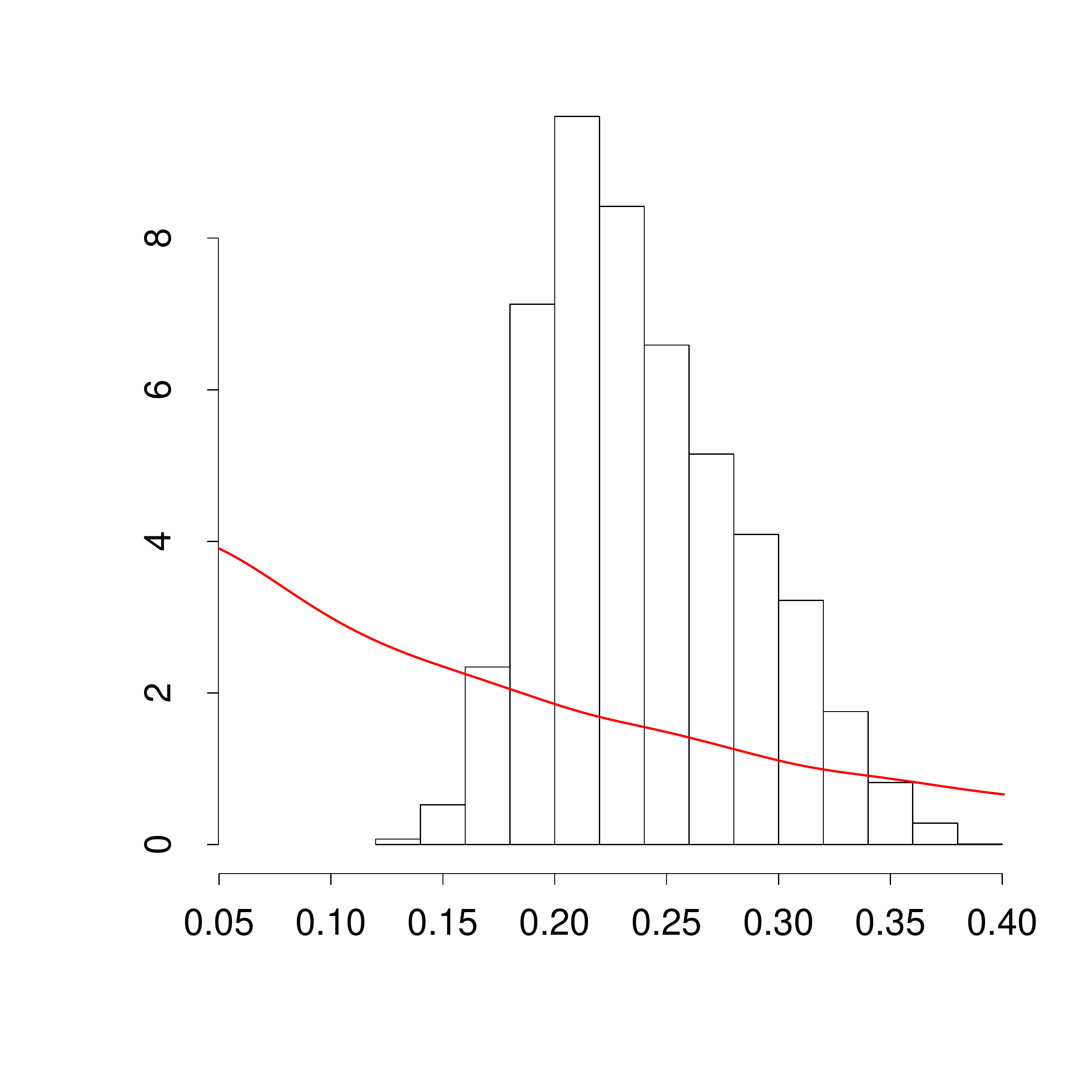} \\
    \begin{sideways} \;\;\;\;\;\;\;\;\;\;\;\  $1/\lambda$  \end{sideways} & 
\includegraphics[width=.33\textwidth]{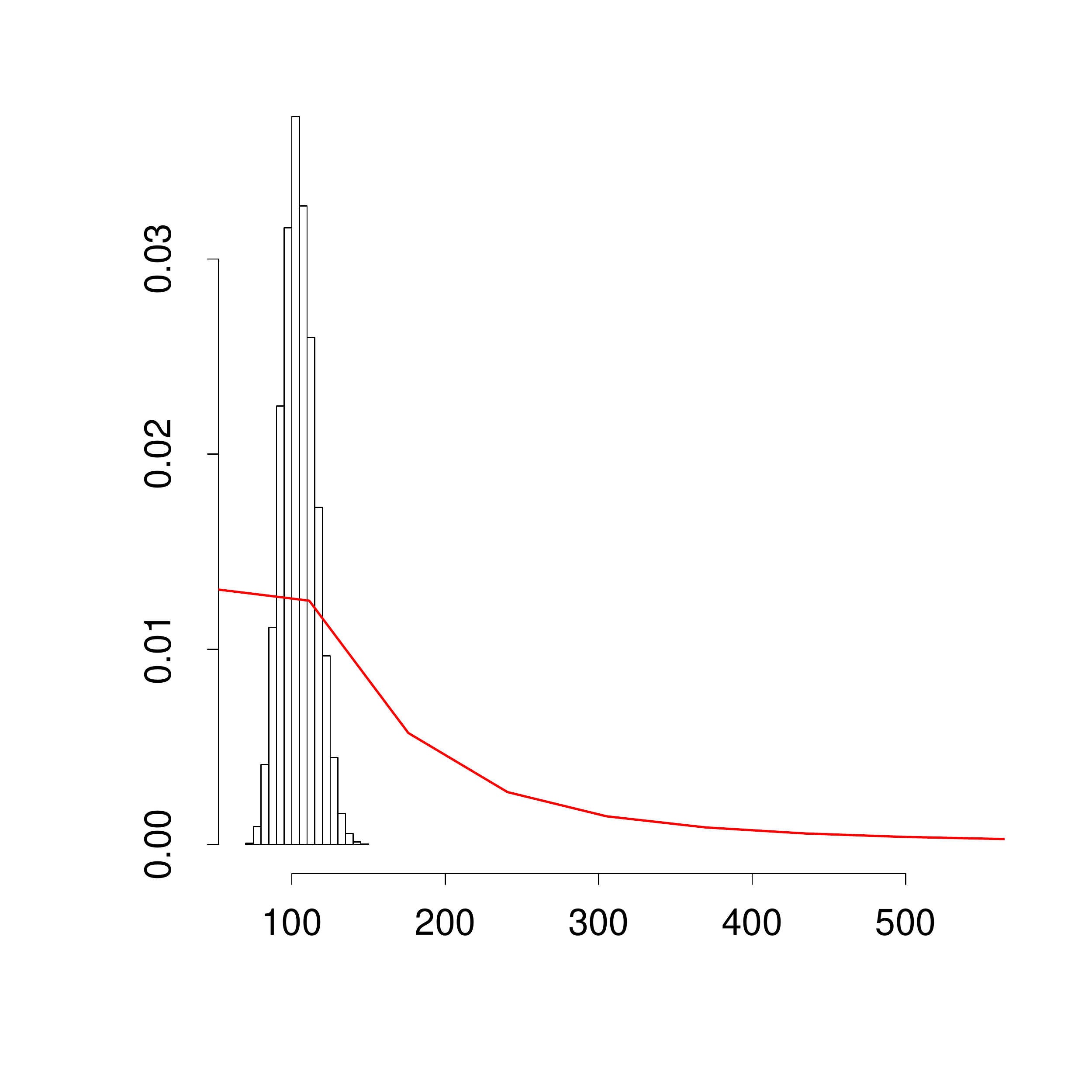}&
    \includegraphics[width=.33\textwidth]{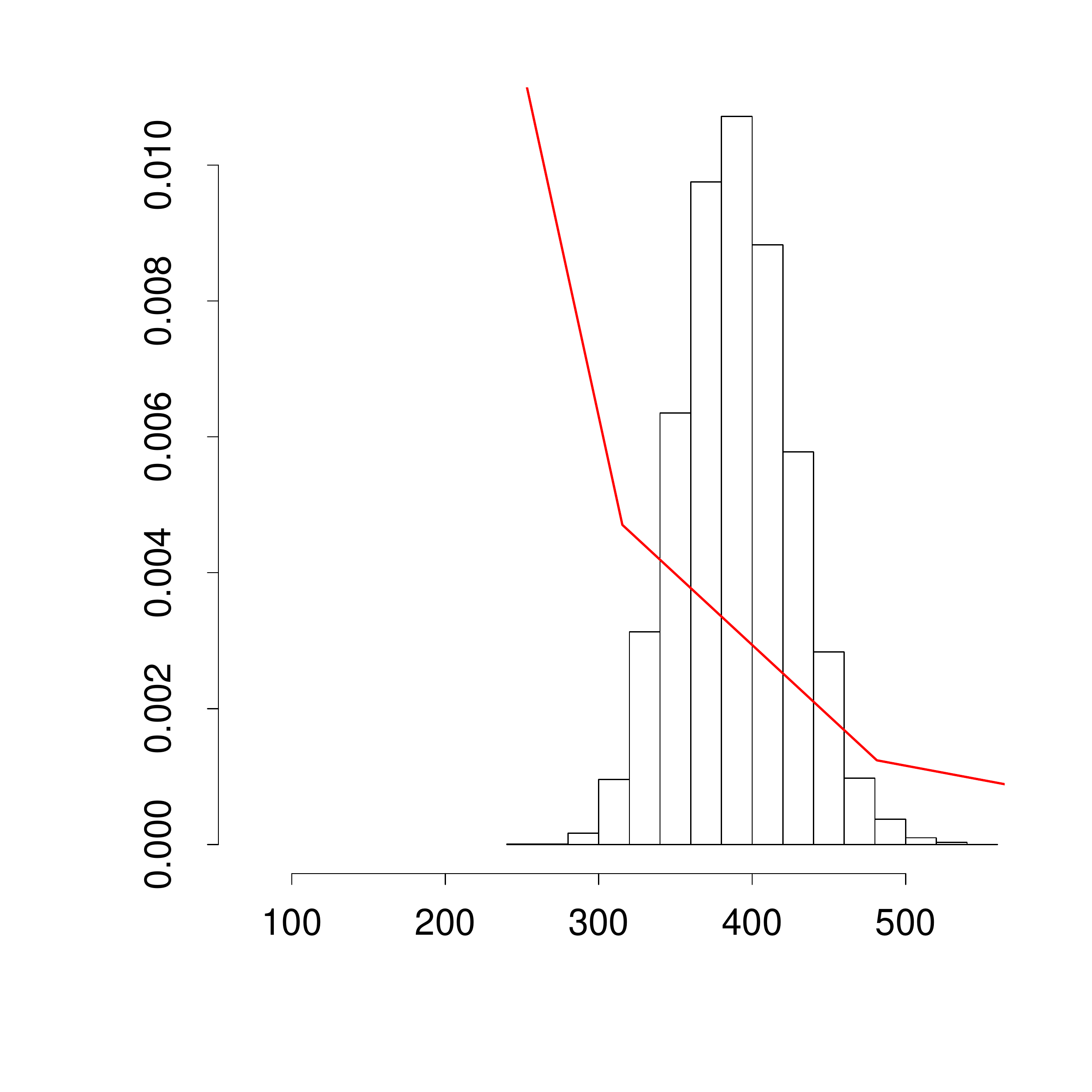} \\
    \end{tabular}
    \caption{Histograms of the posterior samples for the hyperparameters $\omega$, $\tau$ and $1/\lambda$ for the 102\textsuperscript{nd} (left panel) and the 115\textsuperscript{th} (right panel) U.S.\ Senates for $K=3$ and $K=2$, respectively.  The continuous line corresponds to the prior distribution used in the analysis.}
    \label{fi:hyperSenate}
\end{figure}

Figure \ref{fi:hyperSenate} presents histograms of the posterior samples for the hyperparameters $\omega$, $\tau$ and $1/\lambda$ for each of the two Senates.  Note that, as with the simulations, the posterior distributions differ substantially from the priors.  Furthermore, the model prefers relatively smaller values of $\omega$ and relatively larges values of $\tau$ and $1/\lambda$ for the 115\textsuperscript{th} Senate when compared with the 102\textsuperscript{nd}.  

\begin{figure}
    \centering
    \subfigure[102\textsuperscript{nd} Senate]{\includegraphics[width=0.45\textwidth]{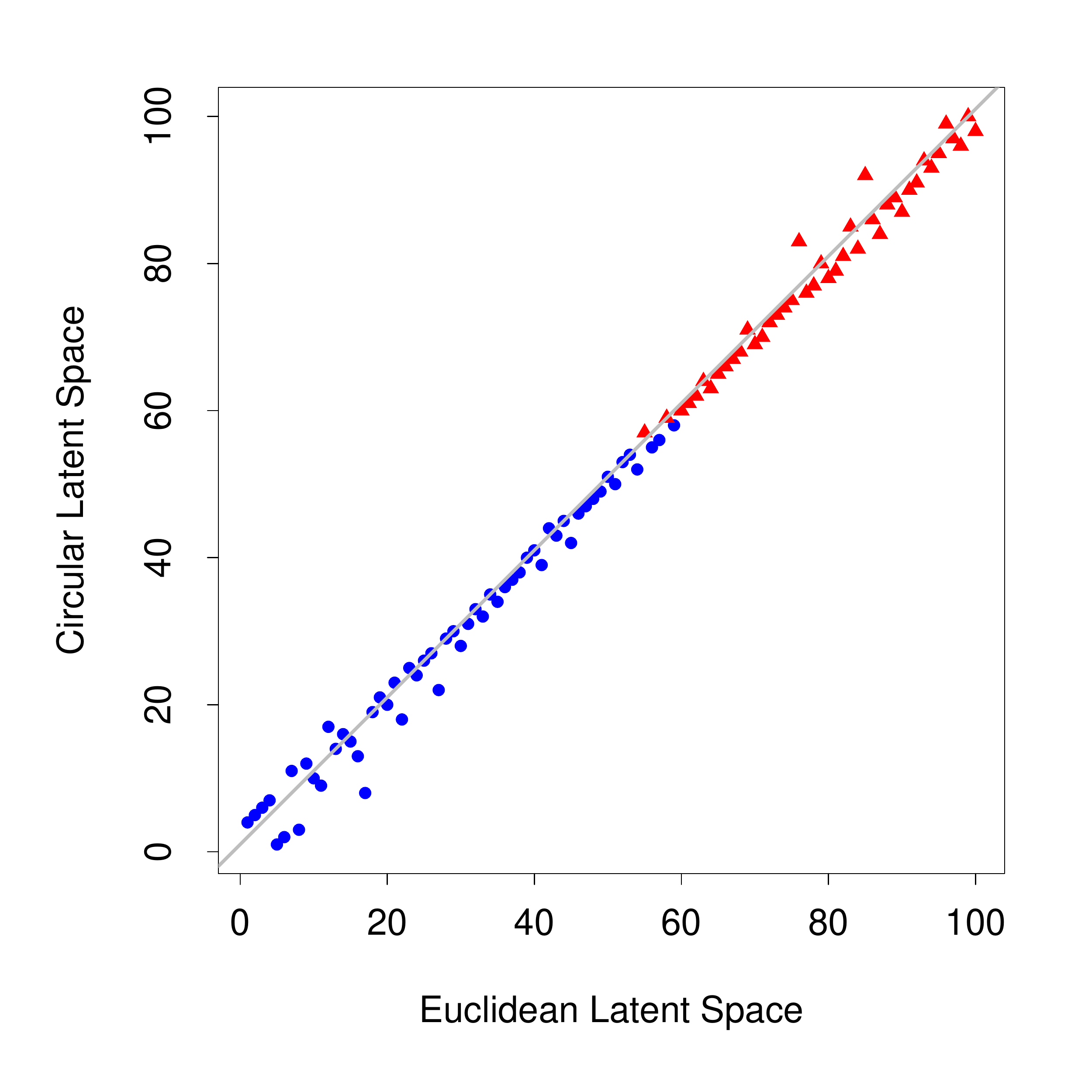}}
    \subfigure[115\textsuperscript{th} Senate]{\includegraphics[width=0.45\textwidth]{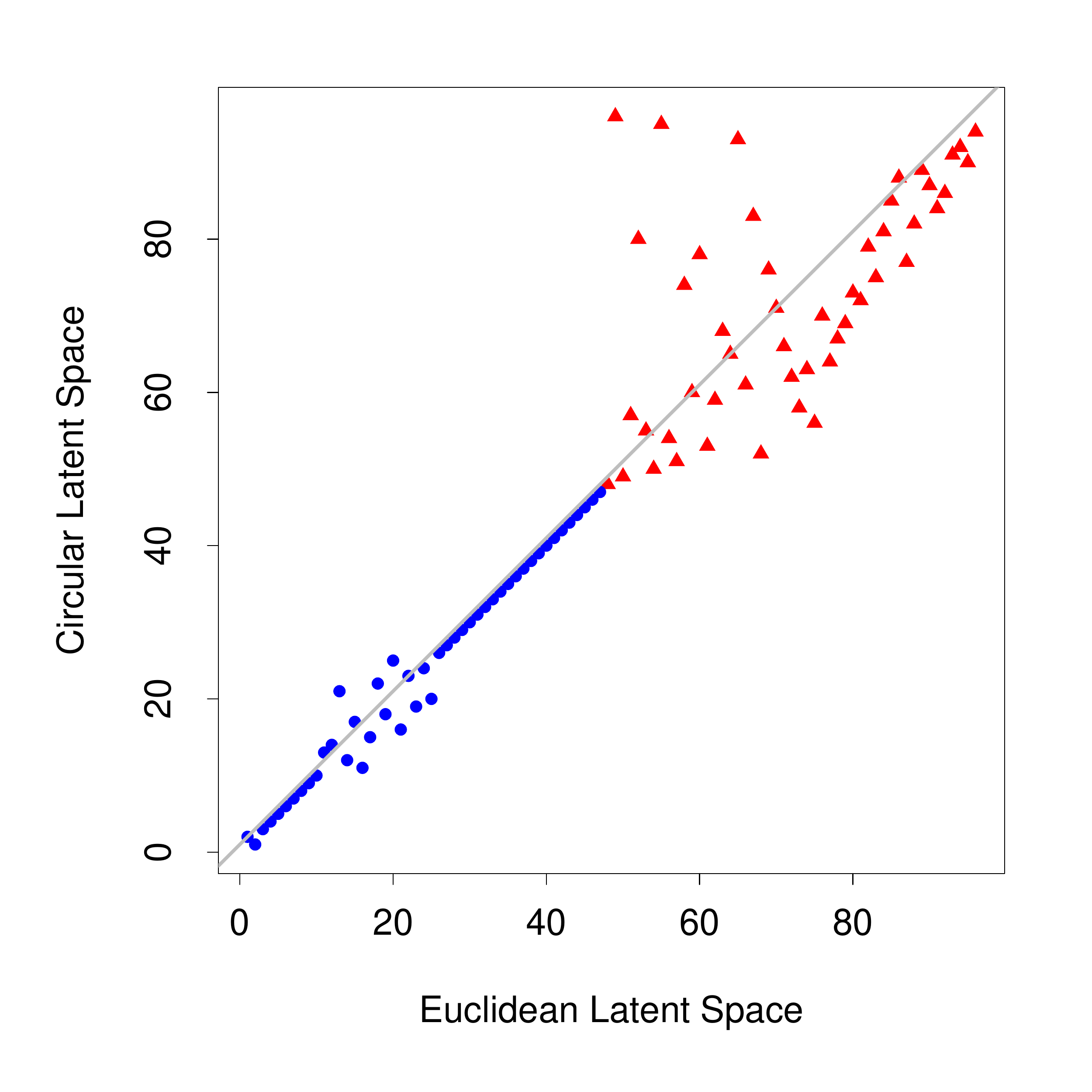}}
        \caption{Comparison of the rank order of legislators between the 1D Euclidean and circular models for the 102\textsuperscript{nd} (left panel) and the 115\textsuperscript{th} (right panel) Senates.  Red triangles correspond to Republican Senators, while blue circles indicate Democrats, as well as Independents who caucus with Democrats.
        }
    \label{fi:ranks102and115}
\end{figure}

To conclude this Section, we focus our attention on the one-dimensional versions of the Euclidean and spherical models.  While one-dimensional models are not preferred by the DIC criteria in our datasets, researchers often use them in practice because the resulting ranking of legislators often reflects their ordering in a liberal-conservative (left-right) scale.  Figure \ref{fi:ranks102and115} compares the median rank order of legislators estimated under the one-dimensional Euclidean and one-dimensional spherical (circular) models for the 102\textsuperscript{nd} (left panel) and the 115\textsuperscript{th} (right panel) Senates.  To generate the ranks under the circular model, we ``unwrap'' the circle and compute the ranks on the basis of the associated angles (which live in the interval $[-\pi,\pi]$).  For the 102\textsuperscript{nd} Senate (where DIC would prefer the one-dimensional Euclidean model over the circular model), the median ranks of legislators under both models are very similar.  This result provides support to our argument that the spherical model can provide a very good approximation to the Euclidean model when there is no evidence of sphericity in the data.  On the other hand, for the 115\textsuperscript{th} (where the circular model would be preferred over the one-dimensional Euclidean model), the ranking of Republican legislators differs substantially between both models.  Digging a little bit deeper, we can see that the five  legislators for which the ranks differ the most are Rand Paul (KY), Mike Lee (UT), Jeff Flake (AZ), Bob Corker (TN) and John Neely Kennedy (LA), all known for being hard line conservatives, but also for bucking their party and voting with (left wing) Democrats on some specific issues.

\section{Discussion}\label{se:conclusion}

We have developed a new flexible class of factor analysis models for multivariate binary data that embeds the observations on spherical latent spaces.  The results from our illustrations suggest that (i) it is possible to distinguish between different geometries and dimensions for the embedding space, (ii) our model can closely approximate traditional factor models when the latent space is Euclidean, (iii) the use spherical latent spaces can be justified, both theoretically and empirically, in applications related to choice models.

The assumption that of a spherical latent space might not be appropriate in every application.  For example, we find it hard to justify their use in settings such as educational testing.  However, we believe that this class of models can have broader applications beyond the analysis of roll call data discussed in Section \ref{se:senate_102_115}.  One such area of application is marketing, where choice models are widely used to understand consumer behavior.  In this context, spherical models could serve to explain the apparent lack of transitivity of preferences that is sometimes present in real marketing data.

While the focus of this paper has been on latent spherical spaces, it is clear that our approach can be extended to other classes of embedding manifolds.  The main challenge associated with this kind of extension is the construction of prior distributions, specially if we aim to estimate the intrinsic dimension of the space.  Similarly, this class of models can be extended beyond binary data to nominal and categorical observations using similar random utility formulations.  This is ongoing work in our group.

\section{Supplementary material}

\subsection{Stability of priors in the Euclidean factor model}\label{ap:priorconvergence}

Let $\mu_j \sim N(0,1/2)$, $\alpha_{j,k} \sim N(0,1/2)$, $\beta_{j,k} \sim N(0,6/[\pi
k]^2)$, and $z_{i,j}(K) = \mu_j + \sum_{k=1}^K \alpha_{j,k}  \beta_{i,k}$.  Because $z_{i,j}(K)$ is the sum of $K+1$ random variables with finite second moments, a simple application of the central limit theorem indicates that $\{z_{i,j}(K) - \E(z_{i,j}(K)) \}/\Var(z_{i,j}(K))$ converges in distribution to standard normal distribution as $K\to\infty$.  Now, note that $(z_{i,j}(K)) = 0$ by construction, and that,
\begin{align*}
\Var \left\{ z_{i,j}(K)\right\} &= \Var(\mu_j) + \sum_{k=1}^K \Var(\alpha_{j,k}) \Var(\beta_{i,k}) \\
&= \frac{1}{2} + \frac{1}{2} \sum_{k=1}^K \frac{6}{\pi^2}\frac{1}{k^2}
= \frac{1}{2} + \frac{1}{2} \frac{6}{\pi^2} \frac{\pi^2}{6} = 1
\end{align*}

As a consequence, $\theta_{i,j} = \Phi\left(z_{i,j}(K) \right)$ converges in distribution to a uniform distribution in $[0,1]$.
\subsection{Derivation of the gradient of log prior and log Jacobian}\label{ap:gradient}

The gradient with respect to the log of Hausdorff measure may appear forbidding to derive, especially the prior component of the full conditional because the expression of the gradient changes with the predetermined maximum dimension. Hence, we develop an automatic method to facilitate the computation of the gradient. Interestingly, the GHMC algorithm projects all the gradient to the tangent space through the momentum update, which allows the gradient computed with or without the constraint to be the same. Nevertheless, it is generally recommended to derive the gradients under the unit norm constraint since they are invariant to the reparameterization with respect to the constraint.

\begin{align}\label{eq:gradient_all}
\nabla \log p_{\mathcal{H}}(\bfx_{\beta_i} \mid \dots) = \nabla \log p_{\mathcal{H}_l} + \nabla \log p_{\mathcal{H}_\pi} + \nabla \log p_{J},
\end{align}
where $\log p_{\mathcal{H}_l}$, $\log p_{\mathcal{H}_\pi}$ and $\nabla \log p_{J}$ is the gradient of log-likelihood, log of prior and log of Jacobian respectively with respect to the Hausdorff measure. 
For $\emph{K}=1$, the model becomes the circular factor model and we focus on the derivation for $K>1$ in this paper. Let $x_{\beta_{i,1}},\dots,  x_{\beta_{i,K}}$ be the independent random variable and $\bfx_{\beta_{i,K+1}}$ be the dependent random variable. As a result, the gradient associated with the last dimension $x_{\beta_{i,K+1}}$ is 0. The first $K$ components of the gradient from the prior and Jacobian are shown as follows,

\begin{multline}\label{eq:prior_jacobian}
 \nabla \log p_{\mathcal{H}_{\pi}} +  \nabla \log p_{\mathcal{H}_{J}} = 4\omega_K  \begin{bmatrix}
x_1\\
x_2\\
\vdots \\
x_{K}
\end{bmatrix} 
+\omega_{1}x_2(x_1^2+x_2^2)^{-\frac{3}{2}}
\begin{bmatrix}
x_2\\
-x_1\\
0\\
\dots
\end{bmatrix} \\
 + 4
 \begin{bmatrix}
x_1\\
x_2\\
\vdots \\
x_{K}
\end{bmatrix}
\circ   
\Sigma_{\text{-}2}
 \begin{bmatrix}
   \omega_2 x_{3}^2/\left(\sum\limits_{t=1}^{3}{x_t^2}\right)^2\\
 \vdots\\
  \omega_{K-2} x_{K-1}^2/\left(\sum\limits_{t=1}^{K-1}{x_t^2}\right)^2\\
  \omega_{K-1} x_{K}^2/\left(\sum\limits_{t=1}^{K}{x_t^2}\right)^2
\end{bmatrix}
-4\begin{bmatrix}
0\\
0\\
\frac{\omega_{2}x_3} {\sum\limits_{t=1}^{3}{x_t^2}}\\
\vdots\\
\frac{\omega_{K-1}x_{K}} {\sum\limits_{t=1}^{K}{x_t^2}}
\end{bmatrix} 
-  \begin{bmatrix}
x_1\\
x_2\\
\vdots \\
x_{K}
\end{bmatrix}
\circ   \Sigma_{\text{-}1} \begin{bmatrix}
\frac{1}{\sum\limits_{t=1}^{2}{x_t^2}}\\
\frac{1}{\sum\limits_{t=1}^{3}{x_t^2}}\\
\vdots\\
\frac{1}{\sum\limits_{t=1}^{K}{x_t^2}}\\
\end{bmatrix},
\end{multline}
where $\Sigma_{-t}$ is an upper triangular matrix of size $\emph{K}-t$ without the first $t$ columns and all non-zero entries are 1. As a special case when $K = 2$, $\Sigma_{-2}$ becomes a 0 matrix and $\Sigma_{-1}$ becomes scalar 1.
Once the maximum dimension is specified, $\Sigma_{-1}$, $\Sigma_{-2}$ can be generated and all the gradients can then be vectored easily during the implementation without explicitly deriving the expression for different maximum dimension. Recall that, $\bfx_{\beta_i}, \bfx_{\zeta_j}$ and $\bfx_{\psi_j}$ share the same prior and hence the corresponding gradient expression will also be the same.

\subsection{Hausdorff measure used in the Geometric Hamiltonian Monte Carlo and their gradients}\label{ap:Hausdorff}
\subsubsection{Hausdorff measure}
\begin{align*}
p(\bsphi \mid \bsomega)& = \left(\frac{1}{2\pi}\right)^{K} 2^{K-1} \frac{1}{ I_0(\omega_1)} \exp\left\{ \omega_1 \cos \phi_1 \right\}
\prod_{k=2}^{K} \frac{1}{I_0(\omega_k)} \exp\left\{ \omega_k \cos 2\phi_k \right\}\\
 &=\frac{1}{2\pi I_0(\omega_1)} \exp\left\{ \omega_1 \cos \phi_1 \right\}
\left\{\prod_{k=2}^K \frac{1}{\pi  I_o(\omega_k)} \exp\left( \omega_k \cos 2\phi_k \right)\right\}\\
& =\frac{1}{2\pi I_0(\omega_1)} \exp\left\{ \omega_1 \cos \phi_1 \right\}\left\{\prod_{k=2}^K \frac{1}{\pi  I_o(\omega_k)} \exp\left( \omega_k ( 2\cos^2 \phi_{k} -1)\right)\right\}\\
& =\frac{1}{2\pi I_0(\omega_1)} \exp\left\{ \omega_1 \cos \phi_1 \right\}\left\{\prod_{k=2}^K \frac{1}{\pi  I_o(\omega_k)} \right\} \exp\left\{\sum_{k=2}^K \omega_k \left(2\cos^2 \phi_{k} -1 \right) \right\}.
\end{align*}

Using the hyperspherical coordinates in (5) from the main paper, we can express our prior proposed prior in terms of Hausdorff measure, 
\begin{multline*}
p(\bfx \mid \bsomega) = |\bfJ| \times \; p\left(\bsphi = g(\bfx) \mid \bsomega\right)  =  \left\{\frac{1}{\prod_{k=1}^K\sqrt{\sum_{t=1}^{k+1} 
x_{t}^2}} \right\} \frac{1}{2\pi  I_o(\omega_1)} \exp\left\{\omega_1 \frac{x_{1}}{\sqrt{x_{1}^2+x_{2}^2}}\right\} \\
\left\{\prod_{k=2}^K \frac{1}{\pi  I_o(\omega_k)}\right\}\exp\left\{ -\sum_{k=2}^K \omega_k \left(2 \frac{x_{k+1}^2}{ \sum_{t=1}^{k+1} x_{t}^2} -1 \right)\right\} 
,\quad \quad \bfx^T \bfx = 1.
\end{multline*}

\subsubsection{Gradients of the loglikelihood}
Recall from Section 4 from the main paper, the gradients with respect to the Hausdorff measure for $\bfx_{\beta_i}, \bfx_{\zeta_j}, \bfx_{\psi_j}$ are derived under the unit norm constraints in which their first $K$ components are independent variables while the last dimension is dependent. Therefore the gradient for the last dimension is always 0 and the gradients shown in this Section are for the first $K$ dimension.

The gradient of log conditional density of $\bfx_{\beta_i}$ under the Hausdorff measure is,
\begin{align*}
\nabla \log p_{\mathcal{H}}(\bfx_{\beta_i} \mid \dots) = \nabla \log p_{\mathcal{H}_l} + \nabla \log p_{\mathcal{H}_\pi} + \nabla \log p_{J},
\end{align*}
where $\nabla \log p_{\mathcal{H}_\pi} + \nabla \log p_{J}$ is defined in \eqref{eq:prior_jacobian}.
The density of the associated Hausdorff measure with respect to the likelihood component is given by
\begin{align*}
p(\bfx_{\beta_i} \mid \cdots ) =
\prod_{j=1}^{J}  \left [G_{\kappa_j}\left(\{\arccos( \bfx_{\zeta_j}^T \bfx_{\beta_i})\}^2-\{\arccos(\bfx_{\psi_j}^T \bfx_{\beta_i})\}^2\right)\right]^{y_{i,j}} \\
\left [1-G_{\kappa_j}\left(\{\arccos( \bfx_{\zeta_j}^T \bfx_{\beta_i})\}^2-\{\arccos(\bfx_{\psi_j}^T \bfx_{\beta_i})\}^2\right)\right]^{1-y_{i,j}},  \quad \bfx_{\beta_i}^T\bfx_{\beta_i} &= 1 ,
\end{align*}
where $\bfx_{\beta_i} = \left(x_{\beta_i,1},x_{\beta_i,2},\dots,x_{\beta_i,K+1} \right), \bfx_{\zeta_j} = \left(x_{\zeta_j,1},x_{\zeta_j,2},\dots,x_{\zeta_j,K+1} \right), 
\bfx_{\psi_j} = \left(x_{\psi_j,1},x_{\psi_j,2},\dots,x_{\psi_j,K+1} \right) $
Hence, the gradient of the likelihood under the Hausdorff measure is simply
\begin{align*}
\nabla \log_{\mathcal{H}_l}p(\bfx_{\beta_i} \mid \cdots)=  \left(\begin{array}{c}
      \sum_{j=1}^J \left\{y_{i,j} e_{i,j,1}^\prime  \frac{g_{\kappa_j}(e_{i,j})}{G_{\kappa_j}(e_{i,j})}-(1-y_{i,j}) e_{i,j,1}^\prime \frac{g_{\kappa_j}(e_{i,j})}{1-G_{\kappa_j}(e_{i,j})}\right\} \\
      \vdots\\
    \sum_{j=1}^J \left\{y_{i,j} e_{i,j,K}^\prime  \frac{g_{\kappa_j}(e_{i,j})}{G_{\kappa_j}(e_{i,j})}-(1-y_{i,j}) e_{i,j,K}^\prime \frac{g_{\kappa_j}(e_{i,j})}{1-G_{\kappa_j}(e_{i,j})}\right\}
\end{array}\right)
,
\end{align*}
where$$e_{i,j} = \{\arccos( \bfx_{\zeta_j}^T \bfx_{\beta_i})\}^2-\{\arccos(\bfx_{\psi_j}^T \bfx_{\beta_i})\}^2,$$
$$
e_{i,j,t}^\prime=\frac{2\arccos\big(\bfx_{\psi_j}^T \bfx_{\beta_i}\big)}{\sqrt{1-\big(\bfx_{\psi_j}^T \bfx_{\beta_i}\big)^2}}\left(x_{\psi_{j,t}}-x_{\psi_{j,K+1}}\frac{x_{\beta_{i,t}}}{x_{\beta_{i,K+1}}} \right)-\frac{2\arccos\big(\bfx_{\zeta_j}^T \bfx_{\beta_i}\big)}{\sqrt{1-\big(\bfx_{\zeta_j}^T \bfx_{\beta_i}\big)^2}}\left(x_{\zeta_{j,t}}-x_{\zeta_{j,K+1}}\frac{x_{\beta_{i,t}}}{x_{\beta_{i,K+1}}} \right).
$$
Next the gradient of log conditional density of $\bfx_{\zeta_j}$ under the Hausdorff measure is,
\begin{align*}
\nabla \log p_{\mathcal{H}}(\bfx_{\zeta_j} \mid \dots) = \nabla \log p_{\mathcal{H}_l} + \nabla \log p_{\mathcal{H}_\pi} + \nabla \log p_{J},
\end{align*}
where $\nabla \log p_{\mathcal{H}_\pi} + \nabla \log p_{J}$ is defined in \eqref{eq:prior_jacobian}.
The density of the associated Hausdorff measure with respect to the likelihood component is given by
\begin{align*}
p(\bfx_{\zeta_j} \mid \cdots ) =
\prod_{i=1}^{I}  \left [G_{\kappa_j}\left(\{\arccos( \bfx_{\zeta_j}^T \bfx_{\beta_i})\}^2-\{\arccos(\bfx_{\psi_j}^T \bfx_{\beta_i})\}^2\right)\right]^{y_{i,j}} \\
\left [1-G_{\kappa_j}\left(\{\arccos( \bfx_{\zeta_j}^T \bfx_{\beta_i})\}^2-\{\arccos(\bfx_{\psi_j}^T \bfx_{\beta_i})\}^2\right)\right]^{1-y_{i,j}},  \quad \bfx_{\zeta_j}^T\bfx_{\zeta_j} &= 1 ,
\end{align*}
where $\bfx_{\beta_i} = \left(x_{\beta_i,1},x_{\beta_i,2},\dots,x_{\beta_i,K+1} \right), \bfx_{\zeta_j} = \left(x_{\zeta_j,1},x_{\zeta_j,2},\dots,x_{\zeta_j,K+1} \right), 
\bfx_{\psi_j} = \left(x_{\psi_j,1},x_{\psi_j,2},\dots,x_{\psi_j,K+1} \right) $
Hence, the gradient of the likelihood under the Hausdorff measure is simply
\begin{align*}
\nabla \log_{\mathcal{H}_l}p(\bfx_{\zeta_j} \mid \cdots)=  \left(\begin{array}{c}
      \sum_{i=1}^I  \left\{y_{i,j} e_{i,j,1}^\prime  \frac{g_{\kappa_j}(e_{i,j})}{G_{\kappa_j}(e_{i,j})}-(1-y_{i,j}) e_{i,j,1}^\prime \frac{g_{\kappa_j}(e_{i,j})}{1-G_{\kappa_j}(e_{i,j})}\right\} \\
      \vdots\\
    \sum_{i=1}^I  \left\{y_{i,j} e_{i,j,K}^\prime  \frac{g_{\kappa_j}(e_{i,j})}{G_{\kappa_j}(e_{i,j})}-(1-y_{i,j}) e_{i,j,K}^\prime \frac{g_{\kappa_j}(e_{i,j})}{1-G_{\kappa_j}(e_{i,j})}\right\}
\end{array}\right)
,
\end{align*}
where$$e_{i,j} = \{\arccos( \bfx_{\zeta_j}^T \bfx_{\beta_i})\}^2-\{\arccos(\bfx_{\psi_j}^T \bfx_{\beta_i})\}^2,$$
$$
e_{i,j,t}^\prime=-\frac{2\arccos\big(\bfx_{\zeta_j}^T \bfx_{\beta_i}\big)}{\sqrt{1-\big(\bfx_{\zeta_j}^T \bfx_{\beta_i}\big)^2}}\left(x_{\beta_{i,t}}-x_{\beta_{i,K+1}}\frac{x_{\zeta_{j,t}}}{x_{\zeta_{j,K+1}}} \right).
$$
Lastly the gradient of log conditional density of $\bfx_{\psi_j}$ under the Hausdorff measure is,
\begin{align*}
\nabla \log p_{\mathcal{H}}(\bfx_{\psi_j} \mid \dots) = \nabla \log p_{\mathcal{H}_l} + \nabla \log p_{\mathcal{H}_\pi} + \nabla \log p_{J},
\end{align*}
where $\nabla \log p_{\mathcal{H}_\pi} + \nabla \log p_{J}$ is defined in \eqref{eq:prior_jacobian}.
The density of the associated Hausdorff measure with respect to the likelihood component is given by
\begin{align*}
p(\bfx_{\zeta_j} \mid \cdots ) =
\prod_{i=1}^{I}  \left [G_{\kappa_j}\left(\{\arccos( \bfx_{\psi_j}^T \bfx_{\beta_i})\}^2-\{\arccos(\bfx_{\zeta_j}^T \bfx_{\beta_i})\}^2\right)\right]^{y_{i,j}} \\
\left [1-G_{\kappa_j}\left(\{\arccos( \bfx_{\psi_j}^T \bfx_{\beta_i})\}^2-\{\arccos(\bfx_{\zeta_j}^T \bfx_{\beta_i})\}^2\right)\right]^{1-y_{i,j}},  \quad \bfx_{\psi_j}^T\bfx_{\psi_j} &= 1 ,
\end{align*}
where $\bfx_{\beta_i} = \left(x_{\beta_i,1},x_{\beta_i,2},\dots,x_{\beta_i,K+1} \right), \bfx_{\zeta_j} = \left(x_{\zeta_j,1},x_{\zeta_j,2},\dots,x_{\zeta_j,K+1} \right), 
\bfx_{\psi_j} = \left(x_{\psi_j,1},x_{\psi_j,2},\dots,x_{\psi_j,K+1} \right) $
Hence, the gradient of the likelihood under the Hausdorff measure is simply
\begin{align*}
\nabla \log_{\mathcal{H}_l}p(\bfx_{\psi_j} \mid \cdots)=  \left(\begin{array}{c}
      \sum_{i=1}^I  \left\{ y_{i,j} e_{i,j,1}^\prime  \frac{g_{\kappa_j}(e_{i,j})}{G_{\kappa_j}(e_{i,j})}-(1-y_{i,j}) e_{i,j,1}^\prime \frac{g_{\kappa_j}(e_{i,j})}{1-G_{\kappa_j}(e_{i,j})} \right \}\\
      \vdots\\
     \sum_{i=1}^I  \left\{y_{i,j} e_{i,j,K}^\prime  \frac{g_{\kappa_j}(e_{i,j})}{G_{\kappa_j}(e_{i,j})}-(1-y_{i,j}) e_{i,j,K}^\prime \frac{g_{\kappa_j}(e_{i,j})}{1-G_{\kappa_j}(e_{i,j})}\right \}
\end{array}\right)
,
\end{align*}
where$$e_{i,j} = \{\arccos( \bfx_{\zeta_j}^T \bfx_{\beta_i})\}^2-\{\arccos(\bfx_{\psi_j}^T \bfx_{\beta_i})\}^2,$$
$$
e_{i,j,t}^\prime=\frac{2\arccos\big(\bfx_{\psi_j}^T \bfx_{\beta_i}\big)}{\sqrt{1-\big(\bfx_{\psi_j}^T \bfx_{\beta_i}\big)^2}}\left(x_{\beta_{i,t}}-x_{\beta_{i,K+1}}\frac{x_{\psi_{j,t}}}{x_{\psi_{j,K+1}}} \right).
$$

\bibliographystyle{bka}
\bibliography{paper2}

\begin{thebibliography}{50}
\providecommand{\natexlab}[1]{#1}
\expandafter\ifx\csname urlstyle\endcsname\relax
  \providecommand{\doi}[1]{doi:\discretionary{}{}{}#1}\else
  \providecommand{\doi}{doi:\discretionary{}{}{}\begingroup
  \urlstyle{rm}\Url}\fi

\bibitem[{Albert \& Chib(1993)}]{albert1993bayesian}
\textsc{Albert, J.~H.} \& \textsc{Chib, S.} (1993).
\newblock Bayesian analysis of binary and polychotomous response data.
\newblock \emph{Journal of the American statistical Association} \textbf{88},
  669--679.

\bibitem[{Ansari \& Jedidi(2000)}]{ansari2000bayesian}
\textsc{Ansari, A.} \& \textsc{Jedidi, K.} (2000).
\newblock Bayesian factor analysis for multilevel binary observations.
\newblock \emph{Psychometrika} \textbf{65}, 475--496.

\bibitem[{Bafumi et~al.(2005)Bafumi, Gelman, Park \& Kaplan}]{bafumi13andrew}
\textsc{Bafumi, J.}, \textsc{Gelman, A.}, \textsc{Park, D.~K.} \&
  \textsc{Kaplan, N.} (2005).
\newblock Practical issues in implementing and understanding bayesian ideal
  point estimation.
\newblock \emph{Political Analysis} \textbf{13}, 171--87.

\bibitem[{Bartholomew(1984)}]{bartholomew1984scaling}
\textsc{Bartholomew, D.~J.} (1984).
\newblock Scaling binary data using a factor model.
\newblock \emph{Journal of the Royal Statistical Society: Series B
  (Methodological)} \textbf{46}, 120--123.

\bibitem[{Bartlett(1957)}]{bartlett1957comment}
\textsc{Bartlett, M.~S.} (1957).
\newblock Comment on ‘a statistical paradox’by dv lindley.
\newblock \emph{Biometrika} \textbf{44}, 533--534.

\bibitem[{Belkin \& Niyogi(2002)}]{belkin2002laplacian}
\textsc{Belkin, M.} \& \textsc{Niyogi, P.} (2002).
\newblock Laplacian eigenmaps and spectral techniques for embedding and
  clustering.
\newblock In \emph{Advances in neural information processing systems}, pp.
  585--591.

\bibitem[{Beskos et~al.(2013)Beskos, Pillai, Roberts, Sanz-Serna, Stuart
  et~al.}]{beskos2013optimal}
\textsc{Beskos, A.}, \textsc{Pillai, N.}, \textsc{Roberts, G.},
  \textsc{Sanz-Serna, J.-M.}, \textsc{Stuart, A.} et~al. (2013).
\newblock Optimal tuning of the hybrid monte carlo algorithm.
\newblock \emph{Bernoulli} \textbf{19}, 1501--1534.

\bibitem[{Betancourt et~al.(2014)Betancourt, Byrne \&
  Girolami}]{betancourt2014optimizing}
\textsc{Betancourt, M.}, \textsc{Byrne, S.} \& \textsc{Girolami, M.} (2014).
\newblock Optimizing the integrator step size for hamiltonian monte carlo.
\newblock \emph{arXiv preprint arXiv:1411.6669} .

\bibitem[{Byrne \& Girolami(2013)}]{byrne2013geodesic}
\textsc{Byrne, S.} \& \textsc{Girolami, M.} (2013).
\newblock Geodesic monte carlo on embedded manifolds.
\newblock \emph{Scandinavian Journal of Statistics} \textbf{40}, 825--845.

\bibitem[{Child(2006)}]{child2006essentials}
\textsc{Child, D.} (2006).
\newblock \emph{The essentials of factor analysis}.
\newblock Bloomsbury Academic, 3rd edition.

\bibitem[{Clinton et~al.(2004)Clinton, Jackman \&
  Rivers}]{clinton2004statistical}
\textsc{Clinton, J.}, \textsc{Jackman, S.} \& \textsc{Rivers, D.} (2004).
\newblock The statistical analysis of roll call data.
\newblock \emph{American Political Science Review} \textbf{98}, 355--370.

\bibitem[{Davidson et~al.(2018)Davidson, Falorsi, De~Cao, Kipf \&
  Tomczak}]{davidson2018hyperspherical}
\textsc{Davidson, T.~R.}, \textsc{Falorsi, L.}, \textsc{De~Cao, N.},
  \textsc{Kipf, T.} \& \textsc{Tomczak, J.~M.} (2018).
\newblock Hyperspherical variational auto-encoders.
\newblock \emph{arXiv preprint arXiv:1804.00891} .

\bibitem[{Dryden(2005)}]{dryden2005}
\textsc{Dryden, I.~L.} (2005).
\newblock Statistical analysis on high-dimensional spheres and shape spaces.
\newblock \emph{Ann. Statist.} \textbf{33}, 1643--1665.
\newblock \doi{10.1214/009053605000000264}.

\bibitem[{Fletcher et~al.(2004)Fletcher, Lu, Pizer \&
  Joshi}]{fletcher2004principal}
\textsc{Fletcher, P.~T.}, \textsc{Lu, C.}, \textsc{Pizer, S.~M.} \&
  \textsc{Joshi, S.} (2004).
\newblock Principal geodesic analysis for the study of nonlinear statistics of
  shape.
\newblock \emph{IEEE transactions on medical imaging} \textbf{23}, 995--1005.

\bibitem[{Gelman et~al.(2013)Gelman, Carlin, Stern, Dunson, Vehtari \&
  Rubin}]{gelman2013bayesian}
\textsc{Gelman, A.}, \textsc{Carlin, J.~B.}, \textsc{Stern, H.~S.},
  \textsc{Dunson, D.~B.}, \textsc{Vehtari, A.} \& \textsc{Rubin, D.~B.} (2013).
\newblock \emph{Bayesian data analysis}.
\newblock CRC press.

\bibitem[{Gelman et~al.(2014)Gelman, Hwang \&
  Vehtari}]{gelman-2014-information}
\textsc{Gelman, A.}, \textsc{Hwang, J.} \& \textsc{Vehtari, A.} (2014).
\newblock Understanding predictive information criteria for {B}ayesian models.
\newblock \emph{Statistics and Computing} \textbf{24}, 997--1016.

\bibitem[{Gelman \& Rubin(1992)}]{GeRu92}
\textsc{Gelman, A.} \& \textsc{Rubin, D.} (1992).
\newblock Inferences from iterative simulation using multiple sequences.
\newblock \emph{Statistical Science} \textbf{7}, 457--472.

\bibitem[{Geweke \& Singleton(1981)}]{geweke1981maximum}
\textsc{Geweke, J.~F.} \& \textsc{Singleton, K.~J.} (1981).
\newblock Maximum likelihood" confirmatory" factor analysis of economic time
  series.
\newblock \emph{International Economic Review} pp. 37--54.

\bibitem[{Griffiths \& Ghahramani(2011)}]{griffiths2011indian}
\textsc{Griffiths, T.~L.} \& \textsc{Ghahramani, Z.} (2011).
\newblock The indian buffet process: An introduction and review.
\newblock \emph{Journal of Machine Learning Research} \textbf{12}.

\bibitem[{Hare \& Poole(2014)}]{hare2014polarization}
\textsc{Hare, C.} \& \textsc{Poole, K.~T.} (2014).
\newblock The polarization of contemporary american politics.
\newblock \emph{Polity} \textbf{46}, 411--429.

\bibitem[{Hoff et~al.(2002)Hoff, Raftery \& Handcock}]{hoff-2002}
\textsc{Hoff, P.~D.}, \textsc{Raftery, A.~E.} \& \textsc{Handcock, M.~S.}
  (2002).
\newblock Latent space approaches to social network analysis.
\newblock \emph{Journal of the american Statistical association} \textbf{97},
  1090--1098.

\bibitem[{Jackman(2001)}]{jackman2001multidimensional}
\textsc{Jackman, S.} (2001).
\newblock Multidimensional analysis of roll call data via bayesian simulation:
  Identification, estimation, inference, and model checking.
\newblock \emph{Political Analysis} \textbf{9}, 227--241.

\bibitem[{Jessee(2012)}]{jessee2012ideology}
\textsc{Jessee, S.~A.} (2012).
\newblock \emph{Ideology and spatial voting in American elections}.
\newblock Cambridge University Press.

\bibitem[{Jung et~al.(2012)Jung, Dryden \& Marron}]{jung2012analysis}
\textsc{Jung, S.}, \textsc{Dryden, I.~L.} \& \textsc{Marron, J.} (2012).
\newblock Analysis of principal nested spheres.
\newblock \emph{Biometrika} \textbf{99}, 551--568.

\bibitem[{Kingma et~al.(2019)Kingma, Welling et~al.}]{kingma2019introduction}
\textsc{Kingma, D.~P.}, \textsc{Welling, M.} et~al. (2019).
\newblock An introduction to variational autoencoders.
\newblock \emph{Foundations and Trends{\textregistered} in Machine Learning}
  \textbf{12}, 307--392.

\bibitem[{Kramer(1991)}]{kramer1991nonlinear}
\textsc{Kramer, M.~A.} (1991).
\newblock Nonlinear principal component analysis using autoassociative neural
  networks.
\newblock \emph{AIChE journal} \textbf{37}, 233--243.

\bibitem[{Lawrence(2005)}]{lawrence2005probabilistic}
\textsc{Lawrence, N.} (2005).
\newblock Probabilistic non-linear principal component analysis with gaussian
  process latent variable models.
\newblock \emph{Journal of machine learning research} \textbf{6}, 1783--1816.

\bibitem[{Legler et~al.(1997)Legler, Ryan \& Ryan}]{legler1997latent}
\textsc{Legler, J.~M.}, \textsc{Ryan, L.~M.} \& \textsc{Ryan, L.~M.} (1997).
\newblock Latent variable models for teratogenesis using multiple binary
  outcomes.
\newblock \emph{Journal of the American Statistical Association} \textbf{92},
  13--20.

\bibitem[{Lewis(2019{\natexlab{a}})}]{Lewis2019b}
\textsc{Lewis, J.} (2019{\natexlab{a}}).
\newblock Why are {Ocasio-Cortez, Omar, Pressley, and Talib} estimated to be
  moderates by {NOMINATE}?
\newblock
  \url{https://voteview.com/articles/Ocasio-Cortez_Omar_Pressley_Tlaib}.

\bibitem[{Lewis(2019{\natexlab{b}})}]{Lewis2019a}
\textsc{Lewis, J.} (2019{\natexlab{b}}).
\newblock Why is {Alexandria Ocasio-Cortez} estimated to be a moderate by
  {NOMINATE}?
\newblock \url{https://voteview.com/articles/ocasio_cortez}.

\bibitem[{Liu \& Wu(1999)}]{liu1999parameter}
\textsc{Liu, J.~S.} \& \textsc{Wu, Y.~N.} (1999).
\newblock Parameter expansion for data augmentation.
\newblock \emph{Journal of the American Statistical Association} \textbf{94},
  1264--1274.

\bibitem[{Marschak(1960)}]{marschak1960binary}
\textsc{Marschak, J.} (1960).
\newblock Binary-choice constraints and random utility indicators.
\newblock In \emph{Stanford Symposium on Mathematical Methods in the Social
  Sciences}. Stanford, CA: Stanford University Press.

\bibitem[{McFadden(1973)}]{mcfadden1973conditional}
\textsc{McFadden, D.} (1973).
\newblock Conditional logit analysis of qualitative choice behavior.
\newblock In \emph{Frontiers of Economics}, Ed. P.~Zarembka, pp. 105--142.
  Institute of Urban and Regional Development, University of California.

\bibitem[{Moser et~al.(2020)Moser, Rodriguez \& Lofland}]{moser2020multiple}
\textsc{Moser, S.}, \textsc{Rodriguez, A.} \& \textsc{Lofland, C.~L.} (2020).
\newblock Multiple ideal points: Revealed preferences in different domains.
\newblock \emph{Political Analysis} \textbf{Forthcoming}.

\bibitem[{Mulaik(2009)}]{mulaik2009foundations}
\textsc{Mulaik, S.~A.} (2009).
\newblock \emph{Foundations of factor analysis}.
\newblock CRC press, 2nd edition.

\bibitem[{Pierre(2002)}]{pierre2002siecle}
\textsc{Pierre, F.~J.} (2002).
\newblock Le si{\`e}cle des id{\'e}ologies.

\bibitem[{Poole \& Rosenthal(2000)}]{poole2000congress}
\textsc{Poole, K.~T.} \& \textsc{Rosenthal, H.} (2000).
\newblock \emph{Congress: A political-economic history of roll call voting}.
\newblock Oxford University Press on Demand.

\bibitem[{Ragusa \& Gaspar(2016)}]{ragusa2016s}
\textsc{Ragusa, J.~M.} \& \textsc{Gaspar, A.} (2016).
\newblock Where’s the tea party? an examination of the tea party’s voting
  behavior in the house of representatives.
\newblock \emph{Political Research Quarterly} \textbf{69}, 361--372.

\bibitem[{Rodr{\'\i}guez \& Moser(2015)}]{rodriguez2015measuring}
\textsc{Rodr{\'\i}guez, A.} \& \textsc{Moser, S.} (2015).
\newblock Measuring and accounting for strategic abstentions in the us senate,
  1989--2012.
\newblock \emph{Journal of the Royal Statistical Society: Series C (Applied
  Statistics)} \textbf{64}, 779--797.

\bibitem[{Roweis \& Saul(2000)}]{roweis2000nonlinear}
\textsc{Roweis, S.~T.} \& \textsc{Saul, L.~K.} (2000).
\newblock Nonlinear dimensionality reduction by locally linear embedding.
\newblock \emph{science} \textbf{290}, 2323--2326.

\bibitem[{Rubin \& Thomas(2001)}]{rubin2001using}
\textsc{Rubin, D.~B.} \& \textsc{Thomas, N.} (2001).
\newblock Using parameter expansion to improve the performance of the em
  algorithm for multidimensional irt population-survey models.
\newblock In \emph{Essays on item response theory}, pp. 193--204. Springer.

\bibitem[{Sommer et~al.(2010)Sommer, Lauze \& Nielsen}]{sommer2010differential}
\textsc{Sommer, S.}, \textsc{Lauze, F.} \& \textsc{Nielsen, M.} (2010).
\newblock The differential of the exponential map, jacobi fields and exact
  principal geodesic analysis.
\newblock \emph{CoRR, abs/1008.1902} .

\bibitem[{Spiegelhalter et~al.(2014)Spiegelhalter, Best, Carlin \& Van~der
  Linde}]{spiegelhalter2014deviance}
\textsc{Spiegelhalter, D.~J.}, \textsc{Best, N.~G.}, \textsc{Carlin, B.~P.} \&
  \textsc{Van~der Linde, A.} (2014).
\newblock The deviance information criterion: 12 years on.
\newblock \emph{Journal of the Royal Statistical Society: Series B (Statistical
  Methodology)} \textbf{76}, 485--493.

\bibitem[{Spiegelhalter et~al.(2002)Spiegelhalter, Best, Carlin \& Van
  Der~Linde}]{spiegelhalter2002bayesian}
\textsc{Spiegelhalter, D.~J.}, \textsc{Best, N.~G.}, \textsc{Carlin, B.~P.} \&
  \textsc{Van Der~Linde, A.} (2002).
\newblock Bayesian measures of model complexity and fit.
\newblock \emph{Journal of the royal statistical society: Series b (statistical
  methodology)} \textbf{64}, 583--639.

\bibitem[{Taylor(2006)}]{taylor2006did}
\textsc{Taylor, J.} (2006).
\newblock \emph{Where did the party go?: William Jennings Bryan, Hubert
  Humphrey, and the Jeffersonian legacy}.
\newblock University of Missouri Press.

\bibitem[{Tenenbaum et~al.(2000)Tenenbaum, De~Silva \&
  Langford}]{tenenbaum2000global}
\textsc{Tenenbaum, J.~B.}, \textsc{De~Silva, V.} \& \textsc{Langford, J.~C.}
  (2000).
\newblock A global geometric framework for nonlinear dimensionality reduction.
\newblock \emph{Science} \textbf{290}, 2319--2323.

\bibitem[{Weisberg(1974)}]{weisberg1974dimensionland}
\textsc{Weisberg, H.~F.} (1974).
\newblock Dimensionland: An excursion into spaces.
\newblock \emph{American Journal of Political Science} pp. 743--776.

\bibitem[{Xu \& Durrett(2018)}]{xu2018spherical}
\textsc{Xu, J.} \& \textsc{Durrett, G.} (2018).
\newblock Spherical latent spaces for stable variational autoencoders.
\newblock \emph{arXiv preprint arXiv:1808.10805} .

\bibitem[{Zhang \& Fletcher(2013)}]{NIPS2013_5133}
\textsc{Zhang, M.} \& \textsc{Fletcher, T.} (2013).
\newblock Probabilistic principal geodesic analysis.
\newblock In \emph{Advances in Neural Information Processing Systems 26}, Eds.
  C.~J.~C. Burges, L.~Bottou, M.~Welling, Z.~Ghahramani \& K.~Q. Weinberger,
  pp. 1178--1186. Curran Associates, Inc.

\bibitem[{Zhang \& Zha(2004)}]{Zhang04principalmanifolds}
\textsc{Zhang, Z.} \& \textsc{Zha, H.} (2004).
\newblock Principal manifolds and nonlinear dimensionality reduction via
  tangent space alignment.
\newblock \emph{SIAM Journal on Scientific Computing} pp. 313--338.

\end{thebibliography}
\end{document}